\documentclass[%
reprint,
aps,
superscriptaddress,
nofootinbib,
nobibnotes,
]{revtex4-1}
\usepackage{color,amsmath,amssymb,graphicx,latexsym,subfigure}
\usepackage{threeparttable,multirow,txfonts,makecell,tabularx}
\DeclareMathAlphabet{\mathcal}{OMS}{cmsy}{m}{n}
\usepackage{appendix}
\usepackage{hyperref}
\usepackage{verbatim}
\newcommand{\nc}{\newcommand}
\nc{\beqa}{\begin{eqnarray}}  
\nc{\eeqa}{\end{eqnarray}}

\begin{document}

\title{Cosmic Simulations of Axion String-Wall Networks: Probing Dark Matter and Gravitational Waves for Discovery}

\author{Yang Li}
\affiliation{
CAS Key Laboratory of Theoretical Physics, Institute of Theoretical Physics, Chinese Academy of Sciences, Beijing 100190, China}
\affiliation{School of Physical Sciences, University of Chinese Academy of Sciences, No. 19A Yuquan Road, Beijing 100049, China}

\author{ Ligong Bian }\email{lgbycl@cqu.edu.cn}

\affiliation{Department of Physics and Chongqing Key Laboratory for Strongly Coupled Physics, Chongqing University, Chongqing 401331, P. R. China}
\affiliation{ Center for High Energy Physics, Peking University, Beijing 100871, China}

\author{Rong-Gen Cai}\email{cairg@itp.ac.cn}
\affiliation{
CAS Key Laboratory of Theoretical Physics, Institute of Theoretical Physics, Chinese Academy of Sciences, Beijing 100190, China}
\affiliation{
School of Physical Science and Technology, Ningbo University, Ningbo 315211, China }
\affiliation{School of Fundamental Physics and Mathematical Sciences
Hangzhou Institute for Advanced Study, University of Chinese Academy of Sciences, Hangzhou 310024, China}

\author{Jing Shu}\email{jshu@pku.edu.cn}
\affiliation{School of Physics and State Key Laboratory of Nuclear Physics and Technology, Peking University, Beijing 100871, China}
\affiliation{ Center for High Energy Physics, Peking University, Beijing 100871, China}
\affiliation{Beijing Laser Acceleration Innovation Center, Huairou, Beijing, 101400, China}

\begin{abstract}

We simultaneously study gravitational waves (GWs) and free axions emitted from axionic string-wall networks in the early universe using advanced 3D lattice simulations. Our simulations start before the Peccei-Quinn phase transition and end with the destruction of string-wall networks below the QCD scale. The axion dark matter (DM) relic abundance radiated from string-wall networks are updated and refined for the scenarios of $N_{\rm DW}>1$. In this scenario, we observe that the GW spectrum is almost independent of the bias term and $N_{\rm DW}$, and $\Omega_{\rm GW}h^2\propto f^{1.29}(f^{-0.43})$ in the IR and middle-frequency regions. After considering the constraints from DM relic abundance, we found that the QCD axion model predicts undetectable GW emissions, and the axion-like particles model allows for a detectable GW signal in the nano-Hertz to the milli-Hertz frequency range corresponding to axion masses range from KeV to TeV. For $N_{\rm DW}=1$, the GW energy density appears undetectable for QCD axions and axion-like particles.

\end{abstract}

\maketitle

\noindent{\it \bfseries  Introduction.}
Axions are ultralight pseudo-scalar particles realized by the Peccei-Quinn (PQ) mechanism originally introduced to solve the Strong CP problem in quantum chromodynamics (QCD)~\cite{Peccei:1977hh,Peccei:1977ur,Wilczek:1977pj}. Afterward, the QCD axion and more generally the axion-like particles (ALPs), as dark matter (DM) candidates gave rise to widespread interest~\cite{Weinberg:1977ma, Abbott:1982af, Dine:1982ah, Preskill:1982cy}. Moreover, the post-inflationary PQ symmetry-breaking scenario with accompanying axion strings and domain walls (DWs), will radiate axions~\cite{Kawasaki:2014sqa} and have an impact on the form of axion relic abundance as well as the cosmic structure formation~\cite{Hogan:1988mp,Kolb:1993zz,Kolb:1993hw,Kolb:1994fi,Zurek:2006sy,Enander:2017ogx,Vaquero:2018tib,Buschmann:2019icd,Chang:2023rll}. Besides, cosmic strings and DWs are known to be the stochastic gravitational wave backgrounds (SGWBs) sources that are important scientific goals of many gravitational wave (GW) detection experiments, such as pulsar timing array (PTA)\cite{Blasi:2020mfx,Samanta:2020cdk,Ellis:2020ena, Bian:2022tju,Gouttenoire:2023ftk,Gouttenoire:2023gbn,NANOGrav:2023hvm,Bian:2020urb,Bian:2023dnv,
Bian:2022qbh,Ferreira:2022zzo}, LISA\cite{Auclair:2023brk,Boileau:2021gbr}, and LIGO~\cite{LIGOScientific:2021nrg}. The thorough studies on those subjects may provide viable search avenues for axion models~\cite{Marsh:2015xka,Borsanyi:2015cka,Hlozek:2014lca,Kawasaki:2013ae,Chang:2019mza,Chang:2021afa,LISACosmologyWorkingGroup:2022jok,Brzeminski:2022haa,Agrawal:2020euj,Jain:2021shf,Jain:2022jrp,Agrawal:2019lkr,Dessert:2021bkv}, given that the direct experimental detection~\cite{Shokair:2014rna,ADMX:2018gho,Brubaker:2016ktl,AlKenany:2016trt,Brubaker:2017rna,Caldwell:2016dcw,Kahn:2016aff,Ouellet:2018beu,Ouellet:2019tlz,Chaudhuri:2014dla,Silva-Feaver:2016qhh} based on the axion particle per se is extremely difficult.

After the spontaneous breaking of the $U(1)_{\rm PQ}$ symmetry, axion strings form and eventually evolve into the scaling regime~\cite{Kibble:1976sj,Shellard:1998mi,Sikivie:2006ni,Kim:2008hd,Yamaguchi:1998gx,Yamaguchi:2002sh,Hiramatsu:2010yu}, and is expected to generate a SGWB with a very wide and nearly scale-invariant GW spectrum~\cite{Gouttenoire:2019kij,Auclair:2019wcv}. As the universe cools, the $U(1)$ symmetry will be explicitly broken at a lower scale, and axion acquires a mass through non-perturbative effects~\cite{Dine:1981rt,Zhitnitsky:1980tq,Kim:1986ax}. Meanwhile, DWs bounded by strings are formed. The fate of the string-wall system depends on the specific axion model and is characterized by the DW number $N_{\rm DW}$. In the KSVZ model~\cite{Kim:1979if,Shifman:1979if} with $N_{\rm DW}=1$, only one DW is attached to each string. The DWs are short-lived and quickly annihilate when $H(T_{\rm dec}) \simeq m$ is satisfied, thus strings is generally expected to dominate the SGWB of the system. 
In the $N_{\rm DW}>1$ case realized by DFSZ model~\cite{Dine:1981rt,Zhitnitsky:1980tq}, $N_{\rm DW}$ walls attached to a string, balance the tension among themselves and lead to stable DWs. However, the stable DWs would eventually dominate the universe and therefore conflict with cosmological observations~\cite{Sikivie:1982qv,Zeldovich:1974uw}. The conventional way to solve this DW problem is to introduce a bias term~\cite{Gelmini:1988sf,Larsson:1996sp} to explicitly break the $Z_{N_{\rm DW}}$ discrete symmetry so that the walls will annihilate under the pressure differences between different vacua before they overclose the universe. The bias term may originate from the fundamental physics at Planck scale~\cite{Kamionkowski:1992mf,Holman:1992us,Dobrescu:1996jp,Barr:1992qq,Dine:1992vx}, and should lead to long-lived DWs due to limitations from the degree of CP violation~\cite{Barr:1992qq}. The long-lived DWs will generate an SGWB which is expect to be much stronger than that of short-lived DWs, which may have sufficient observable amplitudes~\cite{Hiramatsu:2010yn,Hiramatsu:2012sc}. 


To investigate the axion dark matter (DM) possibility from string-wall networks, there are several semi-analytical and numerical studies with $N_{\rm DW}=1$~\cite{Hiramatsu:2012gg,Kawasaki:2014sqa, Buschmann:2019icd} and $N_{\rm DW}>1$~\cite{Saikawa:2017hiv, Hiramatsu:2010yn, Gelmini:2021yzu, Hiramatsu:2012sc, Chang:2023rll, Kawasaki:2014sqa}.
For $N_{\rm DW}>1$, previous 2D lattice simulations investigated the dynamics of the string-wall system including the bias term, where the corresponding GW spectrum is not measured~\cite{Hiramatsu:2010yn,Kawasaki:2014sqa}. Ref.~\cite{Hiramatsu:2012sc} performed 3D lattice simulations to obtain the spectra of free axions and GWs radiated by long-lived DWs, however, the bias term has not been considered. 

In this Letter, we simultaneously study the axion DM and SGWBs produced by the axion string-wall networks. We perform high-resolution 3D lattice simulations for both the $N_{\rm DW}=1$ case and the $N_{\rm DW}>1$ case. Our simulation includes the string formation and evolution in the PQ phase transition era and the destruction of the string-wall network in the QCD era below the confinement scale. To the best of our knowledge, this is the first two-stage simulation on the $N_{\rm DW}>1$ scenarios.
Our study provides the axion DM relic abundance for $N_{\rm DW}>1$ and show that it's possible to probe or exclude axion models with current and forthcoming GW detectors.

\noindent{\it \bfseries  The simulation framework.}
Consider the Lagrangian of the PQ complex scalar field $\varphi$,
\begin{flalign}{
\mathcal{L}=& \frac{1}{2}\partial_\mu\varphi^*\partial^\mu\varphi-\frac{1}{4}\lambda(|\varphi|^2-v^2)^2-\frac{\lambda}{6}T^2|\varphi|^2 \nonumber \\ \hspace{1.8mm} &-\frac{m^2(T)v^2}{N_{\rm DW}^2}(1-{\rm cos}(N_{\rm DW}\theta))+\Xi v^3(\varphi e^{-i\delta}+{\rm h.c.}), \label{full lagrangian}
}\end{flalign}
with $\lambda$ the PQ coupling constant, $v$ the PQ vacuum, $m(T)$ the temperature-dependent QCD axion mass, $\theta={\rm Arg}(\varphi)$ the dimensionless axion field ($a=\theta \times f_a$ is the axion field), $\Xi$ the coefficient of bias term. For the bias term, $\delta$ only determines the location of the true minimum of the potential in numerical simulations, so we fix $\delta=0$ for all of our simulations which will not affect the quantitative results~\cite{Hiramatsu:2010yn}.
When $T>100$ ${\rm MeV}$, $m(T)$ can be parametrized as~\cite{Wantz:2009it}
\begin{equation}
m(T)^2 = \frac{\alpha_a \Lambda^4}{f_a^2 (T / \Lambda)^{6.68}},
\end{equation}
where $\alpha_a = 1.68 \times 10^{-7}$, $\Lambda = 400 \, \mathrm{MeV}$ and $f_a = v/N_{\rm DW}$. When $T\le100$ ${\rm MeV}$, the axion mass no longer grows and is substituted by a zero-temperature mass $m_0 = 5.707 \times 10^{-5} ( 10^{11} \, \, {\rm GeV}/ f_a)$ eV~\cite{GrillidiCortona:2015jxo}. Without loss of universality, we take the QCD axion scenario as a benchmark to perform simulations, and with our careful simulation setup, the numerical results can be applied to the ALPs with constant mass $m=m_0$ and the same other model parameters.

Our simulation is divided into two stages, namely the PQ era and the QCD era, which correspond to the spontaneous breaking of PQ symmetry near the scale of $T\sim v$~\cite{Kim:1979if,Shifman:1979if,Dine:1981rt,Zhitnitsky:1980tq,Srednicki:1985xd} and the explicit breaking near the QCD scale, respectively. In the PQ era, the second line in Eq.~(\ref{full lagrangian}) caused by QCD effects can be neglected. The critical temperature of the PQ phase transition is $T_{\rm c}=\sqrt{3}v$. Our simulation begins at an initial temperature $T_{i}=2.4T_{\rm c}$, and we set $v=2\times10^{17}$ ${\rm GeV}$ to ensure the simulation box can capture enough Hubble volumes. The initial configuration of $\varphi$ can be obtained by thermal spectrum. The fields are rescaled with the PQ vacuum expectation value $v$, while the comoving lattice spacing $\delta x$ (and time-step) are rescaled by $w_*=R_iH_i$ with $R_i$ ($H_i$) being initial scale factor (Hubble parameter). For convenience, we use the rescaled conformal time $\tilde{\eta}$ $(=\eta/\eta_i)$ with the initial conformal time being $\eta_i=1/(R_iH_i)$ in our simulations. We fix $\tilde{\eta}=1$ to be the initial time at which $T=T_{i}$, and set $\lambda = 0.2$ here for concreteness. The EQUATIONS OF MOTION and INITIAL CONDITIONS for the two-stage lattice simulations are presented in the {\it Supplemental Material}. 

\begin{figure}[!htp]
\includegraphics[width=.4\textwidth]{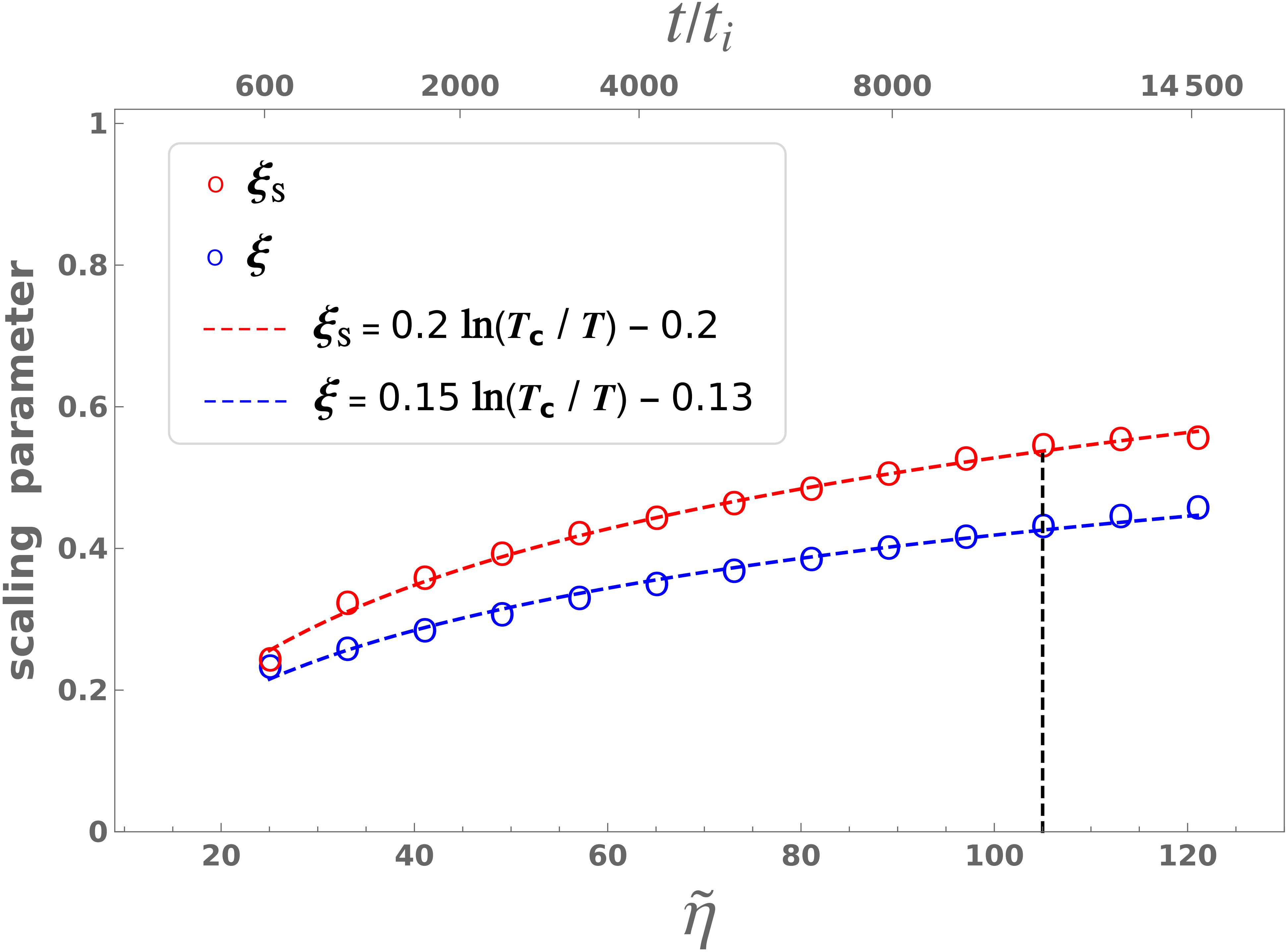} 
  \caption{Evolution of scaling parameters with rescaled conformal time $\tilde{\eta}$ and rescaled cosmic time $\tilde{t}=t/t_i$, with $t_i=1/(2H_i)$ the initial cosmic time. The red and blue dashed lines represent the logarithmic fit of the scaling parameters in two measurement schemes. The black vertical dashed line represents the time we selected as the final time of the PQ era.}  \label{fig:SCinPQprepareforQCD}
\end{figure}

The comoving side length of the simulation box and the number of grid points per side are set to $L_{\rm PQ}=1680/(R_{i}H_{i})$ and $N=2200$. So, the dimensionless comoving lattice spacing is about $\delta \Tilde{x}=0.764$, and the time-step is chosen as $\delta \tilde{\eta}=0.02$. We use the second-order leap-frog algorithm adopted by ${\mathcal CosmoLattice}$~\cite{Figueroa:2021yhd,Figueroa:2020rrl} to evolve the EOM. After the PQ phase transition, the axion string network forms and evolves towards a scaling regime. We continue to evolve the string network until the final moment $\tilde{\eta}_{f}=105$ (for the convenience of calculating scaling parameter near $\tilde{\eta}_{f}=105$, we evolve the string network to $\tilde{\eta}=121$), see FIG.~\ref{fig:SCinPQprepareforQCD}. At $\tilde{\eta}_{f}=105$, the simulation box contains $16^3$ Hubble volumes.

For simulations in the QCD era, it can be extrapolated that the scaling regime of axion string networks in the PQ era has been maintained until the QCD scale~\cite{Hindmarsh:2019csc}. Therefore, we take the field configurations at the final moment of the PQ era as the initial conditions, while keeping the same number of Hubble volumes in the simulation box. In this way, we maintain the correct number density of axion strings per Hubble volume and the status of modes relative to the horizon.

Our simulation in the QCD era starts slightly before the QCD phase transition, with an initial temperature $T_{1}=180$ ${\rm MeV}$. We reinterpret the comoving side length of the simulation box as $L_{\rm QCD}=16/(R_{1}H_{1})$, with $R_{1}$ ($H_{1}$) the initial scale factor (Hubble parameter). Therefore, the simulation box contains $16^3$ Hubble volumes identical to that at the final moment of the PQ era. Due to the scaling property of the axion string network, different PQ symmetry-breaking scales will finally lead to the same dimensionless field configurations after the string networks enter the scaling regime. So we can reinterpret the PQ vacuum as $v_{\rm re}=N_{\rm DW}\cdot f_{a}$ for different $N_{\rm DW}$, with fixed decay constant $f_{ a}=3.5\times 10^{16}$ ${\rm GeV}$ in simulations to ensure that the DW forms shortly before the QCD phase transition. Thus, our simulation will encompass both the QCD phase transition and the state where the axion achieves zero temperature mass for a long time, this will help us to apply the simulation results to ALPs with constant mass.  

We use the rescaled conformal time $\hat{\eta}$ $(=\eta / \eta_1=R/R_1)$ to perform time iteration, with the initial conformal time $\eta_1=1/(R_1H_1)$. Furthermore, we take into account the changes in the number of relativistic degrees of freedom ($g_*(T)$) near the QCD scale~\cite{Wantz:2009it}, which will affect the evolution of the scale factor and the measurement of spectra. The fields are rescaled with the reinterpreted PQ vacuum $v_{\rm re}$, and the comoving lattice spacing (and time-step) are rescaled with $\hat{w}=6R_1 H_1$. So, the dimensionless comoving lattice spacing is $\delta \hat{x} \approx 0.0436$, and the time-step is chosen as $\delta \hat{\eta}=0.0076$. 

To resolve the resolution of the string throughout our simulation, we adjust the string width by defining the parameter $\hat{\lambda} \equiv \lambda v_{\rm re}^2/m(\hat{\eta}_{\rm e})^2$, where $\hat{\eta}_{\rm e}\approx 1.03$ is the time at which $m=H$ is satisfied~\cite{Buschmann:2019icd}. At the initial time, we set $\hat{\lambda}=1300$ so that the string width is about 4.1 times the physical lattice spacing, $\delta_{\rm st}/\delta x_{\rm phy}$=4.1, which is equal to the ratio of the two at the final time of the PQ era. As the physical lattice spacing increases, when $\delta_{\rm st}/\delta x_{\rm phy}=2$ is satisfied, we use the fat string algorithm to adjust $\hat{\lambda}$ to keep this ratio always hold for the rest time. After $\hat{\eta} \simeq 2.39$, the relation $\sigma_{\rm wall} > \mu_{\rm string}(t)/t$ is satisfied, with $\sigma_{\rm wall}\sim 8mf_a^2$ the DW tension, $\mu_{\rm string}(t)=\pi v_{\rm re}^2{\rm ln}(t/\delta_{\rm s})$ the mass-energy density of string, and $\delta_{\rm st} \simeq 1/(\sqrt{\hat{\lambda}}m(\hat{\eta}_{\rm e}))$ the string width, the system dynamics will rapidly dominated by DWs. Therefore, the artificial breaking of the ratio of string width to physical lattice spacing will not have a significant impact on the final simulation results.

When $m(\hat{\eta})=3H(\hat{\eta})$ is satisfied at $\hat{\eta}\approx1.44$, axion begins to feel the pull of its mass, DWs bounded by strings are formed. Afterward, as the temperature decreases to $100$ MeV at $\hat{\eta}\approx 2.47$, the axion mass and DW tension no longer change. We continue to evolve the EOM until the final time $\hat{\eta}_f=20$, at which the temperature is $T_f=14.46$ ${\rm MeV}$. At this time, the physical side length of the simulation box is greater than one Hubble length, and the DW thickness ($\delta_{\rm dw}\sim 1/m_0$) is greater than the physical lattice spacing.

\noindent{\it \bfseries  Results.} 
In the PQ era, we measured the scaling parameters of axion strings in two ways, see FIG.~\ref{fig:SCinPQprepareforQCD}. One is the method explained in Refs.~\cite{Hiramatsu:2012sc} ($\xi$, blue hollow points), we call it the traditional method, and the other is the standard scaling model~\cite{Hindmarsh:2019csc} based on the mean string separation ($\xi_{\rm s}$, red hollow point). At the final time ($\tilde{\eta}_f=105$), the string networks reach scaling regime approximately ($\xi_{\rm s} \approx 0.55$).  We found that the scaling parameters exhibited logarithmic increase behavior under two measurement schemes. Several groups have reported logarithmic increase behavior while using the traditional method~\cite{Gorghetto:2018myk,Kawasaki:2018bzv,Vaquero:2018tib,Buschmann:2019icd}. However, according to reference \cite{Hindmarsh:2019csc}, there should be no more logarithmic increase after adopting the standard scaling model. We preliminarily deduce that this may be caused by insufficient resolution of the Hubble volume since the simulation box contains too much Hubble volume during the PQ phase transition. We have prepared an additional PQ era specifically for measuring GWs radiated by pure global (axion) strings, where the string network stays in the scaling regime for a long time and the scaling parameters do not exhibit a logarithmic increase, see {\it Supplemental Material}.

\begin{figure}[!htp]
\includegraphics[width=.2\textwidth]{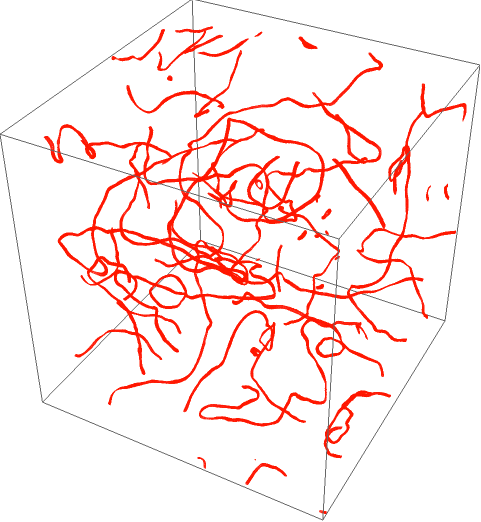} 
\hspace{3mm}
\includegraphics[width=.2\textwidth]{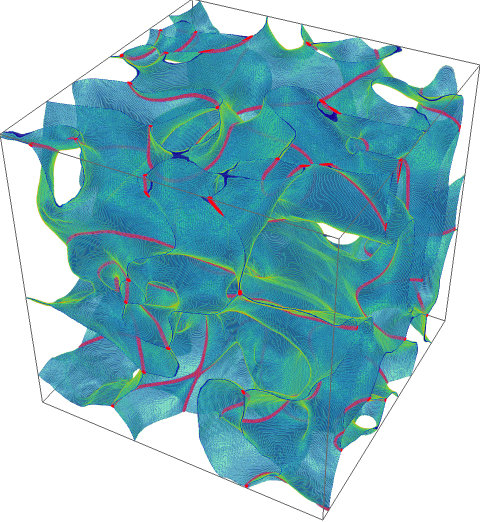}
\hspace{3mm}
\includegraphics[width=.2\textwidth]{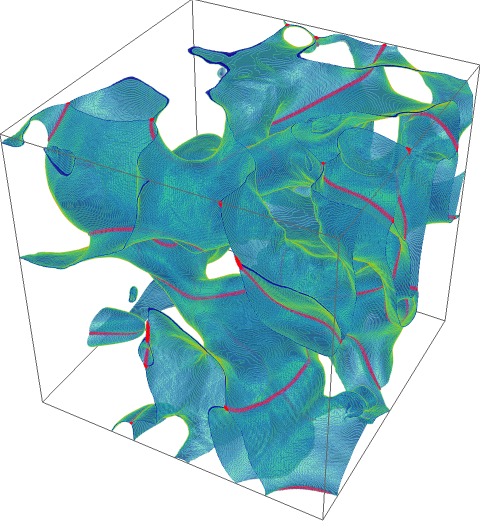}
\hspace{3mm}
\includegraphics[width=.2\textwidth]{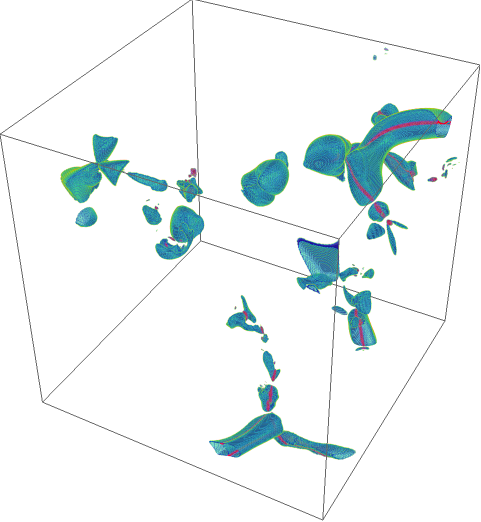}
  \caption{3D snapshots of the string-wall networks at different times in the $550^3$ sub-lattice, with $N_{\rm DW}=3$ and $\Xi=2\times10^{-74}$. The four plots arranged from top-left to bottom-right were recorded at the time  $\hat{\eta}$=1, 2.9, 6.7, and 12.4. The blue and the red region indicate where the DW and string exist, respectively.}
  \vspace{0.1cm}
  \label{fig:3D snapshots of string-wall}
\end{figure}

In the QCD era, for the $N_{\rm DW}>1$ scenario, we fix $N_{\rm DW}=3$ as a benchmark, and choose $\Xi=[0$, $1\times10^{-74}$, $2\times10^{-74}$, $4\times10^{-74}$, $8\times10^{-74}]$ to investigate the effects of the bias term. The four non-zero $\Xi$ values we selected ensure that the DW dominates the system for a sufficiently long time and also can completely decay within our simulation time range. Besides, our choice of $\Xi$ is based on two theoretical constraints with $f_a=3.5\times 10^{16}$ GeV and the corresponding zero-temperature mass $m=m_0$. The first constraint is that the decay of walls is required to occur before wall domination, so there is a lower bound on bias term, $\Xi \geqslant (512/3) \pi  C_d \mathcal{A}^2 G m^2 N_{\rm DW}^{-4}\csc(\pi/N_{\rm DW})^2 $, with $C_d$ being the ratio between the bias term and the wall energy density, i.e., $\Delta V/(\mathcal{A}\sigma_{\rm wall}/t)$), see Eq. (S25) in {\it Supplemental material}. The second constraint comes from that we expect the axion mass to be dominated by non-perturbative effects rather than the bias term~\cite{Hiramatsu:2012sc}: $\Xi \leqslant 2\times 10^{-45}N_{\rm DW}^{-2}(10^{10} {\rm GeV}/f_a)^4$, otherwise, the cosmological history will be significantly modified~\cite{Hiramatsu:2012sc}. We recorded the 3D distribution of the string-wall networks at different times (see FIG. \ref{fig:3D snapshots of string-wall}), where we show that string configurations appear after PQ symmetry breaking (top-left plot), string-wall networks form after $m=3H$ (top-right plot) and decay latter (bottom two plots) as the Universe cools down. 

To obtain the energy density of free axions, we first mask the contamination of the free axion energy density by string core and DW core. We assume the free axions
are all in harmonic mode~\cite{Chang:2023rll} e.g. its total energy takes twice of its kinetic energy~\cite{Hiramatsu:2012gg}, $\rho_a=2\rho_{\rm a,kin}$ $= 2\times \frac{1}{2}\langle \dot{a}_{\rm free}^2(\boldsymbol{\rm x}) \rangle$, where the brackets denote spatial average.
As shown in Sec.IV of the {\it Supplemental material},
the dark matter relic abundance from the axions radiated by the DWs for the ALP and QCD axion scenarios with $N_{\rm DW}>1$ is given by
\begin{eqnarray}
    \Omega^{\rm DW}_a h^2&=&1.02\times 10^{-19}\big(\frac{8}{3}\big)^{\frac{3}{2p}}\bigg(\frac{2p-1}{3-2p}\bigg) C_d^{\frac{3}{2p}-1} \big(\frac{m}{\rm GeV}\big)^{\frac{3}{p}-\frac{3}{2}}\nonumber\\
&& \times\big(\frac{f_a}{\rm GeV}\big)^{4-\frac{3}{p}} \Xi^{1-\frac{3}{2p}} \big(\frac{\csc(\pi/N_{\rm DW})}{N_{\rm DW}^2}\big)^{\frac{3}{p}-2}\;.
\end{eqnarray}

\begin{figure}[!htp]
\includegraphics[width=0.4\textwidth]{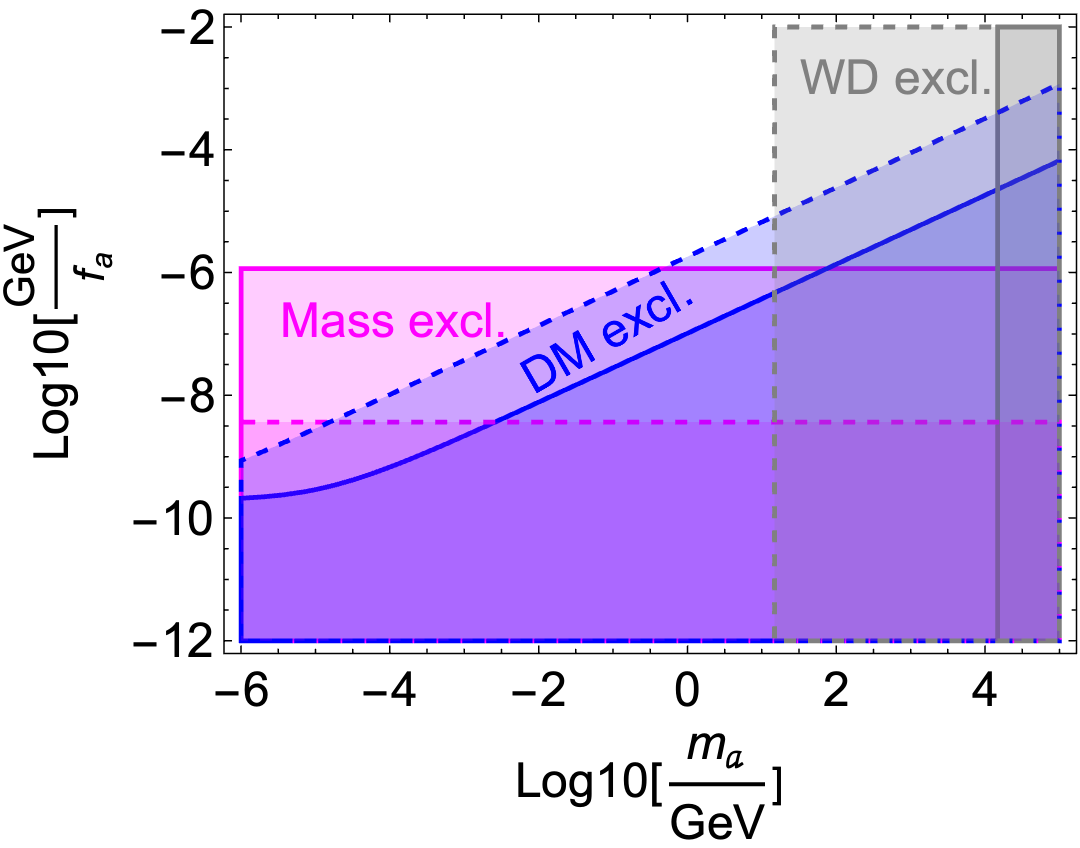}  
  \caption{Constraints on the axion (or ALP) mass versus the PQ scale $f_a$ for the bias term $\Xi=10^{-30}$ (and $\Xi=10^{-40}$) with dashed (and solid) boundaries.  } 
  \vspace{0.1cm}  
  \label{fig:mafa}
\end{figure}

Therefore, the axion relic abundance today for arbitrary $N_{\rm DW}>1$ satisfies  
\begin{flalign} 
\Omega_{a,\rm new}(t_0)h^2 =& \Omega^{\rm DW}_ah^2   +\Omega_{a,0}(t_0)h^2 +  \Omega_{a,\rm st}(t_0)h^2.\label{axion abundance} 
\end{flalign}
We note that the coherent oscillation of the homogeneous axion field ($\Omega_{a,0}(t_0)h^2\simeq 4.63\times10^{-3} (f_{a,\rm new}/10^{10}{\rm GeV})^{1.19}$) and the production of axions from pure axion strings ($\Omega_{a,\rm st}(t_0)h^2\simeq 7.3\times 10^{-3}N_{\rm DW}^2 (f_{a,\rm new}/10^{10}{\rm GeV})^{1.19}$) also contribute to axion cold dark matter (CDM) abundance~\cite{Kawasaki:2014sqa}. 
We show in Fig.~\ref{fig:mafa} the exclusion on the plane of $m_a$ versus $f_a$ considering the requirements of $\Omega_a \leq \Omega_{\rm DM}$ (blue regions), the axion mass not dominated by the bias term (magenta regions), and wall domination (gray regions) with $N_{\rm DW}=6$. As shown in the figure, a larger bias term leads to stronger constraints from dark matter and wall domination and a weaker constraint from the axion mass.

\begin{figure}[!htp]
\includegraphics[width=0.41\textwidth]{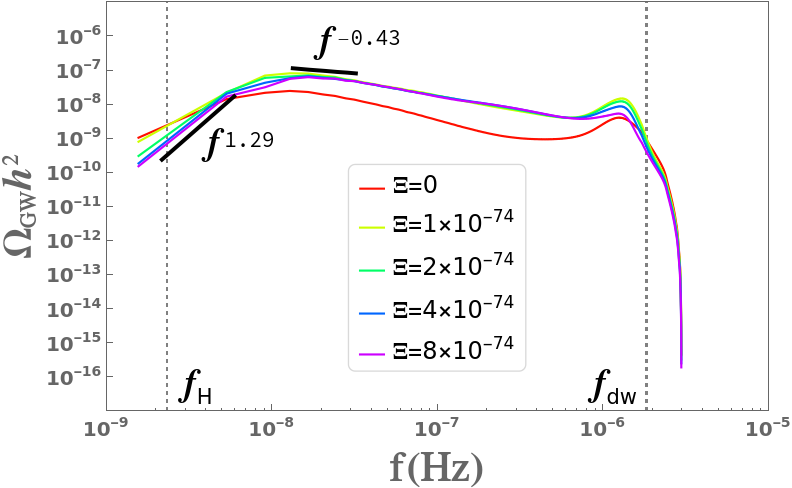}
\includegraphics[width=0.41\textwidth]{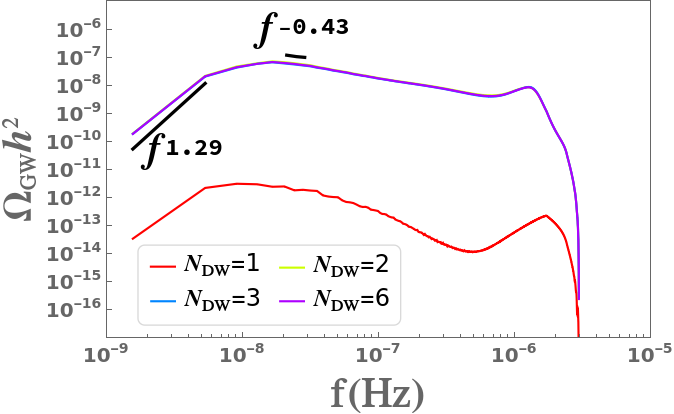}
  \vspace{-0.2cm}
  \caption{Upper panel: GW spectrum today with different coefficients of bias term in the case of $N_{\rm DW}=3$. Lower panel: GW spectrum today with different DW numbers. We set $\Xi=4\times 10^{-74}$ for $N_{\rm DW}>1$ and $\Xi=0$ for $N_{\rm DW}=1$. 
  \vspace{0.05cm}
\label{fig:QCD_GW_spectra}
 }\end{figure}

For the GWs radiated by axion string-wall networks, we investigated the dependence of GW spectrum on $\Xi$ and $N_{\rm DW}$ in FIG.~\ref{fig:QCD_GW_spectra}. The two characteristic scales $f_{\rm H}$ and $f_{\rm dw}$ correspond to the Hubble length ($H^{-1}_{\rm dec}$) and DW thickeness ($m_0^{-1}$) at the final time ($\hat{\eta}_{f}$) in the case of $\Xi=4\times 10^{-74}$. We found that the four curves with $\Xi>0$ almost completely overlap, and have the same slope ($f^{1.29}$and $f^{-0.43}$) near the left peak of the GW spectrum. It can be observed that the amplitudes and shape of the GW spectrum are (almost) independent of $N_{\rm DW}$ in the case of $N_{\rm DW}>1$. The GW spectrum does not depend on $\Xi$ and $N_{\rm DW}$ (for $N_{\rm DW}>1$) is due to 
$\Omega_{\rm gw} \equiv \frac{1}{\rho_{c}} \frac{{\rm d}\rho_{\rm GW}}{{\rm dln}k} \propto \rho_{\rm gw} \propto Gm^2f_a^4$, where $\rho_{c}$ denote the critical energy density, since our simulation suggest that the multiplication of efficiency parameter and area parameter approaches to a constant, i.e., $\epsilon_{\rm gw}\mathcal{A}^2\sim$  0.054, see FIG. S10 in {\it Supplemental Material} for more details.


\begin{figure}[!htp]
\includegraphics[width=.48\textwidth]{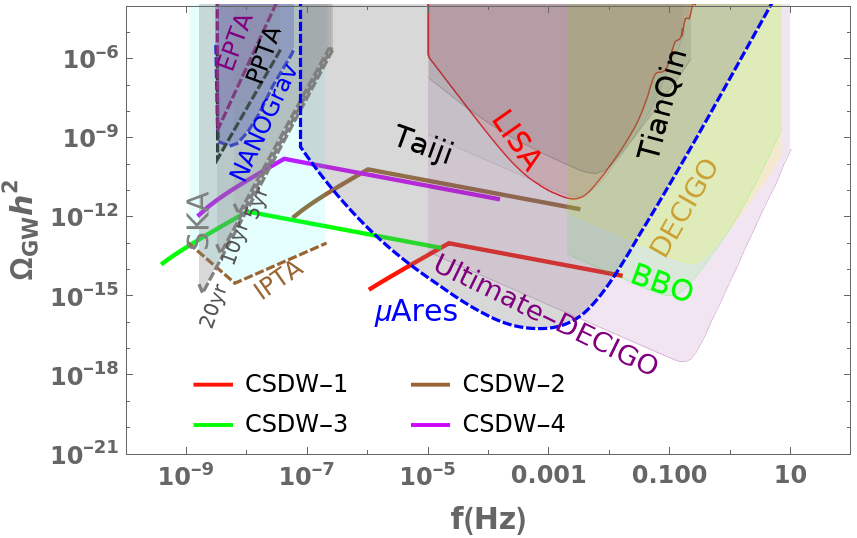}
  \caption{GWs spectra to be probed at present for ALPs. The four solid lines CSDW-1,2,3,4 represent peak amplitude of GWs produced by string-wall networks corresponds to the model parameters ($f_a/{\rm GeV}$, $m/{\rm GeV}$, $\Xi$) = ($5\times10^{8}$, $10^{-6}$, $2\times10^{-41}$), ($10^{6.4}$, $1$, $10^{-33}$),$(10^6,1,10^{-36})$,$(10^5,10^3,10^{-30})$ in the case of $N_{\rm DW}=6$. }\label{fig:GWdetect}
\end{figure}

For ALPs in $N_{\rm DW}>1$ case, we can ignore the nEDM bound, since ALPs can be unrelated to 
the strong CP problem in QCD. Specifically, we found
that there exists a range of parameter spaces that can lead
to detectable GW signals radiated from string-wall networks. We present the
detect-ability of SGWB from axion string-wall networks with various model parameters in FIG.~\ref{fig:GWdetect}, see the four benchmark scenarios of CSDW-$1,2,3,4$ (with $N_{\rm DW}=6$). The reinterpreted peak amplitude of GW spectra can be obtained by $\Omega _{\rm gw,re}h^2=\Omega_{\rm gw,s}h^2 \times m_{\rm re}^2f_{a,\rm re}^4/(m_{s}^2f_{a,s}^4)$, with $\Omega_{\rm gw,s}h^2=6.65\times 10^{-8}$. The subscript "s" denotes the physical quantities in our simulations, while the subscripts "re" represent the reinterpreted parameters. The reinterpreted peak frequency satisfies $f_{\rm peak,re}=7.32\times10^{-8}g_{*}(T_{\rm dec,re})^{1/6}T_{\rm dec,re}$/GeV, with $T_{\rm dec,re}=1.56\times 10^9 (3/8)^{1/(2p)}g_{*}(T_{\rm dec,re})^{-1/4}N_{\rm DW}^{2/p}{\rm GeV^{1/2}}(m_{\rm re})^{1/2-1/p}\Xi^{1/(2p)}f_{a,\rm re}^{1/p}$, which is obtained from Eq.~(S23) in {\it Supplemental material}.

For QCD axion, we require that the axion DM abundance does not exceed the abundance of CDM observed today leading to an upper bound on $f_a$, $f_a\leq 5.1\times 10^{9}$ GeV in the $N_{\rm DW}>1$ case and $f_a\leq7.2\times 10^{10}$GeV in the $N_{\rm DW}=1$ case\cite{Kawasaki:2014sqa}. So, the peak amplitude of GW spectrum today satisfies $(\Omega_{\rm GW} h^2)_{\rm peak}\leq1.42 \times 10^{-21}$ in the $N_{\rm DW}>1$ case and $(\Omega_{\rm GW} h^2)_{\rm peak}\leq5.6 \times 10^{-35}$ in the $N_{\rm DW}=1$ case. The GWs in both cases are too weak to be detected.
For ALPs in $N_{\rm DW}=1$ case, we also have $f_a\leq7.2\times 10^{10}$GeV. So, the GW peak amplitude satisfies $(\Omega_{\rm GW}h^2)_{\rm peak}\leq 5.6\times 10^{-35}$, which is too weak to be probed.

\noindent{\it \bfseries  Conclusion and discussion.}
The cosmic strings and DWs are generally predicted in many particle physics models beyond the Standard Model, both of them are important SGWB sources. In this Letter, we numerically studied the axionic dark matter and SGWB produced by axion string-wall networks with different model parameters.

For both QCD axion and ALPs, we found that the axion energy density $\rho_a$ is proportional to $N_{\rm DW}^{4}$ and gave the DM relic abundance from string-walls decay for the scenarios of $N_{\rm DW}>1$. We also found that axion string-wall networks mainly decay into axions, which have an energy density about one hundred times larger than that of GWs. For the GW radiation of the $N_{\rm DW}>1$ scenarios, through the analysis of the GW energy density and the efficiency of GW production, we observed that the GW spectrum is almost independent of $\Xi$ and $N_{\rm DW}$ (for $\Xi>0$), and is proportional to the square of DW tension. We gave the prediction of the GW spectra shape for string-wall networks with domain wall decay. For the ALP scenarios with $N_{\rm DW}>1$, In part of the parameter spaces where axion mass ranges from KeV to TeV, the string-wall networks can produce detectable GWs in the frequency range $\mathcal{O}(10^{-9}-10^{-3})$ Hz, to be probed by SKA and $\mu$Ares. We also found that the QCD axion predicts undetectable GW radiations when one requires the axion to provide the full DM relic abundance. In $ N_{\rm DW}=1$ case, the GW energy density appears too weak to be probed for QCD axions and ALPs. This provides the possibility of searching for axion models or constraining model parameters through GW detection experiments in the future.



\noindent{\it \bfseries Acknowledgements.}
We are grateful to Adrien Florio and Daniel G. Figueroa for helpful discussions on their public code ${\mathcal CosmoLattice}$. The numerical calculations in this study were carried out on the ORISE Supercomputer.
This work is supported by the National Key Research and Development Program of China under Grant No. 2021YFC2203004 and 2020YFC2201501.
L.B. is supported by the National Natural Science Foundation of China (NSFC) under Grants Nos. 12075041, 12322505, and 12347101.
L.B. also acknowledges Chongqing Natural Science Foundation under Grant
No. CSTB2024NSCQ-JQX0022 and 
Chongqing Talents: Exceptional Young Talents Project No. cstc2024ycjh-bgzxm0020.
R.G.C is supported by the National Key Research and Development Program of China
Grant No. 2020YFC2201502 and 2021YFA0718304
and by the National Natural Science Foundation of China
Grants No. 11821505, No. 11991052, No. 11947302,
No. 12235019.
J.S. is supported by Peking University under Startup Grant No. 7101302974 and the National Natural Science Foundation of China under Grants No. 12025507, and No.12150015, and is supported by the Key Research Program of Frontier Science of the Chinese Academy of Sciences (CAS) under Grants No. ZDBS-LY-7003.

\bibliographystyle{apsrev}

\bibliography{references}

\begin{thebibliography}{102}
\expandafter\ifx\csname natexlab\endcsname\relax\def\natexlab#1{#1}\fi
\expandafter\ifx\csname bibnamefont\endcsname\relax
  \def\bibnamefont#1{#1}\fi
\expandafter\ifx\csname bibfnamefont\endcsname\relax
  \def\bibfnamefont#1{#1}\fi
\expandafter\ifx\csname citenamefont\endcsname\relax
  \def\citenamefont#1{#1}\fi
\expandafter\ifx\csname url\endcsname\relax
  \def\url#1{\texttt{#1}}\fi
\expandafter\ifx\csname urlprefix\endcsname\relax\def\urlprefix{URL }\fi
\providecommand{\bibinfo}[2]{#2}
\providecommand{\eprint}[2][]{\url{#2}}

\bibitem[{\citenamefont{Peccei and Quinn}(1977{\natexlab{a}})}]{Peccei:1977hh}
\bibinfo{author}{\bibfnamefont{R.~D.} \bibnamefont{Peccei}} \bibnamefont{and}
  \bibinfo{author}{\bibfnamefont{H.~R.} \bibnamefont{Quinn}},
  \bibinfo{journal}{Phys. Rev. Lett.} \textbf{\bibinfo{volume}{38}},
  \bibinfo{pages}{1440} (\bibinfo{year}{1977}{\natexlab{a}}).

\bibitem[{\citenamefont{Peccei and Quinn}(1977{\natexlab{b}})}]{Peccei:1977ur}
\bibinfo{author}{\bibfnamefont{R.~D.} \bibnamefont{Peccei}} \bibnamefont{and}
  \bibinfo{author}{\bibfnamefont{H.~R.} \bibnamefont{Quinn}},
  \bibinfo{journal}{Phys. Rev. D} \textbf{\bibinfo{volume}{16}},
  \bibinfo{pages}{1791} (\bibinfo{year}{1977}{\natexlab{b}}).

\bibitem[{\citenamefont{Wilczek}(1978)}]{Wilczek:1977pj}
\bibinfo{author}{\bibfnamefont{F.}~\bibnamefont{Wilczek}},
  \bibinfo{journal}{Phys. Rev. Lett.} \textbf{\bibinfo{volume}{40}},
  \bibinfo{pages}{279} (\bibinfo{year}{1978}).

\bibitem[{\citenamefont{Weinberg}(1978)}]{Weinberg:1977ma}
\bibinfo{author}{\bibfnamefont{S.}~\bibnamefont{Weinberg}},
  \bibinfo{journal}{Phys. Rev. Lett.} \textbf{\bibinfo{volume}{40}},
  \bibinfo{pages}{223} (\bibinfo{year}{1978}).

\bibitem[{\citenamefont{Abbott and Sikivie}(1983)}]{Abbott:1982af}
\bibinfo{author}{\bibfnamefont{L.~F.} \bibnamefont{Abbott}} \bibnamefont{and}
  \bibinfo{author}{\bibfnamefont{P.}~\bibnamefont{Sikivie}},
  \bibinfo{journal}{Phys. Lett. B} \textbf{\bibinfo{volume}{120}},
  \bibinfo{pages}{133} (\bibinfo{year}{1983}).

\bibitem[{\citenamefont{Dine and Fischler}(1983)}]{Dine:1982ah}
\bibinfo{author}{\bibfnamefont{M.}~\bibnamefont{Dine}} \bibnamefont{and}
  \bibinfo{author}{\bibfnamefont{W.}~\bibnamefont{Fischler}},
  \bibinfo{journal}{Phys. Lett. B} \textbf{\bibinfo{volume}{120}},
  \bibinfo{pages}{137} (\bibinfo{year}{1983}).

\bibitem[{\citenamefont{Preskill et~al.}(1983)\citenamefont{Preskill, Wise, and
  Wilczek}}]{Preskill:1982cy}
\bibinfo{author}{\bibfnamefont{J.}~\bibnamefont{Preskill}},
  \bibinfo{author}{\bibfnamefont{M.~B.} \bibnamefont{Wise}}, \bibnamefont{and}
  \bibinfo{author}{\bibfnamefont{F.}~\bibnamefont{Wilczek}},
  \bibinfo{journal}{Phys. Lett. B} \textbf{\bibinfo{volume}{120}},
  \bibinfo{pages}{127} (\bibinfo{year}{1983}).

\bibitem[{\citenamefont{Kawasaki et~al.}(2015)\citenamefont{Kawasaki, Saikawa,
  and Sekiguchi}}]{Kawasaki:2014sqa}
\bibinfo{author}{\bibfnamefont{M.}~\bibnamefont{Kawasaki}},
  \bibinfo{author}{\bibfnamefont{K.}~\bibnamefont{Saikawa}}, \bibnamefont{and}
  \bibinfo{author}{\bibfnamefont{T.}~\bibnamefont{Sekiguchi}},
  \bibinfo{journal}{Phys. Rev. D} \textbf{\bibinfo{volume}{91}},
  \bibinfo{pages}{065014} (\bibinfo{year}{2015}), \eprint{1412.0789}.

\bibitem[{\citenamefont{Hogan and Rees}(1988)}]{Hogan:1988mp}
\bibinfo{author}{\bibfnamefont{C.~J.} \bibnamefont{Hogan}} \bibnamefont{and}
  \bibinfo{author}{\bibfnamefont{M.~J.} \bibnamefont{Rees}},
  \bibinfo{journal}{Phys. Lett. B} \textbf{\bibinfo{volume}{205}},
  \bibinfo{pages}{228} (\bibinfo{year}{1988}).

\bibitem[{\citenamefont{Kolb and Tkachev}(1993)}]{Kolb:1993zz}
\bibinfo{author}{\bibfnamefont{E.~W.} \bibnamefont{Kolb}} \bibnamefont{and}
  \bibinfo{author}{\bibfnamefont{I.~I.} \bibnamefont{Tkachev}},
  \bibinfo{journal}{Phys. Rev. Lett.} \textbf{\bibinfo{volume}{71}},
  \bibinfo{pages}{3051} (\bibinfo{year}{1993}), \eprint{hep-ph/9303313}.

\bibitem[{\citenamefont{Kolb and Tkachev}(1994{\natexlab{a}})}]{Kolb:1993hw}
\bibinfo{author}{\bibfnamefont{E.~W.} \bibnamefont{Kolb}} \bibnamefont{and}
  \bibinfo{author}{\bibfnamefont{I.~I.} \bibnamefont{Tkachev}},
  \bibinfo{journal}{Phys. Rev. D} \textbf{\bibinfo{volume}{49}},
  \bibinfo{pages}{5040} (\bibinfo{year}{1994}{\natexlab{a}}),
  \eprint{astro-ph/9311037}.

\bibitem[{\citenamefont{Kolb and Tkachev}(1994{\natexlab{b}})}]{Kolb:1994fi}
\bibinfo{author}{\bibfnamefont{E.~W.} \bibnamefont{Kolb}} \bibnamefont{and}
  \bibinfo{author}{\bibfnamefont{I.~I.} \bibnamefont{Tkachev}},
  \bibinfo{journal}{Phys. Rev. D} \textbf{\bibinfo{volume}{50}},
  \bibinfo{pages}{769} (\bibinfo{year}{1994}{\natexlab{b}}),
  \eprint{astro-ph/9403011}.

\bibitem[{\citenamefont{Zurek et~al.}(2007)\citenamefont{Zurek, Hogan, and
  Quinn}}]{Zurek:2006sy}
\bibinfo{author}{\bibfnamefont{K.~M.} \bibnamefont{Zurek}},
  \bibinfo{author}{\bibfnamefont{C.~J.} \bibnamefont{Hogan}}, \bibnamefont{and}
  \bibinfo{author}{\bibfnamefont{T.~R.} \bibnamefont{Quinn}},
  \bibinfo{journal}{Phys. Rev. D} \textbf{\bibinfo{volume}{75}},
  \bibinfo{pages}{043511} (\bibinfo{year}{2007}), \eprint{astro-ph/0607341}.

\bibitem[{\citenamefont{Enander et~al.}(2017)\citenamefont{Enander, Pargner,
  and Schwetz}}]{Enander:2017ogx}
\bibinfo{author}{\bibfnamefont{J.}~\bibnamefont{Enander}},
  \bibinfo{author}{\bibfnamefont{A.}~\bibnamefont{Pargner}}, \bibnamefont{and}
  \bibinfo{author}{\bibfnamefont{T.}~\bibnamefont{Schwetz}},
  \bibinfo{journal}{JCAP} \textbf{\bibinfo{volume}{12}}, \bibinfo{pages}{038}
  (\bibinfo{year}{2017}), \eprint{1708.04466}.

\bibitem[{\citenamefont{Vaquero et~al.}(2019)\citenamefont{Vaquero, Redondo,
  and Stadler}}]{Vaquero:2018tib}
\bibinfo{author}{\bibfnamefont{A.}~\bibnamefont{Vaquero}},
  \bibinfo{author}{\bibfnamefont{J.}~\bibnamefont{Redondo}}, \bibnamefont{and}
  \bibinfo{author}{\bibfnamefont{J.}~\bibnamefont{Stadler}},
  \bibinfo{journal}{JCAP} \textbf{\bibinfo{volume}{04}}, \bibinfo{pages}{012}
  (\bibinfo{year}{2019}), \eprint{1809.09241}.

\bibitem[{\citenamefont{Buschmann et~al.}(2020)\citenamefont{Buschmann, Foster,
  and Safdi}}]{Buschmann:2019icd}
\bibinfo{author}{\bibfnamefont{M.}~\bibnamefont{Buschmann}},
  \bibinfo{author}{\bibfnamefont{J.~W.} \bibnamefont{Foster}},
  \bibnamefont{and} \bibinfo{author}{\bibfnamefont{B.~R.} \bibnamefont{Safdi}},
  \bibinfo{journal}{Phys. Rev. Lett.} \textbf{\bibinfo{volume}{124}},
  \bibinfo{pages}{161103} (\bibinfo{year}{2020}), \eprint{1906.00967}.

\bibitem[{\citenamefont{Chang and Cui}(2023)}]{Chang:2023rll}
\bibinfo{author}{\bibfnamefont{C.-F.} \bibnamefont{Chang}} \bibnamefont{and}
  \bibinfo{author}{\bibfnamefont{Y.}~\bibnamefont{Cui}} (\bibinfo{year}{2023}),
  \eprint{2309.15920}.

\bibitem[{\citenamefont{Blasi et~al.}(2021)\citenamefont{Blasi, Brdar, and
  Schmitz}}]{Blasi:2020mfx}
\bibinfo{author}{\bibfnamefont{S.}~\bibnamefont{Blasi}},
  \bibinfo{author}{\bibfnamefont{V.}~\bibnamefont{Brdar}}, \bibnamefont{and}
  \bibinfo{author}{\bibfnamefont{K.}~\bibnamefont{Schmitz}},
  \bibinfo{journal}{Phys. Rev. Lett.} \textbf{\bibinfo{volume}{126}},
  \bibinfo{pages}{041305} (\bibinfo{year}{2021}), \eprint{2009.06607}.

\bibitem[{\citenamefont{Samanta and Datta}(2021)}]{Samanta:2020cdk}
\bibinfo{author}{\bibfnamefont{R.}~\bibnamefont{Samanta}} \bibnamefont{and}
  \bibinfo{author}{\bibfnamefont{S.}~\bibnamefont{Datta}},
  \bibinfo{journal}{JHEP} \textbf{\bibinfo{volume}{05}}, \bibinfo{pages}{211}
  (\bibinfo{year}{2021}), \eprint{2009.13452}.

\bibitem[{\citenamefont{Ellis and Lewicki}(2021)}]{Ellis:2020ena}
\bibinfo{author}{\bibfnamefont{J.}~\bibnamefont{Ellis}} \bibnamefont{and}
  \bibinfo{author}{\bibfnamefont{M.}~\bibnamefont{Lewicki}},
  \bibinfo{journal}{Phys. Rev. Lett.} \textbf{\bibinfo{volume}{126}},
  \bibinfo{pages}{041304} (\bibinfo{year}{2021}), \eprint{2009.06555}.

\bibitem[{\citenamefont{Bian et~al.}(2022{\natexlab{a}})\citenamefont{Bian,
  Shu, Wang, Yuan, and Zong}}]{Bian:2022tju}
\bibinfo{author}{\bibfnamefont{L.}~\bibnamefont{Bian}},
  \bibinfo{author}{\bibfnamefont{J.}~\bibnamefont{Shu}},
  \bibinfo{author}{\bibfnamefont{B.}~\bibnamefont{Wang}},
  \bibinfo{author}{\bibfnamefont{Q.}~\bibnamefont{Yuan}}, \bibnamefont{and}
  \bibinfo{author}{\bibfnamefont{J.}~\bibnamefont{Zong}},
  \bibinfo{journal}{Phys. Rev. D} \textbf{\bibinfo{volume}{106}},
  \bibinfo{pages}{L101301} (\bibinfo{year}{2022}{\natexlab{a}}),
  \eprint{2205.07293}.

\bibitem[{\citenamefont{Gouttenoire and
  Vitagliano}(2024{\natexlab{a}})}]{Gouttenoire:2023ftk}
\bibinfo{author}{\bibfnamefont{Y.}~\bibnamefont{Gouttenoire}} \bibnamefont{and}
  \bibinfo{author}{\bibfnamefont{E.}~\bibnamefont{Vitagliano}},
  \bibinfo{journal}{Phys. Rev. D} \textbf{\bibinfo{volume}{110}},
  \bibinfo{pages}{L061306} (\bibinfo{year}{2024}{\natexlab{a}}),
  \eprint{2306.17841}.

\bibitem[{\citenamefont{Gouttenoire and
  Vitagliano}(2024{\natexlab{b}})}]{Gouttenoire:2023gbn}
\bibinfo{author}{\bibfnamefont{Y.}~\bibnamefont{Gouttenoire}} \bibnamefont{and}
  \bibinfo{author}{\bibfnamefont{E.}~\bibnamefont{Vitagliano}},
  \bibinfo{journal}{Phys. Rev. D} \textbf{\bibinfo{volume}{109}},
  \bibinfo{pages}{123507} (\bibinfo{year}{2024}{\natexlab{b}}),
  \eprint{2311.07670}.

\bibitem[{\citenamefont{Afzal et~al.}(2023)}]{NANOGrav:2023hvm}
\bibinfo{author}{\bibfnamefont{A.}~\bibnamefont{Afzal}} \bibnamefont{et~al.}
  (\bibinfo{collaboration}{NANOGrav}), \bibinfo{journal}{Astrophys. J. Lett.}
  \textbf{\bibinfo{volume}{951}} (\bibinfo{year}{2023}), \eprint{2306.16219}.

\bibitem[{\citenamefont{Bian et~al.}(2021)\citenamefont{Bian, Cai, Liu, Yang,
  and Zhou}}]{Bian:2020urb}
\bibinfo{author}{\bibfnamefont{L.}~\bibnamefont{Bian}},
  \bibinfo{author}{\bibfnamefont{R.-G.} \bibnamefont{Cai}},
  \bibinfo{author}{\bibfnamefont{J.}~\bibnamefont{Liu}},
  \bibinfo{author}{\bibfnamefont{X.-Y.} \bibnamefont{Yang}}, \bibnamefont{and}
  \bibinfo{author}{\bibfnamefont{R.}~\bibnamefont{Zhou}},
  \bibinfo{journal}{Phys. Rev. D} \textbf{\bibinfo{volume}{103}},
  \bibinfo{pages}{L081301} (\bibinfo{year}{2021}), \eprint{2009.13893}.

\bibitem[{\citenamefont{Bian et~al.}(2023)\citenamefont{Bian, Ge, Shu, Wang,
  Yang, and Zong}}]{Bian:2023dnv}
\bibinfo{author}{\bibfnamefont{L.}~\bibnamefont{Bian}},
  \bibinfo{author}{\bibfnamefont{S.}~\bibnamefont{Ge}},
  \bibinfo{author}{\bibfnamefont{J.}~\bibnamefont{Shu}},
  \bibinfo{author}{\bibfnamefont{B.}~\bibnamefont{Wang}},
  \bibinfo{author}{\bibfnamefont{X.-Y.} \bibnamefont{Yang}}, \bibnamefont{and}
  \bibinfo{author}{\bibfnamefont{J.}~\bibnamefont{Zong}}
  (\bibinfo{year}{2023}), \eprint{2307.02376}.

\bibitem[{\citenamefont{Bian et~al.}(2022{\natexlab{b}})\citenamefont{Bian, Ge,
  Li, Shu, and Zong}}]{Bian:2022qbh}
\bibinfo{author}{\bibfnamefont{L.}~\bibnamefont{Bian}},
  \bibinfo{author}{\bibfnamefont{S.}~\bibnamefont{Ge}},
  \bibinfo{author}{\bibfnamefont{C.}~\bibnamefont{Li}},
  \bibinfo{author}{\bibfnamefont{J.}~\bibnamefont{Shu}}, \bibnamefont{and}
  \bibinfo{author}{\bibfnamefont{J.}~\bibnamefont{Zong}}
  (\bibinfo{year}{2022}{\natexlab{b}}), \eprint{2212.07871}.

\bibitem[{\citenamefont{Ferreira et~al.}(2023)\citenamefont{Ferreira, Notari,
  Pujolas, and Rompineve}}]{Ferreira:2022zzo}
\bibinfo{author}{\bibfnamefont{R.~Z.} \bibnamefont{Ferreira}},
  \bibinfo{author}{\bibfnamefont{A.}~\bibnamefont{Notari}},
  \bibinfo{author}{\bibfnamefont{O.}~\bibnamefont{Pujolas}}, \bibnamefont{and}
  \bibinfo{author}{\bibfnamefont{F.}~\bibnamefont{Rompineve}},
  \bibinfo{journal}{JCAP} \textbf{\bibinfo{volume}{02}}, \bibinfo{pages}{001}
  (\bibinfo{year}{2023}), \eprint{2204.04228}.

\bibitem[{\citenamefont{Auclair
  et~al.}(2023{\natexlab{a}})\citenamefont{Auclair, Babak, Quelquejay~Leclere,
  and Steer}}]{Auclair:2023brk}
\bibinfo{author}{\bibfnamefont{P.}~\bibnamefont{Auclair}},
  \bibinfo{author}{\bibfnamefont{S.}~\bibnamefont{Babak}},
  \bibinfo{author}{\bibfnamefont{H.}~\bibnamefont{Quelquejay~Leclere}},
  \bibnamefont{and} \bibinfo{author}{\bibfnamefont{D.~A.} \bibnamefont{Steer}},
  \bibinfo{journal}{Phys. Rev. D} \textbf{\bibinfo{volume}{108}},
  \bibinfo{pages}{043519} (\bibinfo{year}{2023}{\natexlab{a}}),
  \eprint{2305.11653}.

\bibitem[{\citenamefont{Boileau et~al.}(2022)\citenamefont{Boileau, Jenkins,
  Sakellariadou, Meyer, and Christensen}}]{Boileau:2021gbr}
\bibinfo{author}{\bibfnamefont{G.}~\bibnamefont{Boileau}},
  \bibinfo{author}{\bibfnamefont{A.~C.} \bibnamefont{Jenkins}},
  \bibinfo{author}{\bibfnamefont{M.}~\bibnamefont{Sakellariadou}},
  \bibinfo{author}{\bibfnamefont{R.}~\bibnamefont{Meyer}}, \bibnamefont{and}
  \bibinfo{author}{\bibfnamefont{N.}~\bibnamefont{Christensen}},
  \bibinfo{journal}{Phys. Rev. D} \textbf{\bibinfo{volume}{105}},
  \bibinfo{pages}{023510} (\bibinfo{year}{2022}), \eprint{2109.06552}.

\bibitem[{\citenamefont{Abbott et~al.}(2021)}]{LIGOScientific:2021nrg}
\bibinfo{author}{\bibfnamefont{R.}~\bibnamefont{Abbott}} \bibnamefont{et~al.}
  (\bibinfo{collaboration}{LIGO Scientific, Virgo, KAGRA}),
  \bibinfo{journal}{Phys. Rev. Lett.} \textbf{\bibinfo{volume}{126}},
  \bibinfo{pages}{241102} (\bibinfo{year}{2021}), \eprint{2101.12248}.

\bibitem[{\citenamefont{Marsh}(2016)}]{Marsh:2015xka}
\bibinfo{author}{\bibfnamefont{D.~J.~E.} \bibnamefont{Marsh}},
  \bibinfo{journal}{Phys. Rept.} \textbf{\bibinfo{volume}{643}},
  \bibinfo{pages}{1} (\bibinfo{year}{2016}), \eprint{1510.07633}.

\bibitem[{\citenamefont{Borsanyi et~al.}(2016)\citenamefont{Borsanyi, Dierigl,
  Fodor, Katz, Mages, Nogradi, Redondo, Ringwald, and
  Szabo}}]{Borsanyi:2015cka}
\bibinfo{author}{\bibfnamefont{S.}~\bibnamefont{Borsanyi}},
  \bibinfo{author}{\bibfnamefont{M.}~\bibnamefont{Dierigl}},
  \bibinfo{author}{\bibfnamefont{Z.}~\bibnamefont{Fodor}},
  \bibinfo{author}{\bibfnamefont{S.~D.} \bibnamefont{Katz}},
  \bibinfo{author}{\bibfnamefont{S.~W.} \bibnamefont{Mages}},
  \bibinfo{author}{\bibfnamefont{D.}~\bibnamefont{Nogradi}},
  \bibinfo{author}{\bibfnamefont{J.}~\bibnamefont{Redondo}},
  \bibinfo{author}{\bibfnamefont{A.}~\bibnamefont{Ringwald}}, \bibnamefont{and}
  \bibinfo{author}{\bibfnamefont{K.~K.} \bibnamefont{Szabo}},
  \bibinfo{journal}{Phys. Lett. B} \textbf{\bibinfo{volume}{752}},
  \bibinfo{pages}{175} (\bibinfo{year}{2016}), \eprint{1508.06917}.

\bibitem[{\citenamefont{Hlozek et~al.}(2015)\citenamefont{Hlozek, Grin, Marsh,
  and Ferreira}}]{Hlozek:2014lca}
\bibinfo{author}{\bibfnamefont{R.}~\bibnamefont{Hlozek}},
  \bibinfo{author}{\bibfnamefont{D.}~\bibnamefont{Grin}},
  \bibinfo{author}{\bibfnamefont{D.~J.~E.} \bibnamefont{Marsh}},
  \bibnamefont{and} \bibinfo{author}{\bibfnamefont{P.~G.}
  \bibnamefont{Ferreira}}, \bibinfo{journal}{Phys. Rev. D}
  \textbf{\bibinfo{volume}{91}}, \bibinfo{pages}{103512}
  (\bibinfo{year}{2015}), \eprint{1410.2896}.

\bibitem[{\citenamefont{Kawasaki and Nakayama}(2013)}]{Kawasaki:2013ae}
\bibinfo{author}{\bibfnamefont{M.}~\bibnamefont{Kawasaki}} \bibnamefont{and}
  \bibinfo{author}{\bibfnamefont{K.}~\bibnamefont{Nakayama}},
  \bibinfo{journal}{Ann. Rev. Nucl. Part. Sci.} \textbf{\bibinfo{volume}{63}},
  \bibinfo{pages}{69} (\bibinfo{year}{2013}), \eprint{1301.1123}.

\bibitem[{\citenamefont{Chang and Cui}(2020)}]{Chang:2019mza}
\bibinfo{author}{\bibfnamefont{C.-F.} \bibnamefont{Chang}} \bibnamefont{and}
  \bibinfo{author}{\bibfnamefont{Y.}~\bibnamefont{Cui}},
  \bibinfo{journal}{Phys. Dark Univ.} \textbf{\bibinfo{volume}{29}},
  \bibinfo{pages}{100604} (\bibinfo{year}{2020}), \eprint{1910.04781}.

\bibitem[{\citenamefont{Chang and Cui}(2022)}]{Chang:2021afa}
\bibinfo{author}{\bibfnamefont{C.-F.} \bibnamefont{Chang}} \bibnamefont{and}
  \bibinfo{author}{\bibfnamefont{Y.}~\bibnamefont{Cui}},
  \bibinfo{journal}{JHEP} \textbf{\bibinfo{volume}{03}}, \bibinfo{pages}{114}
  (\bibinfo{year}{2022}), \eprint{2106.09746}.

\bibitem[{\citenamefont{Auclair
  et~al.}(2023{\natexlab{b}})}]{LISACosmologyWorkingGroup:2022jok}
\bibinfo{author}{\bibfnamefont{P.}~\bibnamefont{Auclair}} \bibnamefont{et~al.}
  (\bibinfo{collaboration}{LISA Cosmology Working Group}),
  \bibinfo{journal}{Living Rev. Rel.} \textbf{\bibinfo{volume}{26}},
  \bibinfo{pages}{5} (\bibinfo{year}{2023}{\natexlab{b}}), \eprint{2204.05434}.

\bibitem[{\citenamefont{Brzeminski et~al.}(2022)\citenamefont{Brzeminski, Hook,
  and Marques-Tavares}}]{Brzeminski:2022haa}
\bibinfo{author}{\bibfnamefont{D.}~\bibnamefont{Brzeminski}},
  \bibinfo{author}{\bibfnamefont{A.}~\bibnamefont{Hook}}, \bibnamefont{and}
  \bibinfo{author}{\bibfnamefont{G.}~\bibnamefont{Marques-Tavares}},
  \bibinfo{journal}{JHEP} \textbf{\bibinfo{volume}{11}}, \bibinfo{pages}{061}
  (\bibinfo{year}{2022}), \eprint{2203.13842}.

\bibitem[{\citenamefont{Agrawal et~al.}(2022)\citenamefont{Agrawal, Hook,
  Huang, and Marques-Tavares}}]{Agrawal:2020euj}
\bibinfo{author}{\bibfnamefont{P.}~\bibnamefont{Agrawal}},
  \bibinfo{author}{\bibfnamefont{A.}~\bibnamefont{Hook}},
  \bibinfo{author}{\bibfnamefont{J.}~\bibnamefont{Huang}}, \bibnamefont{and}
  \bibinfo{author}{\bibfnamefont{G.}~\bibnamefont{Marques-Tavares}},
  \bibinfo{journal}{JHEP} \textbf{\bibinfo{volume}{01}}, \bibinfo{pages}{103}
  (\bibinfo{year}{2022}), \eprint{2010.15848}.

\bibitem[{\citenamefont{Jain et~al.}(2021)\citenamefont{Jain, Long, and
  Amin}}]{Jain:2021shf}
\bibinfo{author}{\bibfnamefont{M.}~\bibnamefont{Jain}},
  \bibinfo{author}{\bibfnamefont{A.~J.} \bibnamefont{Long}}, \bibnamefont{and}
  \bibinfo{author}{\bibfnamefont{M.~A.} \bibnamefont{Amin}},
  \bibinfo{journal}{JCAP} \textbf{\bibinfo{volume}{05}}, \bibinfo{pages}{055}
  (\bibinfo{year}{2021}), \eprint{2103.10962}.

\bibitem[{\citenamefont{Jain et~al.}(2022)\citenamefont{Jain, Hagimoto, Long,
  and Amin}}]{Jain:2022jrp}
\bibinfo{author}{\bibfnamefont{M.}~\bibnamefont{Jain}},
  \bibinfo{author}{\bibfnamefont{R.}~\bibnamefont{Hagimoto}},
  \bibinfo{author}{\bibfnamefont{A.~J.} \bibnamefont{Long}}, \bibnamefont{and}
  \bibinfo{author}{\bibfnamefont{M.~A.} \bibnamefont{Amin}},
  \bibinfo{journal}{JCAP} \textbf{\bibinfo{volume}{10}}, \bibinfo{pages}{090}
  (\bibinfo{year}{2022}), \eprint{2208.08391}.

\bibitem[{\citenamefont{Agrawal et~al.}(2020)\citenamefont{Agrawal, Hook, and
  Huang}}]{Agrawal:2019lkr}
\bibinfo{author}{\bibfnamefont{P.}~\bibnamefont{Agrawal}},
  \bibinfo{author}{\bibfnamefont{A.}~\bibnamefont{Hook}}, \bibnamefont{and}
  \bibinfo{author}{\bibfnamefont{J.}~\bibnamefont{Huang}},
  \bibinfo{journal}{JHEP} \textbf{\bibinfo{volume}{07}}, \bibinfo{pages}{138}
  (\bibinfo{year}{2020}), \eprint{1912.02823}.

\bibitem[{\citenamefont{Dessert et~al.}(2022)\citenamefont{Dessert, Long, and
  Safdi}}]{Dessert:2021bkv}
\bibinfo{author}{\bibfnamefont{C.}~\bibnamefont{Dessert}},
  \bibinfo{author}{\bibfnamefont{A.~J.} \bibnamefont{Long}}, \bibnamefont{and}
  \bibinfo{author}{\bibfnamefont{B.~R.} \bibnamefont{Safdi}},
  \bibinfo{journal}{Phys. Rev. Lett.} \textbf{\bibinfo{volume}{128}},
  \bibinfo{pages}{071102} (\bibinfo{year}{2022}), \eprint{2104.12772}.

\bibitem[{\citenamefont{Shokair et~al.}(2014)}]{Shokair:2014rna}
\bibinfo{author}{\bibfnamefont{T.~M.} \bibnamefont{Shokair}}
  \bibnamefont{et~al.}, \bibinfo{journal}{Int. J. Mod. Phys. A}
  \textbf{\bibinfo{volume}{29}}, \bibinfo{pages}{1443004}
  (\bibinfo{year}{2014}), \eprint{1405.3685}.

\bibitem[{\citenamefont{Du et~al.}(2018)}]{ADMX:2018gho}
\bibinfo{author}{\bibfnamefont{N.}~\bibnamefont{Du}} \bibnamefont{et~al.}
  (\bibinfo{collaboration}{ADMX}), \bibinfo{journal}{Phys. Rev. Lett.}
  \textbf{\bibinfo{volume}{120}}, \bibinfo{pages}{151301}
  (\bibinfo{year}{2018}), \eprint{1804.05750}.

\bibitem[{\citenamefont{Brubaker
  et~al.}(2017{\natexlab{a}})}]{Brubaker:2016ktl}
\bibinfo{author}{\bibfnamefont{B.~M.} \bibnamefont{Brubaker}}
  \bibnamefont{et~al.}, \bibinfo{journal}{Phys. Rev. Lett.}
  \textbf{\bibinfo{volume}{118}}, \bibinfo{pages}{061302}
  (\bibinfo{year}{2017}{\natexlab{a}}), \eprint{1610.02580}.

\bibitem[{\citenamefont{Al~Kenany et~al.}(2017)}]{AlKenany:2016trt}
\bibinfo{author}{\bibfnamefont{S.}~\bibnamefont{Al~Kenany}}
  \bibnamefont{et~al.}, \bibinfo{journal}{Nucl. Instrum. Meth. A}
  \textbf{\bibinfo{volume}{854}}, \bibinfo{pages}{11} (\bibinfo{year}{2017}),
  \eprint{1611.07123}.

\bibitem[{\citenamefont{Brubaker
  et~al.}(2017{\natexlab{b}})\citenamefont{Brubaker, Zhong, Lamoreaux, Lehnert,
  and van Bibber}}]{Brubaker:2017rna}
\bibinfo{author}{\bibfnamefont{B.~M.} \bibnamefont{Brubaker}},
  \bibinfo{author}{\bibfnamefont{L.}~\bibnamefont{Zhong}},
  \bibinfo{author}{\bibfnamefont{S.~K.} \bibnamefont{Lamoreaux}},
  \bibinfo{author}{\bibfnamefont{K.~W.} \bibnamefont{Lehnert}},
  \bibnamefont{and} \bibinfo{author}{\bibfnamefont{K.~A.} \bibnamefont{van
  Bibber}}, \bibinfo{journal}{Phys. Rev. D} \textbf{\bibinfo{volume}{96}},
  \bibinfo{pages}{123008} (\bibinfo{year}{2017}{\natexlab{b}}),
  \eprint{1706.08388}.

\bibitem[{\citenamefont{Caldwell et~al.}(2017)\citenamefont{Caldwell, Dvali,
  Majorovits, Millar, Raffelt, Redondo, Reimann, Simon, and
  Steffen}}]{Caldwell:2016dcw}
\bibinfo{author}{\bibfnamefont{A.}~\bibnamefont{Caldwell}},
  \bibinfo{author}{\bibfnamefont{G.}~\bibnamefont{Dvali}},
  \bibinfo{author}{\bibfnamefont{B.}~\bibnamefont{Majorovits}},
  \bibinfo{author}{\bibfnamefont{A.}~\bibnamefont{Millar}},
  \bibinfo{author}{\bibfnamefont{G.}~\bibnamefont{Raffelt}},
  \bibinfo{author}{\bibfnamefont{J.}~\bibnamefont{Redondo}},
  \bibinfo{author}{\bibfnamefont{O.}~\bibnamefont{Reimann}},
  \bibinfo{author}{\bibfnamefont{F.}~\bibnamefont{Simon}}, \bibnamefont{and}
  \bibinfo{author}{\bibfnamefont{F.}~\bibnamefont{Steffen}}
  (\bibinfo{collaboration}{MADMAX Working Group}), \bibinfo{journal}{Phys. Rev.
  Lett.} \textbf{\bibinfo{volume}{118}}, \bibinfo{pages}{091801}
  (\bibinfo{year}{2017}), \eprint{1611.05865}.

\bibitem[{\citenamefont{Kahn et~al.}(2016)\citenamefont{Kahn, Safdi, and
  Thaler}}]{Kahn:2016aff}
\bibinfo{author}{\bibfnamefont{Y.}~\bibnamefont{Kahn}},
  \bibinfo{author}{\bibfnamefont{B.~R.} \bibnamefont{Safdi}}, \bibnamefont{and}
  \bibinfo{author}{\bibfnamefont{J.}~\bibnamefont{Thaler}},
  \bibinfo{journal}{Phys. Rev. Lett.} \textbf{\bibinfo{volume}{117}},
  \bibinfo{pages}{141801} (\bibinfo{year}{2016}), \eprint{1602.01086}.

\bibitem[{\citenamefont{Ouellet et~al.}(2019{\natexlab{a}})}]{Ouellet:2018beu}
\bibinfo{author}{\bibfnamefont{J.~L.} \bibnamefont{Ouellet}}
  \bibnamefont{et~al.}, \bibinfo{journal}{Phys. Rev. Lett.}
  \textbf{\bibinfo{volume}{122}}, \bibinfo{pages}{121802}
  (\bibinfo{year}{2019}{\natexlab{a}}), \eprint{1810.12257}.

\bibitem[{\citenamefont{Ouellet et~al.}(2019{\natexlab{b}})}]{Ouellet:2019tlz}
\bibinfo{author}{\bibfnamefont{J.~L.} \bibnamefont{Ouellet}}
  \bibnamefont{et~al.}, \bibinfo{journal}{Phys. Rev. D}
  \textbf{\bibinfo{volume}{99}}, \bibinfo{pages}{052012}
  (\bibinfo{year}{2019}{\natexlab{b}}), \eprint{1901.10652}.

\bibitem[{\citenamefont{Chaudhuri et~al.}(2015)\citenamefont{Chaudhuri, Graham,
  Irwin, Mardon, Rajendran, and Zhao}}]{Chaudhuri:2014dla}
\bibinfo{author}{\bibfnamefont{S.}~\bibnamefont{Chaudhuri}},
  \bibinfo{author}{\bibfnamefont{P.~W.} \bibnamefont{Graham}},
  \bibinfo{author}{\bibfnamefont{K.}~\bibnamefont{Irwin}},
  \bibinfo{author}{\bibfnamefont{J.}~\bibnamefont{Mardon}},
  \bibinfo{author}{\bibfnamefont{S.}~\bibnamefont{Rajendran}},
  \bibnamefont{and} \bibinfo{author}{\bibfnamefont{Y.}~\bibnamefont{Zhao}},
  \bibinfo{journal}{Phys. Rev. D} \textbf{\bibinfo{volume}{92}},
  \bibinfo{pages}{075012} (\bibinfo{year}{2015}), \eprint{1411.7382}.

\bibitem[{\citenamefont{Silva-Feaver et~al.}(2017)}]{Silva-Feaver:2016qhh}
\bibinfo{author}{\bibfnamefont{M.}~\bibnamefont{Silva-Feaver}}
  \bibnamefont{et~al.}, \bibinfo{journal}{IEEE Trans. Appl. Supercond.}
  \textbf{\bibinfo{volume}{27}}, \bibinfo{pages}{1400204}
  (\bibinfo{year}{2017}), \eprint{1610.09344}.

\bibitem[{\citenamefont{Kibble}(1976)}]{Kibble:1976sj}
\bibinfo{author}{\bibfnamefont{T.~W.~B.} \bibnamefont{Kibble}},
  \bibinfo{journal}{J. Phys. A} \textbf{\bibinfo{volume}{9}},
  \bibinfo{pages}{1387} (\bibinfo{year}{1976}).

\bibitem[{\citenamefont{Shellard and Battye}(1998)}]{Shellard:1998mi}
\bibinfo{author}{\bibfnamefont{E.~P.~S.} \bibnamefont{Shellard}}
  \bibnamefont{and} \bibinfo{author}{\bibfnamefont{R.~A.}
  \bibnamefont{Battye}}, \bibinfo{journal}{Phys. Rept.}
  \textbf{\bibinfo{volume}{307}}, \bibinfo{pages}{227} (\bibinfo{year}{1998}),
  \eprint{astro-ph/9808220}.

\bibitem[{\citenamefont{Sikivie}(2008)}]{Sikivie:2006ni}
\bibinfo{author}{\bibfnamefont{P.}~\bibnamefont{Sikivie}},
  \bibinfo{journal}{Lect. Notes Phys.} \textbf{\bibinfo{volume}{741}},
  \bibinfo{pages}{19} (\bibinfo{year}{2008}), \eprint{astro-ph/0610440}.

\bibitem[{\citenamefont{Kim and Carosi}(2010)}]{Kim:2008hd}
\bibinfo{author}{\bibfnamefont{J.~E.} \bibnamefont{Kim}} \bibnamefont{and}
  \bibinfo{author}{\bibfnamefont{G.}~\bibnamefont{Carosi}},
  \bibinfo{journal}{Rev. Mod. Phys.} \textbf{\bibinfo{volume}{82}},
  \bibinfo{pages}{557} (\bibinfo{year}{2010}), \bibinfo{note}{[Erratum:
  Rev.Mod.Phys. 91, 049902 (2019)]}, \eprint{0807.3125}.

\bibitem[{\citenamefont{Yamaguchi et~al.}(1999)\citenamefont{Yamaguchi,
  Kawasaki, and Yokoyama}}]{Yamaguchi:1998gx}
\bibinfo{author}{\bibfnamefont{M.}~\bibnamefont{Yamaguchi}},
  \bibinfo{author}{\bibfnamefont{M.}~\bibnamefont{Kawasaki}}, \bibnamefont{and}
  \bibinfo{author}{\bibfnamefont{J.}~\bibnamefont{Yokoyama}},
  \bibinfo{journal}{Phys. Rev. Lett.} \textbf{\bibinfo{volume}{82}},
  \bibinfo{pages}{4578} (\bibinfo{year}{1999}), \eprint{hep-ph/9811311}.

\bibitem[{\citenamefont{Yamaguchi and Yokoyama}(2003)}]{Yamaguchi:2002sh}
\bibinfo{author}{\bibfnamefont{M.}~\bibnamefont{Yamaguchi}} \bibnamefont{and}
  \bibinfo{author}{\bibfnamefont{J.}~\bibnamefont{Yokoyama}},
  \bibinfo{journal}{Phys. Rev. D} \textbf{\bibinfo{volume}{67}},
  \bibinfo{pages}{103514} (\bibinfo{year}{2003}), \eprint{hep-ph/0210343}.

\bibitem[{\citenamefont{Hiramatsu
  et~al.}(2011{\natexlab{a}})\citenamefont{Hiramatsu, Kawasaki, Sekiguchi,
  Yamaguchi, and Yokoyama}}]{Hiramatsu:2010yu}
\bibinfo{author}{\bibfnamefont{T.}~\bibnamefont{Hiramatsu}},
  \bibinfo{author}{\bibfnamefont{M.}~\bibnamefont{Kawasaki}},
  \bibinfo{author}{\bibfnamefont{T.}~\bibnamefont{Sekiguchi}},
  \bibinfo{author}{\bibfnamefont{M.}~\bibnamefont{Yamaguchi}},
  \bibnamefont{and} \bibinfo{author}{\bibfnamefont{J.}~\bibnamefont{Yokoyama}},
  \bibinfo{journal}{Phys. Rev. D} \textbf{\bibinfo{volume}{83}},
  \bibinfo{pages}{123531} (\bibinfo{year}{2011}{\natexlab{a}}),
  \eprint{1012.5502}.

\bibitem[{\citenamefont{Gouttenoire et~al.}(2020)\citenamefont{Gouttenoire,
  Servant, and Simakachorn}}]{Gouttenoire:2019kij}
\bibinfo{author}{\bibfnamefont{Y.}~\bibnamefont{Gouttenoire}},
  \bibinfo{author}{\bibfnamefont{G.}~\bibnamefont{Servant}}, \bibnamefont{and}
  \bibinfo{author}{\bibfnamefont{P.}~\bibnamefont{Simakachorn}},
  \bibinfo{journal}{JCAP} \textbf{\bibinfo{volume}{07}}, \bibinfo{pages}{032}
  (\bibinfo{year}{2020}), \eprint{1912.02569}.

\bibitem[{\citenamefont{Auclair et~al.}(2020)}]{Auclair:2019wcv}
\bibinfo{author}{\bibfnamefont{P.}~\bibnamefont{Auclair}} \bibnamefont{et~al.},
  \bibinfo{journal}{JCAP} \textbf{\bibinfo{volume}{04}}, \bibinfo{pages}{034}
  (\bibinfo{year}{2020}), \eprint{1909.00819}.

\bibitem[{\citenamefont{Dine et~al.}(1981)\citenamefont{Dine, Fischler, and
  Srednicki}}]{Dine:1981rt}
\bibinfo{author}{\bibfnamefont{M.}~\bibnamefont{Dine}},
  \bibinfo{author}{\bibfnamefont{W.}~\bibnamefont{Fischler}}, \bibnamefont{and}
  \bibinfo{author}{\bibfnamefont{M.}~\bibnamefont{Srednicki}},
  \bibinfo{journal}{Phys. Lett. B} \textbf{\bibinfo{volume}{104}},
  \bibinfo{pages}{199} (\bibinfo{year}{1981}).

\bibitem[{\citenamefont{Zhitnitsky}(1980)}]{Zhitnitsky:1980tq}
\bibinfo{author}{\bibfnamefont{A.~R.} \bibnamefont{Zhitnitsky}},
  \bibinfo{journal}{Sov. J. Nucl. Phys.} \textbf{\bibinfo{volume}{31}},
  \bibinfo{pages}{260} (\bibinfo{year}{1980}).

\bibitem[{\citenamefont{Kim}(1987)}]{Kim:1986ax}
\bibinfo{author}{\bibfnamefont{J.~E.} \bibnamefont{Kim}},
  \bibinfo{journal}{Phys. Rept.} \textbf{\bibinfo{volume}{150}},
  \bibinfo{pages}{1} (\bibinfo{year}{1987}).

\bibitem[{\citenamefont{Kim}(1979)}]{Kim:1979if}
\bibinfo{author}{\bibfnamefont{J.~E.} \bibnamefont{Kim}},
  \bibinfo{journal}{Phys. Rev. Lett.} \textbf{\bibinfo{volume}{43}},
  \bibinfo{pages}{103} (\bibinfo{year}{1979}).

\bibitem[{\citenamefont{Shifman et~al.}(1980)\citenamefont{Shifman, Vainshtein,
  and Zakharov}}]{Shifman:1979if}
\bibinfo{author}{\bibfnamefont{M.~A.} \bibnamefont{Shifman}},
  \bibinfo{author}{\bibfnamefont{A.~I.} \bibnamefont{Vainshtein}},
  \bibnamefont{and} \bibinfo{author}{\bibfnamefont{V.~I.}
  \bibnamefont{Zakharov}}, \bibinfo{journal}{Nucl. Phys. B}
  \textbf{\bibinfo{volume}{166}}, \bibinfo{pages}{493} (\bibinfo{year}{1980}).

\bibitem[{\citenamefont{Sikivie}(1982)}]{Sikivie:1982qv}
\bibinfo{author}{\bibfnamefont{P.}~\bibnamefont{Sikivie}},
  \bibinfo{journal}{Phys. Rev. Lett.} \textbf{\bibinfo{volume}{48}},
  \bibinfo{pages}{1156} (\bibinfo{year}{1982}).

\bibitem[{\citenamefont{Zeldovich et~al.}(1974)\citenamefont{Zeldovich,
  Kobzarev, and Okun}}]{Zeldovich:1974uw}
\bibinfo{author}{\bibfnamefont{Y.~B.} \bibnamefont{Zeldovich}},
  \bibinfo{author}{\bibfnamefont{I.~Y.} \bibnamefont{Kobzarev}},
  \bibnamefont{and} \bibinfo{author}{\bibfnamefont{L.~B.} \bibnamefont{Okun}},
  \bibinfo{journal}{Zh. Eksp. Teor. Fiz.} \textbf{\bibinfo{volume}{67}},
  \bibinfo{pages}{3} (\bibinfo{year}{1974}).

\bibitem[{\citenamefont{Gelmini et~al.}(1989)\citenamefont{Gelmini, Gleiser,
  and Kolb}}]{Gelmini:1988sf}
\bibinfo{author}{\bibfnamefont{G.~B.} \bibnamefont{Gelmini}},
  \bibinfo{author}{\bibfnamefont{M.}~\bibnamefont{Gleiser}}, \bibnamefont{and}
  \bibinfo{author}{\bibfnamefont{E.~W.} \bibnamefont{Kolb}},
  \bibinfo{journal}{Phys. Rev. D} \textbf{\bibinfo{volume}{39}},
  \bibinfo{pages}{1558} (\bibinfo{year}{1989}).

\bibitem[{\citenamefont{Larsson et~al.}(1997)\citenamefont{Larsson, Sarkar, and
  White}}]{Larsson:1996sp}
\bibinfo{author}{\bibfnamefont{S.~E.} \bibnamefont{Larsson}},
  \bibinfo{author}{\bibfnamefont{S.}~\bibnamefont{Sarkar}}, \bibnamefont{and}
  \bibinfo{author}{\bibfnamefont{P.~L.} \bibnamefont{White}},
  \bibinfo{journal}{Phys. Rev. D} \textbf{\bibinfo{volume}{55}},
  \bibinfo{pages}{5129} (\bibinfo{year}{1997}), \eprint{hep-ph/9608319}.

\bibitem[{\citenamefont{Kamionkowski and
  March-Russell}(1992)}]{Kamionkowski:1992mf}
\bibinfo{author}{\bibfnamefont{M.}~\bibnamefont{Kamionkowski}}
  \bibnamefont{and}
  \bibinfo{author}{\bibfnamefont{J.}~\bibnamefont{March-Russell}},
  \bibinfo{journal}{Phys. Lett. B} \textbf{\bibinfo{volume}{282}},
  \bibinfo{pages}{137} (\bibinfo{year}{1992}), \eprint{hep-th/9202003}.

\bibitem[{\citenamefont{Holman et~al.}(1992)\citenamefont{Holman, Hsu, Kephart,
  Kolb, Watkins, and Widrow}}]{Holman:1992us}
\bibinfo{author}{\bibfnamefont{R.}~\bibnamefont{Holman}},
  \bibinfo{author}{\bibfnamefont{S.~D.~H.} \bibnamefont{Hsu}},
  \bibinfo{author}{\bibfnamefont{T.~W.} \bibnamefont{Kephart}},
  \bibinfo{author}{\bibfnamefont{E.~W.} \bibnamefont{Kolb}},
  \bibinfo{author}{\bibfnamefont{R.}~\bibnamefont{Watkins}}, \bibnamefont{and}
  \bibinfo{author}{\bibfnamefont{L.~M.} \bibnamefont{Widrow}},
  \bibinfo{journal}{Phys. Lett. B} \textbf{\bibinfo{volume}{282}},
  \bibinfo{pages}{132} (\bibinfo{year}{1992}), \eprint{hep-ph/9203206}.

\bibitem[{\citenamefont{Dobrescu}(1997)}]{Dobrescu:1996jp}
\bibinfo{author}{\bibfnamefont{B.~A.} \bibnamefont{Dobrescu}},
  \bibinfo{journal}{Phys. Rev. D} \textbf{\bibinfo{volume}{55}},
  \bibinfo{pages}{5826} (\bibinfo{year}{1997}), \eprint{hep-ph/9609221}.

\bibitem[{\citenamefont{Barr and Seckel}(1992)}]{Barr:1992qq}
\bibinfo{author}{\bibfnamefont{S.~M.} \bibnamefont{Barr}} \bibnamefont{and}
  \bibinfo{author}{\bibfnamefont{D.}~\bibnamefont{Seckel}},
  \bibinfo{journal}{Phys. Rev. D} \textbf{\bibinfo{volume}{46}},
  \bibinfo{pages}{539} (\bibinfo{year}{1992}).

\bibitem[{\citenamefont{Dine}(1992)}]{Dine:1992vx}
\bibinfo{author}{\bibfnamefont{M.}~\bibnamefont{Dine}}, in
  \emph{\bibinfo{booktitle}{{Conference on Topics in Quantum Gravity}}}
  (\bibinfo{year}{1992}), \eprint{hep-th/9207045}.

\bibitem[{\citenamefont{Hiramatsu
  et~al.}(2011{\natexlab{b}})\citenamefont{Hiramatsu, Kawasaki, and
  Saikawa}}]{Hiramatsu:2010yn}
\bibinfo{author}{\bibfnamefont{T.}~\bibnamefont{Hiramatsu}},
  \bibinfo{author}{\bibfnamefont{M.}~\bibnamefont{Kawasaki}}, \bibnamefont{and}
  \bibinfo{author}{\bibfnamefont{K.}~\bibnamefont{Saikawa}},
  \bibinfo{journal}{JCAP} \textbf{\bibinfo{volume}{08}}, \bibinfo{pages}{030}
  (\bibinfo{year}{2011}{\natexlab{b}}), \eprint{1012.4558}.

\bibitem[{\citenamefont{Hiramatsu et~al.}(2013)\citenamefont{Hiramatsu,
  Kawasaki, Saikawa, and Sekiguchi}}]{Hiramatsu:2012sc}
\bibinfo{author}{\bibfnamefont{T.}~\bibnamefont{Hiramatsu}},
  \bibinfo{author}{\bibfnamefont{M.}~\bibnamefont{Kawasaki}},
  \bibinfo{author}{\bibfnamefont{K.}~\bibnamefont{Saikawa}}, \bibnamefont{and}
  \bibinfo{author}{\bibfnamefont{T.}~\bibnamefont{Sekiguchi}},
  \bibinfo{journal}{JCAP} \textbf{\bibinfo{volume}{01}}, \bibinfo{pages}{001}
  (\bibinfo{year}{2013}), \eprint{1207.3166}.

\bibitem[{\citenamefont{Hiramatsu et~al.}(2012)\citenamefont{Hiramatsu,
  Kawasaki, Saikawa, and Sekiguchi}}]{Hiramatsu:2012gg}
\bibinfo{author}{\bibfnamefont{T.}~\bibnamefont{Hiramatsu}},
  \bibinfo{author}{\bibfnamefont{M.}~\bibnamefont{Kawasaki}},
  \bibinfo{author}{\bibfnamefont{K.}~\bibnamefont{Saikawa}}, \bibnamefont{and}
  \bibinfo{author}{\bibfnamefont{T.}~\bibnamefont{Sekiguchi}},
  \bibinfo{journal}{Phys. Rev. D} \textbf{\bibinfo{volume}{85}},
  \bibinfo{pages}{105020} (\bibinfo{year}{2012}), \bibinfo{note}{[Erratum:
  Phys.Rev.D 86, 089902 (2012)]}, \eprint{1202.5851}.

\bibitem[{\citenamefont{Saikawa}(2017)}]{Saikawa:2017hiv}
\bibinfo{author}{\bibfnamefont{K.}~\bibnamefont{Saikawa}},
  \bibinfo{journal}{Universe} \textbf{\bibinfo{volume}{3}}, \bibinfo{pages}{40}
  (\bibinfo{year}{2017}), \eprint{1703.02576}.

\bibitem[{\citenamefont{Gelmini et~al.}(2021)\citenamefont{Gelmini, Simpson,
  and Vitagliano}}]{Gelmini:2021yzu}
\bibinfo{author}{\bibfnamefont{G.~B.} \bibnamefont{Gelmini}},
  \bibinfo{author}{\bibfnamefont{A.}~\bibnamefont{Simpson}}, \bibnamefont{and}
  \bibinfo{author}{\bibfnamefont{E.}~\bibnamefont{Vitagliano}},
  \bibinfo{journal}{Phys. Rev. D} \textbf{\bibinfo{volume}{104}},
  \bibinfo{pages}{061301} (\bibinfo{year}{2021}), \eprint{2103.07625}.

\bibitem[{\citenamefont{Wantz and Shellard}(2010)}]{Wantz:2009it}
\bibinfo{author}{\bibfnamefont{O.}~\bibnamefont{Wantz}} \bibnamefont{and}
  \bibinfo{author}{\bibfnamefont{E.~P.~S.} \bibnamefont{Shellard}},
  \bibinfo{journal}{Phys. Rev. D} \textbf{\bibinfo{volume}{82}},
  \bibinfo{pages}{123508} (\bibinfo{year}{2010}), \eprint{0910.1066}.

\bibitem[{\citenamefont{Grilli~di Cortona et~al.}(2016)\citenamefont{Grilli~di
  Cortona, Hardy, Pardo~Vega, and Villadoro}}]{GrillidiCortona:2015jxo}
\bibinfo{author}{\bibfnamefont{G.}~\bibnamefont{Grilli~di Cortona}},
  \bibinfo{author}{\bibfnamefont{E.}~\bibnamefont{Hardy}},
  \bibinfo{author}{\bibfnamefont{J.}~\bibnamefont{Pardo~Vega}},
  \bibnamefont{and}
  \bibinfo{author}{\bibfnamefont{G.}~\bibnamefont{Villadoro}},
  \bibinfo{journal}{JHEP} \textbf{\bibinfo{volume}{01}}, \bibinfo{pages}{034}
  (\bibinfo{year}{2016}), \eprint{1511.02867}.

\bibitem[{\citenamefont{Srednicki}(1985)}]{Srednicki:1985xd}
\bibinfo{author}{\bibfnamefont{M.}~\bibnamefont{Srednicki}},
  \bibinfo{journal}{Nucl. Phys. B} \textbf{\bibinfo{volume}{260}},
  \bibinfo{pages}{689} (\bibinfo{year}{1985}).

\bibitem[{\citenamefont{Figueroa et~al.}(2023)\citenamefont{Figueroa, Florio,
  Torrenti, and Valkenburg}}]{Figueroa:2021yhd}
\bibinfo{author}{\bibfnamefont{D.~G.} \bibnamefont{Figueroa}},
  \bibinfo{author}{\bibfnamefont{A.}~\bibnamefont{Florio}},
  \bibinfo{author}{\bibfnamefont{F.}~\bibnamefont{Torrenti}}, \bibnamefont{and}
  \bibinfo{author}{\bibfnamefont{W.}~\bibnamefont{Valkenburg}},
  \bibinfo{journal}{Comput. Phys. Commun.} \textbf{\bibinfo{volume}{283}},
  \bibinfo{pages}{108586} (\bibinfo{year}{2023}), \eprint{2102.01031}.

\bibitem[{\citenamefont{Figueroa et~al.}(2021)\citenamefont{Figueroa, Florio,
  Torrenti, and Valkenburg}}]{Figueroa:2020rrl}
\bibinfo{author}{\bibfnamefont{D.~G.} \bibnamefont{Figueroa}},
  \bibinfo{author}{\bibfnamefont{A.}~\bibnamefont{Florio}},
  \bibinfo{author}{\bibfnamefont{F.}~\bibnamefont{Torrenti}}, \bibnamefont{and}
  \bibinfo{author}{\bibfnamefont{W.}~\bibnamefont{Valkenburg}},
  \bibinfo{journal}{JCAP} \textbf{\bibinfo{volume}{04}}, \bibinfo{pages}{035}
  (\bibinfo{year}{2021}), \eprint{2006.15122}.

\bibitem[{\citenamefont{Hindmarsh et~al.}(2020)\citenamefont{Hindmarsh,
  Lizarraga, Lopez-Eiguren, and Urrestilla}}]{Hindmarsh:2019csc}
\bibinfo{author}{\bibfnamefont{M.}~\bibnamefont{Hindmarsh}},
  \bibinfo{author}{\bibfnamefont{J.}~\bibnamefont{Lizarraga}},
  \bibinfo{author}{\bibfnamefont{A.}~\bibnamefont{Lopez-Eiguren}},
  \bibnamefont{and}
  \bibinfo{author}{\bibfnamefont{J.}~\bibnamefont{Urrestilla}},
  \bibinfo{journal}{Phys. Rev. Lett.} \textbf{\bibinfo{volume}{124}},
  \bibinfo{pages}{021301} (\bibinfo{year}{2020}), \eprint{1908.03522}.

\bibitem[{\citenamefont{Gorghetto et~al.}(2018)\citenamefont{Gorghetto, Hardy,
  and Villadoro}}]{Gorghetto:2018myk}
\bibinfo{author}{\bibfnamefont{M.}~\bibnamefont{Gorghetto}},
  \bibinfo{author}{\bibfnamefont{E.}~\bibnamefont{Hardy}}, \bibnamefont{and}
  \bibinfo{author}{\bibfnamefont{G.}~\bibnamefont{Villadoro}},
  \bibinfo{journal}{JHEP} \textbf{\bibinfo{volume}{07}}, \bibinfo{pages}{151}
  (\bibinfo{year}{2018}), \eprint{1806.04677}.

\bibitem[{\citenamefont{Kawasaki et~al.}(2018)\citenamefont{Kawasaki,
  Sekiguchi, Yamaguchi, and Yokoyama}}]{Kawasaki:2018bzv}
\bibinfo{author}{\bibfnamefont{M.}~\bibnamefont{Kawasaki}},
  \bibinfo{author}{\bibfnamefont{T.}~\bibnamefont{Sekiguchi}},
  \bibinfo{author}{\bibfnamefont{M.}~\bibnamefont{Yamaguchi}},
  \bibnamefont{and} \bibinfo{author}{\bibfnamefont{J.}~\bibnamefont{Yokoyama}},
  \bibinfo{journal}{PTEP} \textbf{\bibinfo{volume}{2018}},
  \bibinfo{pages}{091E01} (\bibinfo{year}{2018}), \eprint{1806.05566}.

\bibitem[{\citenamefont{Kolb and Turner}(1990)}]{Kolb:1990vq}
\bibinfo{author}{\bibfnamefont{E.~W.} \bibnamefont{Kolb}} \bibnamefont{and}
  \bibinfo{author}{\bibfnamefont{M.~S.} \bibnamefont{Turner}},
  \emph{\bibinfo{title}{{The Early Universe}}}, vol.~\bibinfo{volume}{69}
  (\bibinfo{year}{1990}), ISBN \bibinfo{isbn}{978-0-201-62674-2}.

\bibitem[{\citenamefont{Fleury and Moore}(2016)}]{Fleury:2015aca}
\bibinfo{author}{\bibfnamefont{L.}~\bibnamefont{Fleury}} \bibnamefont{and}
  \bibinfo{author}{\bibfnamefont{G.~D.} \bibnamefont{Moore}},
  \bibinfo{journal}{JCAP} \textbf{\bibinfo{volume}{01}}, \bibinfo{pages}{004}
  (\bibinfo{year}{2016}), \eprint{1509.00026}.

\bibitem[{\citenamefont{Hiramatsu et~al.}(2010)\citenamefont{Hiramatsu,
  Kawasaki, and Saikawa}}]{Hiramatsu:2010yz}
\bibinfo{author}{\bibfnamefont{T.}~\bibnamefont{Hiramatsu}},
  \bibinfo{author}{\bibfnamefont{M.}~\bibnamefont{Kawasaki}}, \bibnamefont{and}
  \bibinfo{author}{\bibfnamefont{K.}~\bibnamefont{Saikawa}},
  \bibinfo{journal}{JCAP} \textbf{\bibinfo{volume}{05}}, \bibinfo{pages}{032}
  (\bibinfo{year}{2010}), \eprint{1002.1555}.

\bibitem[{\citenamefont{Jeong et~al.}(2014)\citenamefont{Jeong, Kawasaki, and
  Takahashi}}]{Jeong:2013oza}
\bibinfo{author}{\bibfnamefont{K.~S.} \bibnamefont{Jeong}},
  \bibinfo{author}{\bibfnamefont{M.}~\bibnamefont{Kawasaki}}, \bibnamefont{and}
  \bibinfo{author}{\bibfnamefont{F.}~\bibnamefont{Takahashi}},
  \bibinfo{journal}{JCAP} \textbf{\bibinfo{volume}{02}}, \bibinfo{pages}{046}
  (\bibinfo{year}{2014}), \eprint{1310.1774}.

\bibitem[{\citenamefont{Weinberg}(2008)}]{Weinberg:2008zzc}
\bibinfo{author}{\bibfnamefont{S.}~\bibnamefont{Weinberg}},
  \emph{\bibinfo{title}{{Cosmology}}} (\bibinfo{year}{2008}), ISBN
  \bibinfo{isbn}{978-0-19-852682-7}.

\bibitem[{\citenamefont{O'Hare}(2020)}]{AxionLimits}
\bibinfo{author}{\bibfnamefont{C.}~\bibnamefont{O'Hare}}
  (\bibinfo{year}{2020}),
  \urlprefix\url{https://cajohare.github.io/AxionLimits/}.

\bibitem[{\citenamefont{Price and Siemens}(2008)}]{Price:2008hq}
\bibinfo{author}{\bibfnamefont{L.~R.} \bibnamefont{Price}} \bibnamefont{and}
  \bibinfo{author}{\bibfnamefont{X.}~\bibnamefont{Siemens}},
  \bibinfo{journal}{Phys. Rev. D} \textbf{\bibinfo{volume}{78}},
  \bibinfo{pages}{063541} (\bibinfo{year}{2008}), \eprint{0805.3570}.

\bibitem[{\citenamefont{Dufaux et~al.}(2007)\citenamefont{Dufaux, Bergman,
  Felder, Kofman, and Uzan}}]{Dufaux:2007pt}
\bibinfo{author}{\bibfnamefont{J.~F.} \bibnamefont{Dufaux}},
  \bibinfo{author}{\bibfnamefont{A.}~\bibnamefont{Bergman}},
  \bibinfo{author}{\bibfnamefont{G.~N.} \bibnamefont{Felder}},
  \bibinfo{author}{\bibfnamefont{L.}~\bibnamefont{Kofman}}, \bibnamefont{and}
  \bibinfo{author}{\bibfnamefont{J.-P.} \bibnamefont{Uzan}},
  \bibinfo{journal}{Phys. Rev. D} \textbf{\bibinfo{volume}{76}},
  \bibinfo{pages}{123517} (\bibinfo{year}{2007}), \eprint{0707.0875}.

\bibitem[{\citenamefont{Easther et~al.}(2008)\citenamefont{Easther, Giblin, and
  Lim}}]{Easther:2007vj}
\bibinfo{author}{\bibfnamefont{R.}~\bibnamefont{Easther}},
  \bibinfo{author}{\bibfnamefont{J.~T.} \bibnamefont{Giblin}},
  \bibnamefont{and} \bibinfo{author}{\bibfnamefont{E.~A.} \bibnamefont{Lim}},
  \bibinfo{journal}{Phys. Rev. D} \textbf{\bibinfo{volume}{77}},
  \bibinfo{pages}{103519} (\bibinfo{year}{2008}), \eprint{0712.2991}.

\bibitem[{\citenamefont{Lopez-Eiguren et~al.}(2017)\citenamefont{Lopez-Eiguren,
  Lizarraga, Hindmarsh, and Urrestilla}}]{Lopez-Eiguren:2017dmc}
\bibinfo{author}{\bibfnamefont{A.}~\bibnamefont{Lopez-Eiguren}},
  \bibinfo{author}{\bibfnamefont{J.}~\bibnamefont{Lizarraga}},
  \bibinfo{author}{\bibfnamefont{M.}~\bibnamefont{Hindmarsh}},
  \bibnamefont{and}
  \bibinfo{author}{\bibfnamefont{J.}~\bibnamefont{Urrestilla}},
  \bibinfo{journal}{JCAP} \textbf{\bibinfo{volume}{07}}, \bibinfo{pages}{026}
  (\bibinfo{year}{2017}), \eprint{1705.04154}.

\bibitem[{\citenamefont{Fenu et~al.}(2009)\citenamefont{Fenu, Figueroa, Durrer,
  and Garcia-Bellido}}]{Fenu:2009qf}
\bibinfo{author}{\bibfnamefont{E.}~\bibnamefont{Fenu}},
  \bibinfo{author}{\bibfnamefont{D.~G.} \bibnamefont{Figueroa}},
  \bibinfo{author}{\bibfnamefont{R.}~\bibnamefont{Durrer}}, \bibnamefont{and}
  \bibinfo{author}{\bibfnamefont{J.}~\bibnamefont{Garcia-Bellido}},
  \bibinfo{journal}{JCAP} \textbf{\bibinfo{volume}{10}}, \bibinfo{pages}{005}
  (\bibinfo{year}{2009}), \eprint{0908.0425}.

\end{thebibliography}

\clearpage

\onecolumngrid
\begin{center}
  \textbf{\large \it Supplemental Material}\\[.2cm]
\end{center}

\onecolumngrid
\setcounter{equation}{0}
\setcounter{figure}{0}
\setcounter{table}{0}
\setcounter{section}{0}
\setcounter{page}{1}
\makeatletter
\renewcommand{\theequation}{S\arabic{equation}}
\renewcommand{\thefigure}{S\arabic{figure}}

This supplemental material includes the details of our simulation and explanations of the results presented in the bulk text, as well as some extra results. We begin with a detailed description of our numerical scheme, including equations of motion and initial conditions. Then, we introduce the methods of identifying topological defects on a discrete lattice, together with the calculation of the scaling parameters of string and the comoving area density of the domain wall (DW). Next, we explained how to measure the power spectrum and energy density of free axions. Besides, we explain the procedure of calculating the power spectrum of gravitational waves (GWs). Finally, we present the simulation results for the spectrum of GWs in the case of $N_{\rm DW}=1$ and $N_{\rm DW}>1$ in the QCD axion scenario.

\section{Equations of motion}\label{sec:SupEOM}
We conduct simulations in the radiation-dominated universe with FLRW metric
\begin{equation}
{\rm d}s^2=g_{\mu\nu}{\rm d}x^{\mu}{\rm d}x^{\nu}=R^2(\eta)(-{\rm d}\eta^2+{\rm d}\boldsymbol{x}^2),
\end{equation}
with $\eta=\int{\frac{{\rm d}t}{R(t)}}$ the conformal time, and $R$ denote the scale factor.
The field equations of motion can be obtained by varying the following action
\begin{equation}
S=-\int{d^4x\sqrt{-g}\left\{ \frac{1}{2}\partial_\mu\varphi^*\partial^\mu\varphi+\frac{1}{4}\lambda(|\varphi|^2-v^2)^2+\frac{\lambda}{6}T^2|\varphi|^2+\frac{m^2(T)v^2}{N_{\rm DW}^2}(1-\frac{|\varphi|}{v}{\rm cos}(N_{\rm DW}\theta))-\Xi v^3(\varphi e^{-i\delta}+{\rm h.c.}) \right\}},
\end{equation}
with $\varphi=\phi_1+i\phi_2$ the PQ complex scalar. Note, that we multiply an additional factor $|\varphi|$ in front of the cosine term of the original QCD potential to avoid the singularity at $|\varphi|=0$. This modification prevents numerical instabilities in the simulation and does not affect the quantitative behavior of topological defects to a large extent~\cite{Hiramatsu:2012gg,Hiramatsu:2012sc}. In the form of the above action, we can get the equations of motion of $\phi_1$ and $\phi_2$ in curved space
\begin{eqnarray}
&&\phi_1''+2\frac{R'}{R}\phi_1'-\nabla^2 \phi_1 \nonumber\\
 &&\ =-R^2\left\{ \lambda\phi_1(\phi_1^2+\phi_2^2-v^2+\frac{1}{3}T^2)-\frac{m^2(T)v}{N_{\rm DW}^2}({\rm cos}\theta {\rm cos}N_{\rm DW}\theta+N_{\rm DW}{\rm sin}\theta{\rm sin}N_{\rm DW}\theta)-2\Xi v^3{\rm cos}\delta \right\}, \\
&&\phi_2''+2\frac{R'}{R}\phi_2'-\nabla^2 \phi_2 \nonumber\\
&&\ =-R^2\left\{ \lambda\phi_2(\phi_1^2+\phi_2^2-v^2+\frac{1}{3}T^2)-\frac{m^2(T)v}{N_{\rm DW}^2}({\rm sin}\theta {\rm cos}N_{\rm DW}\theta-N_{\rm DW}{\rm cos}\theta{\rm sin}N_{\rm DW}\theta)-2\Xi v^3{\rm sin}\delta \right\},
\end{eqnarray}
with $'={\rm d}/{\rm d}\eta$ and $\nabla_i=\partial/\partial x^i$.

We use two parameters $f_*$ and $w_*$ to do the rescale for physical quantities
\begin{equation}
\tilde{\phi}_i=\phi_i/f_*, \quad {\rm d}\tilde{\eta}=\frac{1}{R}w_*{\rm d}t, \quad {\rm d}\tilde{x}^i=w_*{\rm d}x^i, \quad \tilde{S}=\left(\frac{w_*}{f_*}\right)^2S(f_*\tilde{\varphi}). \label{dimensionless rescale} 
\end{equation}

Then, the equations of motion expressed by dimensionless fields and space-time variables follow immediately from varying the dimensionless action $\tilde{S}$  
\begin{eqnarray}
&&\tilde{\phi}_1''+2\frac{R'}{R}\tilde{\phi}_1'-\tilde{\nabla}^2 \tilde{\phi}_1 \nonumber\\
 &&\ =-R^2\left\{ \lambda\frac{f_*^2}{w_*^2}\tilde{\phi}_1(\tilde{\phi}_1^2+\tilde{\phi}_2^2-\frac{v^2}{f_*^2}+\frac{1}{3}\frac{T^2}{f_*^2})-\frac{m^2(T)v}{N_{\rm DW}^2f_*w_*^2}({\rm cos}\theta {\rm cos}N_{\rm DW}\theta+N_{\rm DW}{\rm sin}\theta{\rm sin}N_{\rm DW}\theta)-2\Xi \frac{v^3}{f_*w_*^2}{\rm cos}\delta \right\}, \\
&&\tilde{\phi}_2''+2\frac{R'}{R}\tilde{\phi}_2'-\tilde{\nabla}^2 \tilde{\phi}_2 \nonumber\\
 &&\ =-R^2\left\{ \lambda\frac{f_*^2}{w_*^2}\tilde{\phi}_2(\tilde{\phi}_1^2+\tilde{\phi}_2^2-\frac{v^2}{f_*^2}+\frac{1}{3}\frac{T^2}{f_*^2})-\frac{m^2(T)v}{N_{\rm DW}^2f_*w_*^2}({\rm sin}\theta {\rm cos}N_{\rm DW}\theta-N_{\rm DW}{\rm cos}\theta{\rm sin}N_{\rm DW}\theta)-2\Xi \frac{v^3}{f_*w_*^2}{\rm sin}\delta \right\}\label{S7}
\end{eqnarray}
with $'={\rm d}/{\rm d}\tilde{\eta}$ and $\tilde{\nabla}_i=\partial/\partial\tilde{x}^i$ here.

We use the rescaled conformal time $\eta_{\rm re}$ in our simulations of the PQ era and QCD era, 
\begin{equation}
\eta_{\rm re}=\frac{\eta}{\eta_i}=\frac{R(\eta)}{R_i}=\frac{g_*(T)^{-1/3}T_i}{g_{*}(T_i)^{-1/3}T(\eta)}. \label{rescaled conformal time}
\end{equation}
Where the quantities with subscript "$i$" indicate that it was defined at the initial time of the PQ era or QCD era, and $g_*(T)$ denotes the number of temperature-dependent relativistic degrees of freedom. The evolution of scale factor in a radiation-dominated universe is generally determined by the Friedmann equations,
\begin{equation}
R(\eta)=R_i\left(1+ \frac{1}{p}R_iH_i\times(\eta-\eta_i) \right)^p,
\label{scale factor}
\end{equation}
where $p\equiv 2/(3(1+w)-2)$ is the pressure density of the background fluid, and $w=1/3$ is the equation of state. Combining Eq.~(\ref{rescaled conformal time}) and Eq.~(\ref{scale factor}), the initial conformal time must be chosen as $\eta_i=1/(R_iH_i)$. To distinguish, we denote the rescaled conformal time of the PQ era and the QCD era as $\tilde{\eta}$ and $\hat{\eta}$, respectively.

In the PQ era, the critical temperature of PQ phase transition is $T_{\rm crit}=\sqrt{3}v$, and we fix $\tilde{\eta}=1$ to be the initial rescaled conformal time which satisfying $T(\tilde{\eta}=1)=2.4T_{\rm crit}$. For convenience, we set $f_*=v$ and $w_*=R_iH_i$ to do the rescale in Eq.~(\ref{dimensionless rescale}). Since the temperature in the PQ era is much higher than the QCD energy scale $(T \gg \Lambda_{\rm QCD})$, the temperature-dependent axion mass is severely suppressed, so the axion mass and the bias term responsible for domain wall formation and decay, respectively, can be neglected. For simplicity, we fix the number of relativistic degrees of freedom in the PQ era to a typical value at high temperatures, that is $g_*=81$.

In the QCD era, we use the rescaled conformal time $\hat{\eta}$ $(=\eta / \eta_1=R/R_1)$ to conduct iteration, with the initial conformal time $\eta_1=1/(R_1H_1)$ and $R_1$($H_1$) denote the initial scale factor (Hubble parameter). The fields are rescaled with the reinterpreted PQ vacuum $v_{\rm re}$, and the comoving lattice spacing (and time-step) are rescaled with $\hat{w}=6R_1 H_1$. This means that we use $f_*=v_{\rm re}$ and $w_*=\hat{w}$ to do the rescale from Eq.~(\ref{dimensionless rescale}) to Eq.~(\ref{S7}). So, the dimensionless comoving lattice spacing is $\delta \hat{x} \approx 0.0436$, and the time-step is chosen as $\delta \hat{\eta}=0.0076$.

Since $T \sim \Lambda_{\rm QCD} \ll v$ in QCD era, the contribution of temperature to the effective mass of complex scalar can be neglected, so we set $T=0$ throughout the simulation. But we still consider the temperature-dependent axion mass $m(T)$ and the number of relativistic degrees of freedom $g_*(T)=e^{\gamma(T)}$~\cite{Wantz:2009it}, with
\begin{eqnarray}
\gamma(T)\approx&&1.21+0.572*(1+{\rm Tanh}[({\rm ln}(T/{\rm GeV})+8.77)/0.682])\\&&+\ 0.33*(1+{\rm Tanh}[({\rm ln}(T/{\rm GeV})+2.95)/1.01])\nonumber\\&&+\ 0.579*(1+{\rm Tanh}[({\rm ln}(T/{\rm GeV})+1.8)/0.165])\nonumber\\&&+\ 0.138*(1+{\rm Tanh}[({\rm ln}(T/{\rm GeV})+0.162)/0.934])\nonumber\\&&+\ 0.108*(1+{\rm Tanh}[({\rm ln}(T/{\rm GeV})-3.76)/0.869]).\nonumber
\end{eqnarray}
We numerically solve the following equations to evolve scale factor and Hubble parameter~\cite{Kolb:1990vq}
\begin{eqnarray}
&&R \approx 3.699\times10^{-10}g_*(T)^{-1/3}\frac{\rm Mev}{T},\\
&&H \approx 1.66g_*(T)^{1/2}\frac{T^2}{M_{\rm pl}},\nonumber\\
&&t=1/(2H) \approx 0.3012g_*(T)^{-1/2}\frac{M_{\rm pl}}{T^2},\nonumber
\end{eqnarray}
with $M_{\rm pl}=1.22\times10^{\rm 19} \ \text{GeV}$ the full planck mass. Then we can obtain the following relationships
\begin{eqnarray}
R(T) =R(T_1) \frac{g_*(T)^{-1/3}T_1}{g_{*}(T_1)^{-1/3}T}, \quad  H(T) = H(T_1) \frac{g_*(T)^{1/2}T^2}{g_{*}(T_1)^{1/2}T_1^2}.
\end{eqnarray}
We first set the initial temperature $T_1$ and scale factor $R(T_1)$, and can obtain the scale factor and Hubble parameter at any temperature through the above equations.

\section{Initial Conditions} 
\label{sec:initial}
Our simulation is divided into two stages, the PQ era and the QCD era. At the start time of the PQ era, the temperature is higher than the critical temperature of the PQ phase transition, so the complex scalar field can be considered to be in thermal equilibrium. Thus, we can use the following thermal spectrum to describe the amplitude and momentum distribution of each scalar component in momentum space~\cite{Hiramatsu:2010yu}
\begin{equation}
\mathcal{P}_{\phi_1}(k)=\mathcal{P}_{\phi_2}(k)=\frac{n_k}{w_k}=\frac{1}{w_k}\frac{1}{e^{w_k/T}-1}, \quad \mathcal{P}_{\dot{\phi}_1}(k)=\mathcal{P}_{\dot{\phi}_2}(k)=n_k w_k=\frac{w_k}{e^{w_k/T}-1},
\end{equation}
where $n_k$ denotes the occupation number of the Bose-Einstein distribution, $w_k=\sqrt{k^2/R^2+m_{\rm eff}^2}$ and $k$ are physical frequency and comoving momenta respectively, $m_{\rm eff}^2=\lambda(T^2/3-v^2)$ is the initial effective mass square of each scalar field and overdots represent differentiation with respect to cosmic time $t$.

For continuum, the two-point correlation functions can be written as
\begin{align}
\langle \phi_i (\boldsymbol{{\rm k}}) \phi_j(\boldsymbol{{\rm k}}') \rangle &= (2\pi)^3\mathcal{P}_{\phi}(k)\delta(\boldsymbol{{\rm k}}-\boldsymbol{{\rm k}}') \delta_{ij},\\
\langle \dot{\phi}_i (\boldsymbol{{\rm k}}) \dot{\phi}_j(\boldsymbol{{\rm k}}') \rangle &= (2\pi)^3\mathcal{P}_{\dot{\phi}}(k)\delta(\boldsymbol{{\rm k}}-\boldsymbol{{\rm k}}') \delta_{ij}, \nonumber \\
\langle \phi_i(\boldsymbol{{\rm k}}) \dot{\phi}_j(\boldsymbol{{\rm k}}') \rangle &= 0 \nonumber \,.
\end{align}
With proper rescale~\cite{Figueroa:2020rrl}, we can recreate the correlation functions equivalent to that in the continuum on the discrete lattice, which does not depend explicitly on the volume 
\begin{align}
\qquad \langle | \phi_i(\boldsymbol{{\rm k}}) |^2 \rangle &= \left(\frac{N}{\delta x_{\rm phy}}\right)^3\mathcal{P}_{\phi_i}(k), \qquad \langle \phi_i(\boldsymbol{{\rm k}})  \rangle = 0, \\
\qquad \langle | \dot \phi_i(\boldsymbol{{\rm k}}) |^2 \rangle &= \left(\frac{N}{\delta x_{\rm phy}}\right)^3\mathcal{P}_{\dot{\phi}_i}(k),\qquad \langle \dot \phi_i(\boldsymbol{{\rm k}}) \rangle = 0, \nonumber
\end{align}
where $N$ denote the number of points per side and $\delta x_{\rm phy}$ denote the physical lattice spacing. We first generate $\phi_i(\boldsymbol{{\rm k}})$ and $\dot \phi_i(\boldsymbol{{\rm k}})$ following Gaussian random distribution in momentum space, which contain all modes from infrared truncation $k_{\rm IR}$ to maximum momentum $k_{\rm max}=\sqrt{3}k_{\rm UV}$ of the simulation box, with $k_{\rm UV}$ the ultraviolet truncation in one direction. Then, by applying the discrete inverse Fourier transform to the field in momentum space, we can acquire the field in three-dimensional coordinate space.

\section{Identification of string and domain wall}

\begin{figure*}[htb]
\includegraphics[width=.33\textwidth]{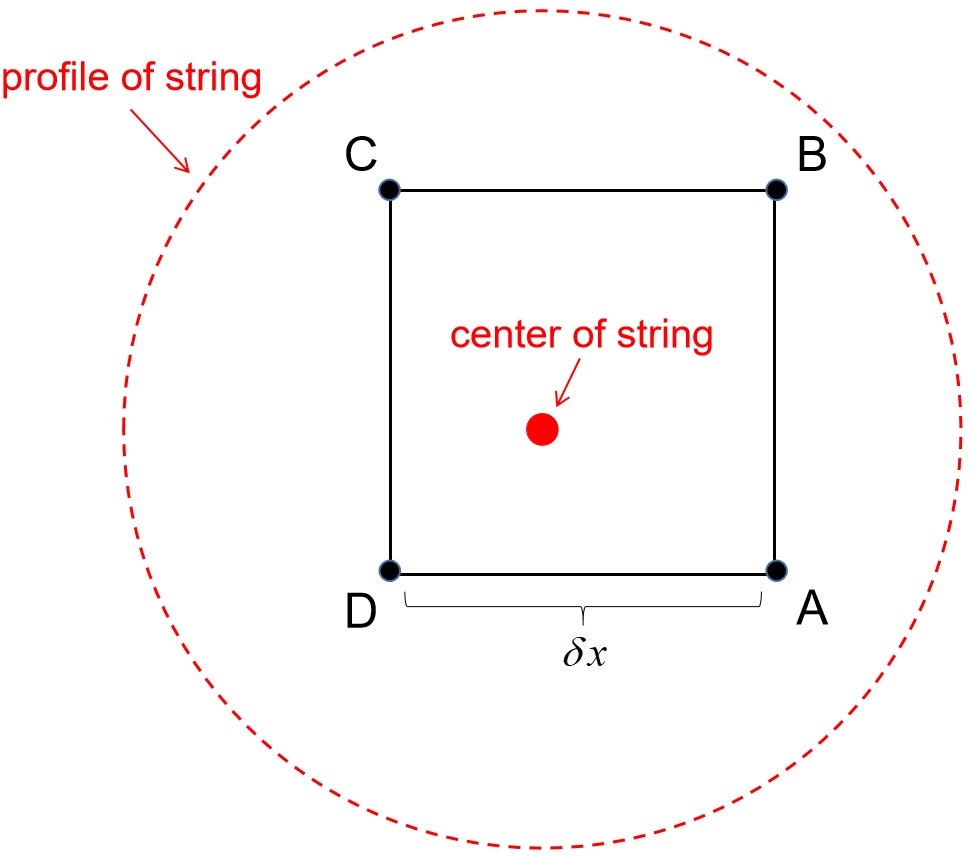} \hspace{5mm}
\includegraphics[width=.31\textwidth]{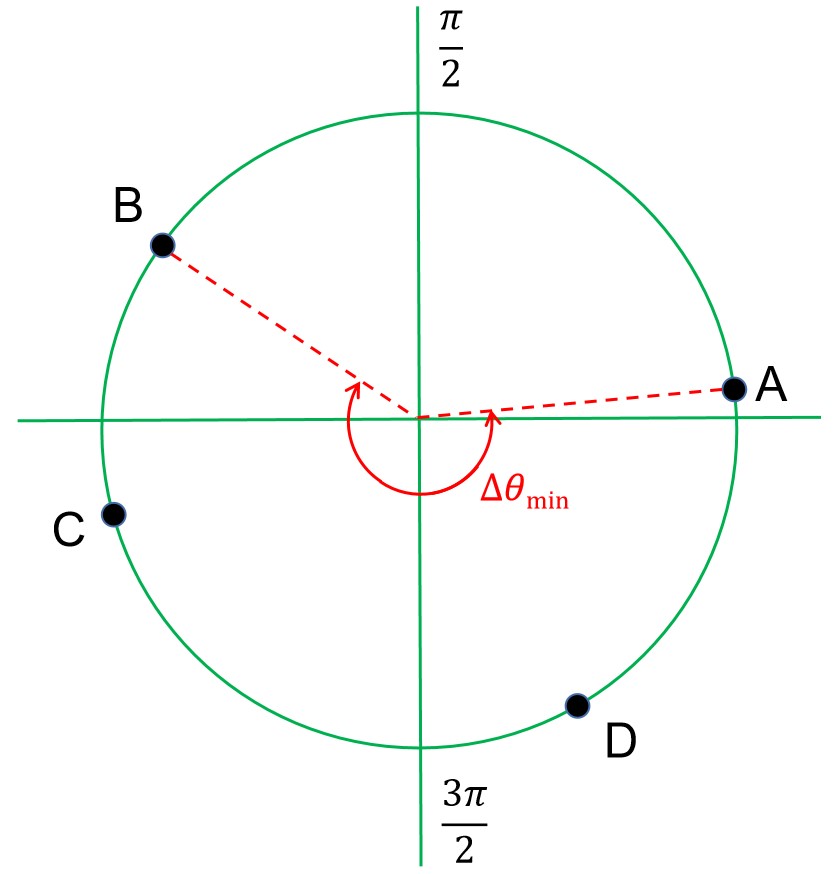} \hspace{5mm}
\includegraphics[width=.28\textwidth]{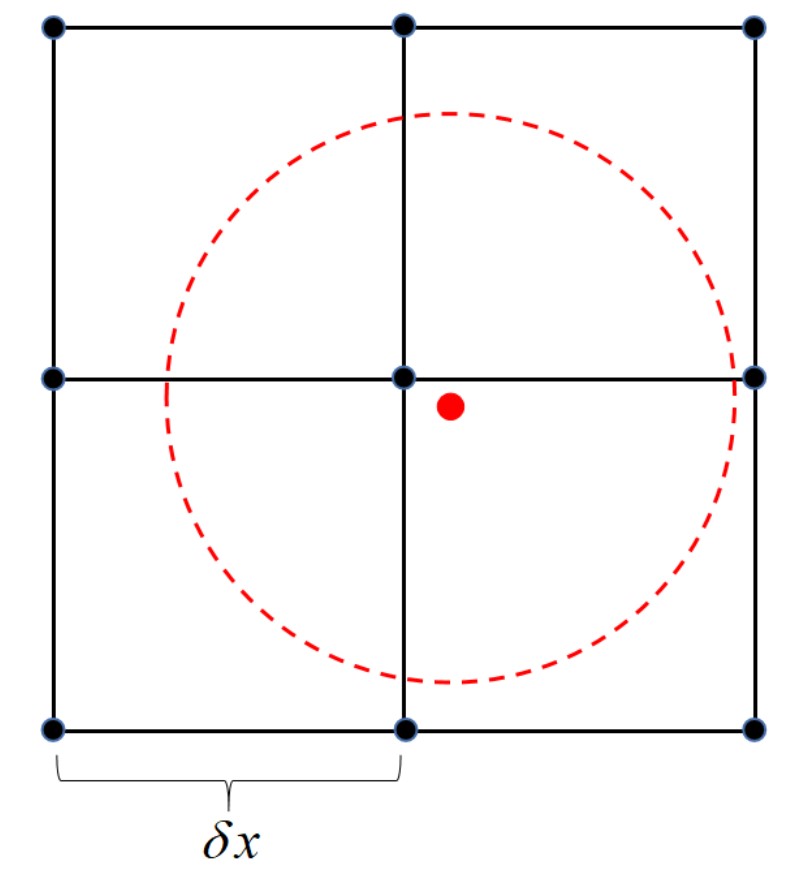}
  \caption{Illustration of the scheme for identifying global string. The left panel shows that a string penetrates a square loop on the lattice. If the center of the string is inside the plaquette and the four points on the plaquette are within the width of the string core, the phase change of the loop around the plaquette should be $2\pi$ due to the property of vortex-like defects. The middle panel shows the phase distribution of the complex scalar field at four points when the string penetrates the square loop, where there is a direction that allows the phase to change continuously and the minimum phase range containing four points is greater than $\pi \ (\Delta \theta_{\rm min}>\pi)$. The right panel illustrates that when the shrinking comoving width of the string is smaller than twice the comoving lattice spacing, it is not reliable to identify string only by phase.}
  \vspace{0.1cm}
  \label{fig:StringIdentification}
\end{figure*}

For identification of string core, we calculate the phase winding of each square loop with a side length of $\delta x$ in the comoving simulation box. The phase of $\varphi$ at four points of the square loop is defined in the range of $(0,2\pi]$. String penetrates the square loop if the minimum phase range which contains the four points is greater than $\pi$ and the phase changes continuously (see Fig.~\ref{fig:StringIdentification}). For a specific square loop, assuming that the minimum phase at four points is $\theta_{\rm min}$, and the above criteria for string penetration can be converted into the following five sub-criteria: \\
(1) $\theta_{\rm min}<\pi$. \\
(2) There exists at least one phase at another point minus $\theta_{\rm min}$ is greater than $\pi$. \\
(3) There exists at least one phase at another point minus $\theta_{\rm min}$ is smaller than $\pi$. \\
(4) Denote the phase closest to $\pi$ in all phases greater than $\pi$ as $\theta_{a}$,
and denote the phase closest to $\pi$ in all phases smaller than $\pi$ as $\theta_{b}$, it is required to meet $\theta_a - \theta_b < \pi$. \\
(5) Calculate the difference between the phases at each of two adjacent points in a counterclockwise direction, the multiplication of the four differences is required to be negative. 

    We denote the number of square loops that meet the above five criteria simultaneously as $n_{\rm c}$. The total physical string length is the number of square loops penetrated by strings multiplied by physical lattice spacing and can be expressed as $l_{\rm phy}=(2/3)n_{\rm c}(R\delta x)$. We multiply a factor 2/3 to counteract the Manhattan effect~\cite{Fleury:2015aca}. Note, to ensure that the string identification scheme is always effective throughout the simulation, the physical string width should be at least twice greater than the physical lattice spacing at the final time (see right panel of Fig.~\ref{fig:StringIdentification}).

Now we will introduce two methods for calculating scaling parameters. The first method we will refer to as the traditional method, based on the following formula~\cite{Hiramatsu:2012sc}
\begin{equation}
\xi=\frac{\rho_{\rm st}t^2}{\mu_{\rm st}}, \quad \text{with} \ \rho_{\rm st}=\frac{\mu_{\rm st}l}{R^2V},
\end{equation}
where $\rho_{\rm st}$ is the energy density of long straight strings, $\mu_{\rm st} \simeq \pi v^2 {\rm ln}(t/\delta_{\rm st})$ is the average mass energy of strings per unit length, $\delta_{\rm st} \simeq 1/(\sqrt{\lambda}v)$ is the string core width, and $t$ is the cosmic time, $l$ and $V$ are total comoving length of string and comoving volume of the simulation box, respectively. Then, the scaling parameters can be simplified as 
\begin{equation}
\xi=\frac{lt^2}{R^2V}.\label{traditional way}
\end{equation}

The second method for calculating scaling parameters is mainly based on the mean string separation $L_{\rm m}$~\cite{Hindmarsh:2019csc},
\begin{equation}
L_{\rm m}=\sqrt{V_{\rm phy}/l_{\rm phy}},
\end{equation}
where $V_{\rm phy}$ and $l_{\rm phy}$ are the physical volume of the simulation box and the total physical length of the string, respectively. After the string networks enter the scaling regime, $L_{\rm m}$ increases linearly with $t$, and the scaling parameter $\xi_{\rm s}$ tends to be a constant. Note that we use $\xi_{\rm s}$ to denote the scaling parameters here to distinguish it from the scaling parameters obtained in the first scheme. However, in numerical simulations, the formation and initial relaxation of string networks will introduce a timescale $t_0$, which can be considered to be the $t$-axis intercept of a linear fit to $L_{\rm m}(t)$. Therefore, $L_{\rm m}=\kappa(t-t_0)$ is satisfied, with $\kappa$ is the slope obtained through linear fitting. Next, the scaling parameters can be expressed as 
\begin{equation}
\xi_{\rm s}=\frac{l_{\rm phy}t^2}{V_{\rm phy}}=\frac{t^2}{L_{\rm m}^2}=\frac{1}{\kappa^2}\frac{t^2}{(t-t_0)^2}\rightarrow \frac{1}{\kappa^2}. \label{standard scaling model}
\end{equation}
The final approximation can eliminate the influence of $t_0$ on the initial evolution of string networks since $t_0/t\rightarrow0$ is satisfied over cosmological time scales.

\begin{figure*}[!htp]
\includegraphics[width=.34\textwidth]{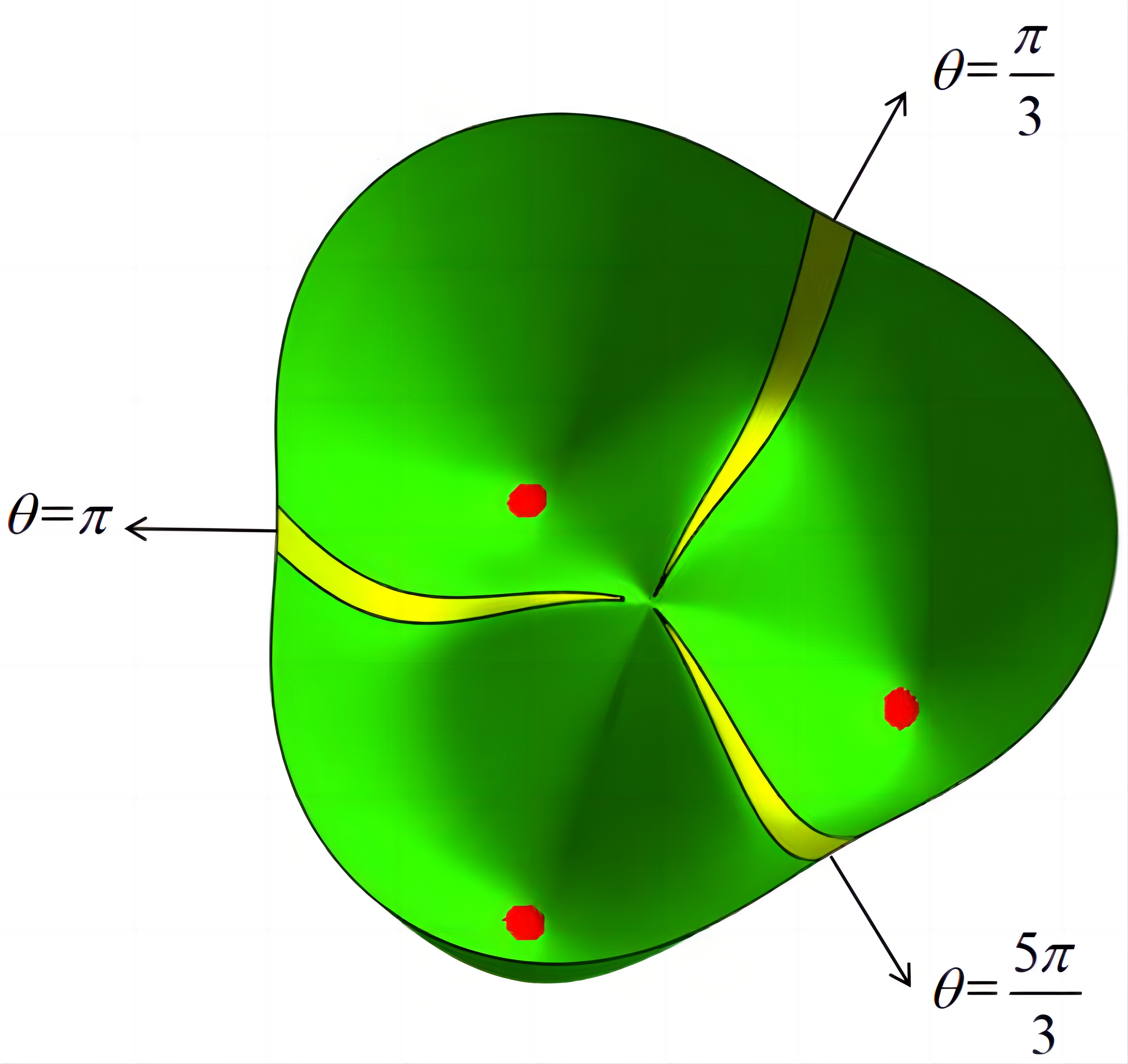} 
\hspace{18mm}
\includegraphics[width=.34\textwidth]{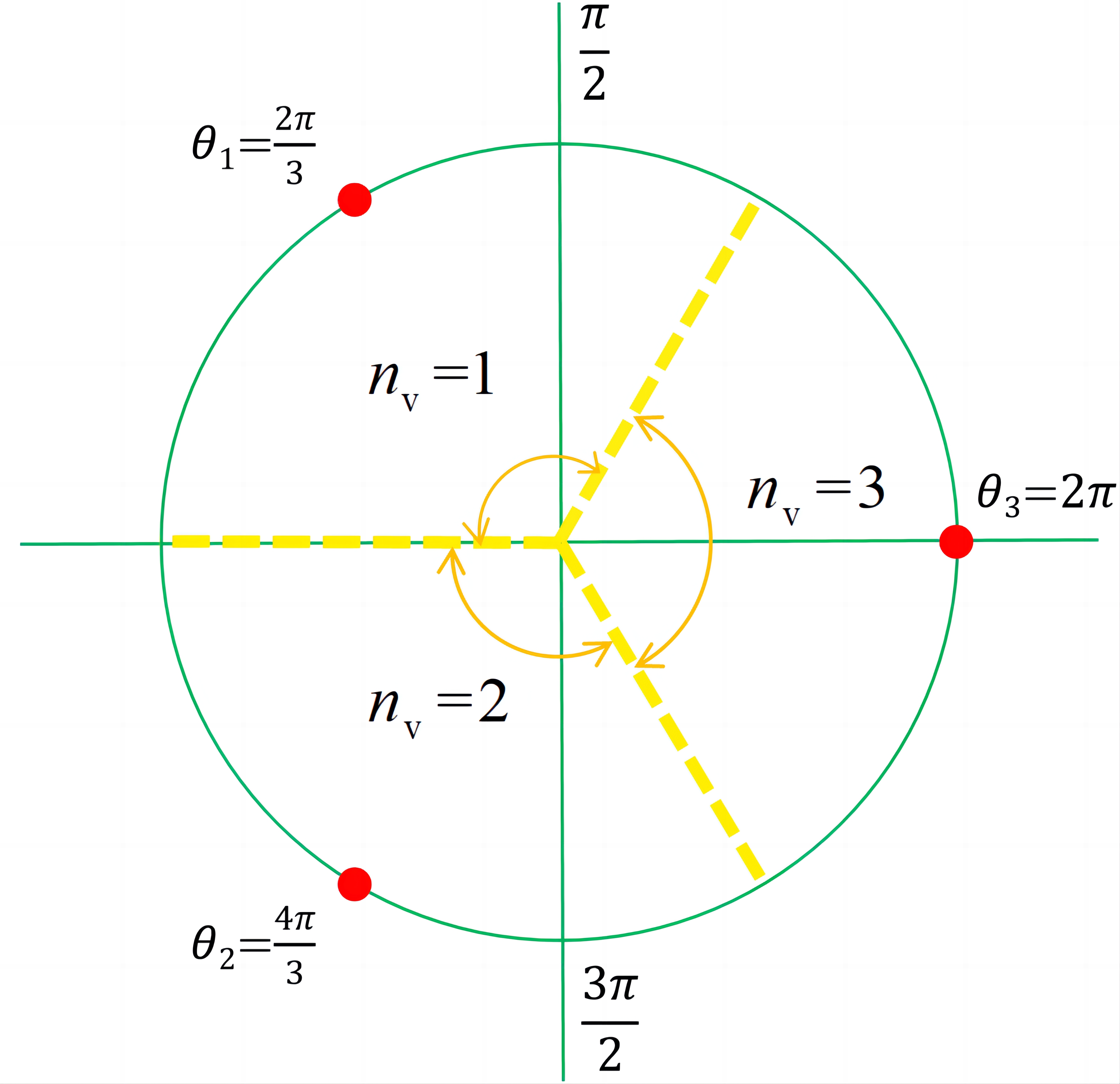}
\caption{Illustration of the identification of DWs in the case of $N_{\rm DW}=3$. The left panel shows a top view near the origin of potential, with the red dot indicating the position of potential minimums, and the yellow region representing the location of the DW, which is the boundary between the two potential minimums. The right panel interprets how to assign vacuum numbers according to the phase. The potential minimums are located at $\theta_1$, $\theta_2$ and $\theta_3$. Then, divide the entire phase region equally around each minimum, and assign different vacuum numbers to each divided region. }
  \vspace{0.1cm}
\label{fig:DWidentification}
\end{figure*}

We can also identify where the DW exists according to the phase of the complex scalar field. When considering the non-zero axion mass and $N_{\rm DW}>1$, there will be $N_{\rm DW}$ minima of potential energy located at where $\theta/(2\pi/N_{\rm DW})$ is an integer, with $\theta \in (0,2\pi]$ the phase of $\varphi$~\cite{Hiramatsu:2012sc}. The DW is located at the boundary with higher potential energy between two minima. Firstly, we assign an integer vacuum number $n_{\rm v}$ to each lattice point to indicate which phase region they belong to. The vacuum number depends on which of the $N_{\rm DW}$ phases that minimize the potential energy is closest to the phase on the lattice point. For example, in the case of $N_{\rm DW}=3$, the three phases that minimize the potential energy are $2\pi/3$, $4\pi/3$, and $2\pi$ in ascending order (see Fig.~\ref{fig:DWidentification}). If the phase of $\varphi$ on a lattice point is $5\pi/6$, it is closest to the first phase which minimizes the potential energy, so the vacuum number is 1. After all the lattice points have been assigned the vacuum number, the DW intersects the link between the two adjacent lattice points when the vacuum number of the two lattice points is different. 

In the case of $N_{\rm DW}=1$, the identification of DW is relatively simple, domain wall intersects the link if $\phi_{1}<0$ and $\phi_2$ have different signs at the two ends of the link~\cite{Hiramatsu:2010yn}.

To calculate the comoving area density of the DW, we first define the quantity $\delta$ that takes the value 1 at both ends of the link intersecting with the DW and equals 0 elsewhere. Then, the comoving area density can be expressed as~\cite{Hiramatsu:2010yz}
\begin{equation}
A/V=C\sum\limits_{\rm links} \delta \frac{|\nabla \theta|}{|\theta_{,x}|+|\theta_{,y}|+|\theta_{,z}|},
\end{equation}
where $\theta_{,i} \ (i=x,y,z)$ denote the spatial derivatives of the dimensionless axion field $\theta(\boldsymbol{{\rm x}})$, and $C$ is a parameter chosen to satisfy $A/V=1$ when all the links have the value $\delta=1$, $A$ and $V$ are the comoving area of DW and the comoving volume of the simulation box, respectively.

The area parameter $\mathcal{A}$ of DW can be expressed as 
\begin{equation}
\mathcal{A}=\frac{\rho_{\rm wall}}{\sigma_{\rm wall}}t, \ \ \text{with} \ \rho_{\rm wall}=\frac{\sigma_{\rm wall}A}{R(t)V},
\end{equation}
where $\rho_{\rm wall}$ is the energy density of the DW, $\sigma_{\rm wall}$ is the surface mass density of the DW. The simplified area parameter can be written as $\mathcal{A}=\frac{At}{R(t)V}$,
where the comoving area is defined as 
\begin{equation}
A=\Delta A \sum\limits_{\rm links} \delta \frac{|\nabla \theta|}{|\theta_{,x}|+|\theta_{,y}|+|\theta_{,z}|},
\end{equation}
where $\Delta A=(\delta x)^2$ is the comoving area of one grid surface.

\begin{figure*}[!htp]
\includegraphics[width=0.4\textwidth]{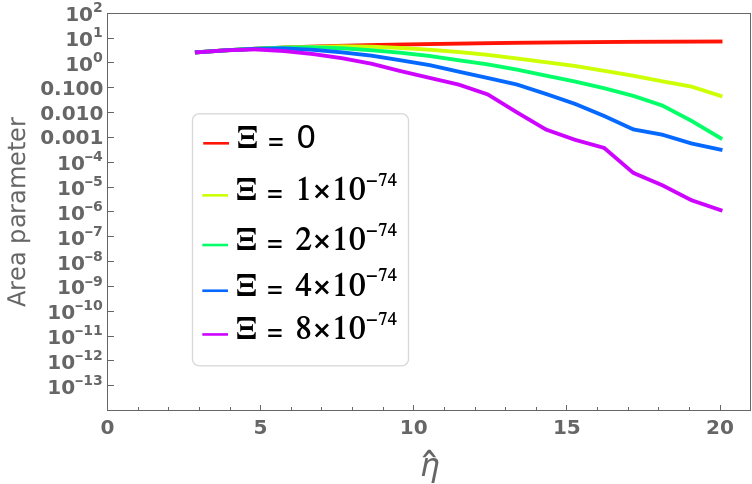} 
\hspace{10mm}
\includegraphics[width=0.4\textwidth]{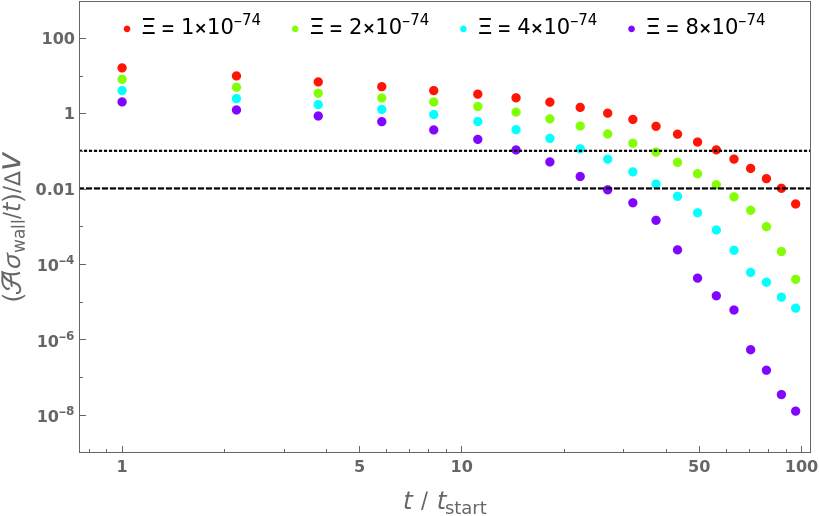}
\includegraphics[width=0.4\textwidth]{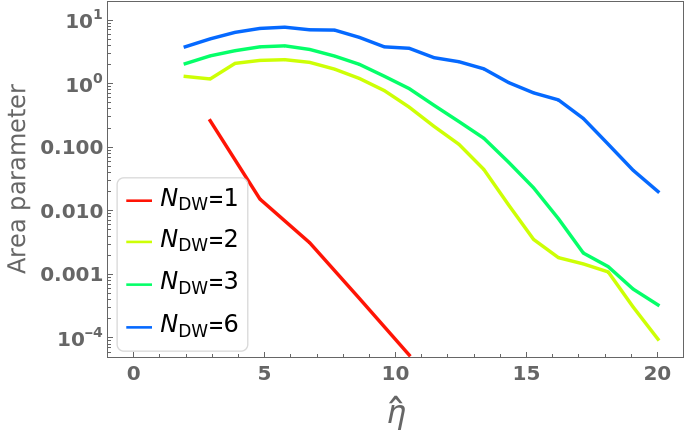} 
\hspace{10mm}
\includegraphics[width=0.4\textwidth]{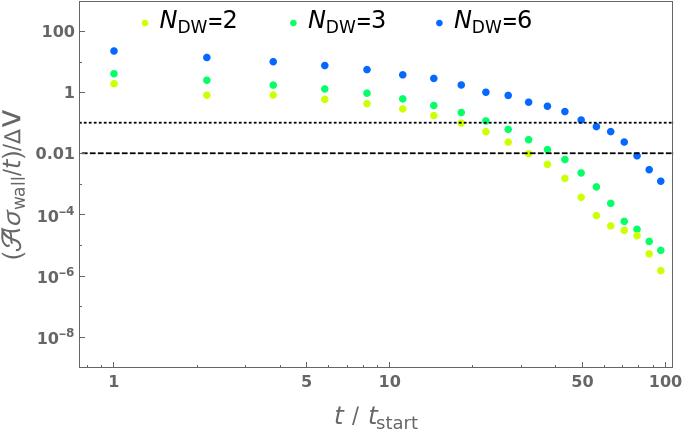}    
  \caption{Top-Left: Time evolution of area parameter of DWs with different bias terms when $N_{\rm DW}=3$, Top-Right: Time evolution of $\rho_{\rm wall}/\Delta V$ with different $\Xi$ for $N_{\rm DW}=3$; Bottom-Left: Time evolution of area parameter of DWs with  $N_{\rm DW}>1$ and $\Xi=4\times10^{-74}$, Bottom-Right: Time evolution of $\rho_{\rm wall}/\Delta V$ with $N_{\rm DW}>1$ and $\Xi=4\times 10^{-74}$.} 
  \vspace{0.1cm}  
  \label{fig:gw spectrum and area parameter}
\end{figure*}

\begin{figure*}[!htp]
\includegraphics[width=0.5\textwidth]{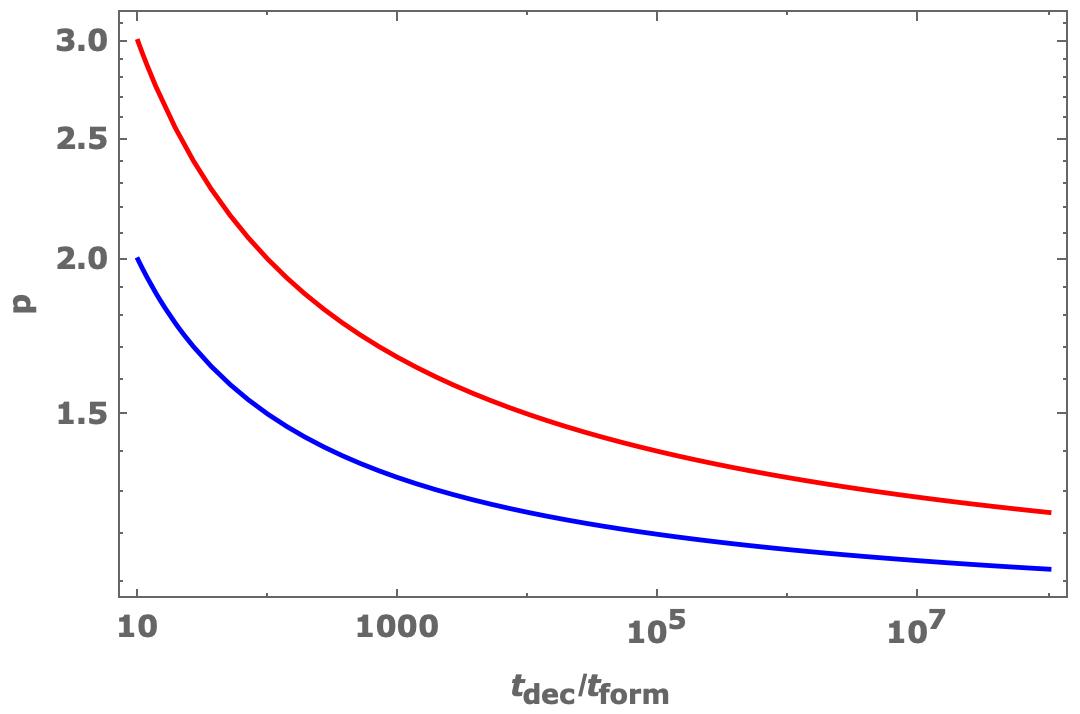} 
\hspace{10mm}
\caption{ The $p$ factor as a function of $t_{\rm dec}/t_{\rm form}$. Where the blue (red) curve corresponds to the case of $C_d=10$ ($C_d=100$). }.
 \label{fig:ptdec}
\end{figure*}
We measured the time evolution of area parameter $\mathcal{A}$ of DWs in the QCD era, see left plots of Fig.~\ref{fig:gw spectrum and area parameter}. We found that area parameter decreased faster as $\Xi$ increased and $N_{
\rm DW
}$ ($N_{\rm DW}>1$) decreased. In the right plots of Fig.~\ref{fig:gw spectrum and area parameter}, $\rho_{\rm wall}=\mathcal{A}\sigma_{\rm wall}/t$ is the energy density of DWs, $\Delta V=2\Xi v^4(1-{\rm cos}(2\pi/N_{\rm DW}))$ is a volume pressure acting on DWs due to the exist of the bias term, $t$ and $t_{\rm start}$ denote the cosmic time and initial cosmic time. The DWs decay time can be obtained when the DWs tension is comparable with the bias term, i.e., $\mathcal{A}\sigma_{\rm wall} /t
\sim\Delta V$. For the scenarios where the DWs deviate from the scaling regime at the decay time, one needs to consider the departure of the area parameter of the DWs from that of the scaling regime~\cite{Jeong:2013oza,Kawasaki:2014sqa}.
We fix the $C_d=\Delta V/(\mathcal{A}\sigma_{\rm wall}/ t)\sim \mathcal{O}( 10^{2})$ to obtain the decay time of the DWs when the area parameter almost decrease to its percent level (or ten percent). The decay time can therefore be parameterized as: 
\begin{equation}\label{eq:tdec}
t_{\rm dec}=t_{\rm form}\left(\frac{C_d \mathcal{A_{\rm form} \sigma_{\rm wall}}}{t_{\rm form} \Xi N_{\rm DW}^4 f_a^4 (1-\cos(2\pi/N_{\rm DW}))}\right)^{1/p} \;,
\end{equation}
with $p=1+\ln{C_d}/\ln{(t_{\rm dec}/t_{\rm form})}$ denoting the deviation factor of the area parameter from the scaling case and the DWs formation time being $t_{\rm form}$ when $m=3 H$. As illustated in Fig.~\ref{fig:ptdec}, for the QCD axion in the scenario of $C_d=10$, the $p\geq 1.25$ considering that $t_{\rm dec}/t_{\rm form}\leq \mathcal{O}(10^4)$. For the ALP axion under the situation of $C_d=100$, the $p\geq 1.25$ for $t_{\rm dec}/t_{\rm form}\leq \mathcal{O}(10^8)$.

Since the DW energy density evolves as 
$\rho_{wall}\sim \sigma_{\rm wall}/H^{-1}\propto R^{-2}$, its dilution is much slower than that of the matter ($\sim R^{-3}$) and radiation ($\sim R^{-4}$) as the Universe expands. To avoid overclose the universe, the DWs should decay early enough before they dominate the universe, this places an upper-bound on the DWs decay time $t_{\rm dec}$, i.e., $t_{\rm dec}< t_{\rm WD}$. 
with 
\begin{equation}
t_{\rm WD}=\frac{3}{16\pi G\sigma_{\rm wall}}\;.
\end{equation}
Based on such consideration, Ref.~\cite{Hiramatsu:2012sc} gives a lower bound on the $\Xi$ based on a rough estimation of the $\Delta V$. And, Ref.~\cite{Kawasaki:2014sqa} assumes the DWs can fully decay before they dominate the universe.
We indeed found that $t_{\rm dec}< t_{\rm WD}$ leads to,
\begin{equation}\label{eq:wd}
\Xi>\frac{512\pi \mathcal{A}_{\rm dec}^2 C_d G m^2\csc(\pi/N_{\rm DW})^2 }{3N_{\rm DW}^4}\;.
\end{equation} 
we denote the bound as {\it WD excl.} in the following.
We also note that to avoid the bias term dominating over the QCD instanton effect in the contributions to axion mass term, one can place an upper-bound on the bias term as~\cite{Hiramatsu:2012sc}
\begin{equation}\label{eq:bam}
\Xi < 2\times 10^{-45}N_{\rm DW}^{-2}( \frac{\rm 10^{10} GeV}{f_a})^4\;,
\end{equation} 
we denote the limit as {\it Mass excl.} condition.

\section{Measurement of power spectra and energy density of free axions}
We compute the power spectrum $P(k)$ of axions, which is defined by
\begin{equation}
\frac{1}{2}\langle \dot{a}(\boldsymbol{{\rm k}})^* \dot{a}(\boldsymbol{{\rm k'}}) \rangle = (2\pi)^3\frac{2\pi^2}{k^3}P(k)\delta(\boldsymbol{\rm k}-\boldsymbol{\rm k'}) ,
\end{equation}
where $ \dot{a}(\boldsymbol{{\rm k}})=\int d^3 \boldsymbol{{\rm x}}\dot{a}(\boldsymbol{{\rm x}})e^{i\boldsymbol{{\rm k}}\boldsymbol{{\rm x}}}$ denote the Fourier component of the time derivative of the axion field, and $\langle...\rangle$ represents an ensemble average. The value of $\dot{a}(\boldsymbol{{\rm x}})$ can be obtained from the data of $\phi_i$ and $\dot{\phi}_i$ $(i=1,2)$
\begin{equation}
\dot{a}(\boldsymbol{{\rm x}})=f_a \frac{\phi_1 \dot{\phi_2}-\dot{\phi_1}\phi_2 }{
\phi_1^2+\phi_2^2}.
\end{equation}
The time derivative of the axion field $\dot{a}(\boldsymbol{\rm{x}})$
contain the contaminations from the
string core and the DW core. In order to get the time derivative of the free axion field $\dot{a}_{\rm free}(\boldsymbol{\rm{x}})$, we need to mask the contribution of the string core and DW core by the screening function defined below
\begin{equation}
W(\boldsymbol{\rm x})=W_{\rm st}(\boldsymbol{\rm x})\times W_{\rm dw}(\boldsymbol{\rm x}), \quad \quad \dot{a}_{\rm free}(\boldsymbol{\rm x})=W(\boldsymbol{\rm x})\dot{a}(\boldsymbol{\rm x}),
\end{equation}
where $W_{\rm st}(\boldsymbol{\rm x})$ and $W_{\rm dw}(\boldsymbol{\rm x})$ represent the screening functions of the string and DW, respectively. 

\begin{figure*}[!htp]
\includegraphics[width=.38\textwidth]{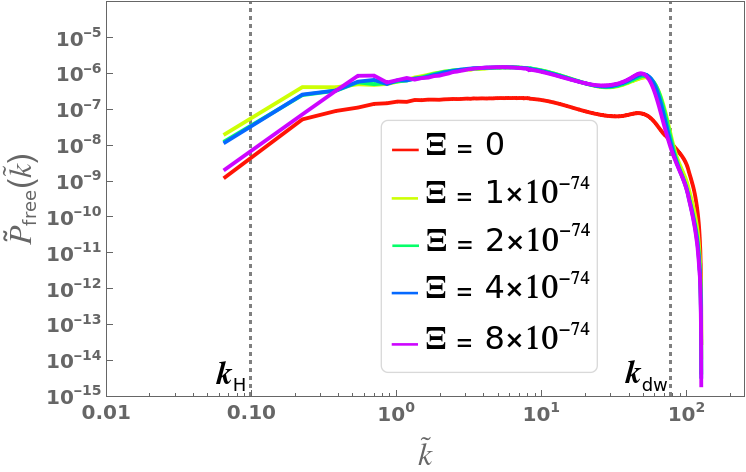} 
\hspace{20mm}
\includegraphics[width=.38\textwidth]{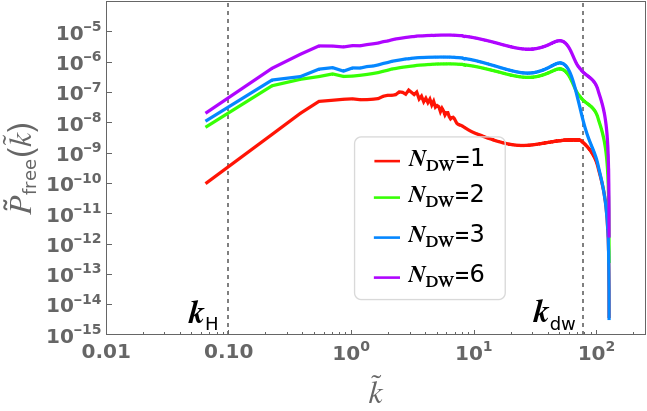}
  \caption{Left panel: Dimensionless spectrum of axions ($\tilde{P}_{\rm free}(\tilde{k}$)) with different coefficients of bias term in the case of $N_{\rm DW}=3$, where $\tilde{k}$ denote the dimensionless comoving momentum.  Right panel: Free axion spectrum with different DW numbers. We set $\Xi=4\times 10^{-74}$ for $N_{\rm DW}>1$ and $\Xi=0$ for $N_{\rm DW}=1$.   }
\vspace{0.1cm}\label{fig:rho_a}
\end{figure*}

Firstly, We decompose the complex scalar as $\varphi(\boldsymbol{\rm x})=(r(\boldsymbol{\rm x})+v)e^{i\frac{a(\boldsymbol{\rm x})}{f_a}}$, where $r(\boldsymbol{\rm x})$ denote the radial mode, $a(\boldsymbol{\rm x})$ denote the axion field. For the $W_{\rm st}(\boldsymbol{\rm x})$, we follow the scheme in Refs.~\cite{Gorghetto:2018myk}. We take advantage of the automatic masking of string cores given by the factor:$W_{\rm st}(\boldsymbol{\rm x})=(1+r(\boldsymbol{\rm x})/v)$, where $W_{\rm st}(\boldsymbol{\rm x})$ takes 0 near the false vacuum and 1 near the true vacuum. For the $W_{\rm dw}(\boldsymbol{\rm x})$, We imitate the automatic masking scheme of the string cores, and we use the phase of the complex scalar field ($\theta=a/f_a$) to obtain the masking function: $W_{\rm dw}(\boldsymbol{\rm x})=| (\theta \quad{\rm mod} \quad (2\pi / N_{\rm DW})) - (\pi/N_{\rm DW}) | / (\pi/N_{\rm DW}), $ where $"{\rm mod}"$ represents the remainder after dividing two numbers. Thus, $W_{\rm dw}  (\boldsymbol{\rm x})$ will take 0 near the DW core and 1 near the true vacuum. Eventually, we get the time derivative of a smooth and continuously varying free axion field $\dot{a}_{\rm free}(\boldsymbol{\rm x})=W_{\rm st}(\boldsymbol{\rm x})\times W_{\rm dw}(\boldsymbol{\rm x})  \dot{a}(\boldsymbol{\rm x})$. Similarly, we can get the power spectrum of free axions through 
\begin{equation}
\frac{1}{2}\langle \dot{a}_{\rm free}(\boldsymbol{{\rm k}})^* \dot{a}_{\rm free}(\boldsymbol{{\rm k'}}) \rangle = (2\pi)^3\frac{2\pi^2}{k^3}P_{\rm free}(k)\delta(\boldsymbol{\rm k}-\boldsymbol{\rm k'}).
\end{equation}
In Fig.~\ref{fig:rho_a}, we plot the dimensionless free axion spectrum with different $\Xi$ (left panel) and $N_{\rm DW}$ (right panel). We found that the free axion spectrum is almost independent of $\Xi$ and is proportional to $N_{\rm DW}$.

\begin{figure*}[!htp]
\includegraphics[width=0.36\textwidth]{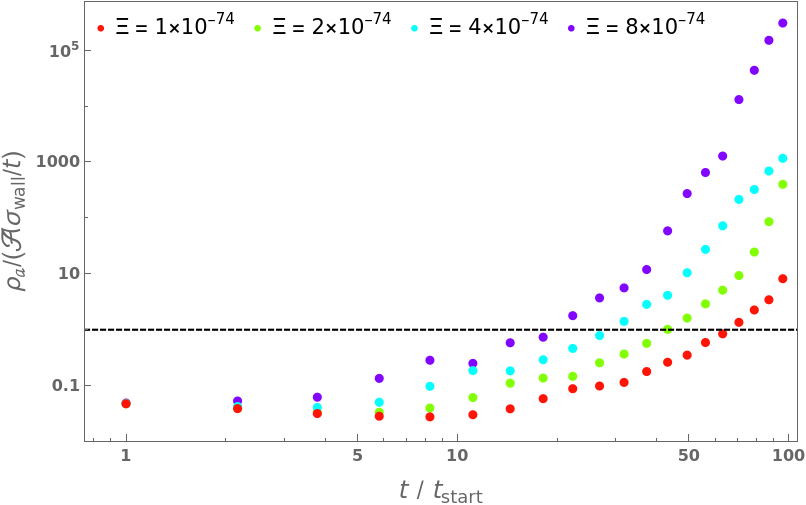}
\hspace{10mm}
\includegraphics[width=0.36\textwidth]{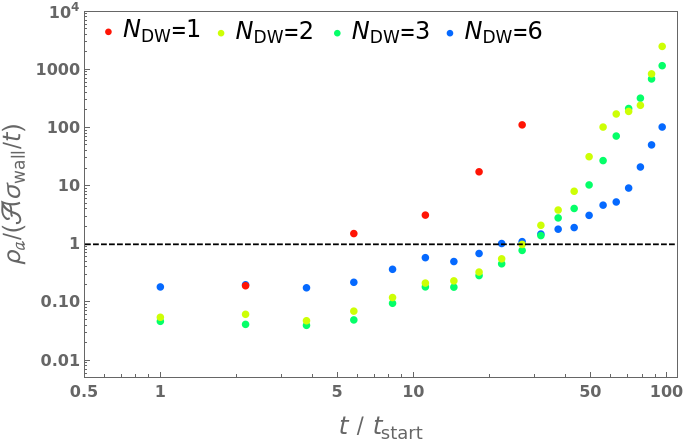}
\caption{ Left panel: time evolution of the ratio of $\rho_a$ of $\mathcal{A}\sigma_{\rm wall}/t$ under different $\Xi$ in the case of $N_{\rm DW}=3$. Right panel: time evolution of the ratio of $\rho_a$ of $\mathcal{A}\sigma_{\rm wall}/t$ with different $N_{\rm DW}$, we set $\Xi=0$ for $N_{\rm DW}=1$ and $\Xi=4\times 10^{-74}$ for $N_{\rm DW}>1$. }
\vspace{0.1cm}\label{fig:rho_aasigmawallt}
\end{figure*}

We further estimate the total energy of the
emitted free axions, we assume the free axions
are all in harmonic mode~\cite{Chang:2023rll} e.g. its total energy takes twice of its kinetic energy~\cite{Hiramatsu:2012gg}, $\rho_a=2\rho_{\rm a,kin}$. So $\rho_{a}\simeq 2\times \frac{1}{2}\langle \dot{a}_{\rm free}^2(\boldsymbol{\rm x}) \rangle$, where the brackets denote spatial average. 
We plot the time evolution of the ratio of $\rho_a$ of $\rho_{\rm wall} =\mathcal{A}\sigma_{\rm wall}/t$ with different $\Xi$ in left panel of Fig.~\ref{fig:rho_aasigmawallt}. It can be found that as $\Xi$ increases, the ratio of $\rho_a$ to $\rho_{\rm wall}$ also gradually increases, indicating that $\rho_a/\rho_{\rm wall}$ is proportional to $\Xi$ and $t$. In right panel of Fig.~\ref{fig:rho_aasigmawallt}, we plot the time evolution of the ratio of $\rho_a$ over $\rho_{\rm wall} =\mathcal{A}\sigma_{\rm wall}/t$ with different $N_{\rm DW}$. Indeed, the DWs decay mostly to axions, we have $\rho_a\sim \mathcal{A} \sigma_{\rm wall}/t\sim 8  \mathcal{A}  m f_a^2/t_{\rm dec}$ around the $t_{\rm dec}$. 

\begin{figure*}[!htp]
\includegraphics[width=0.4\textwidth]{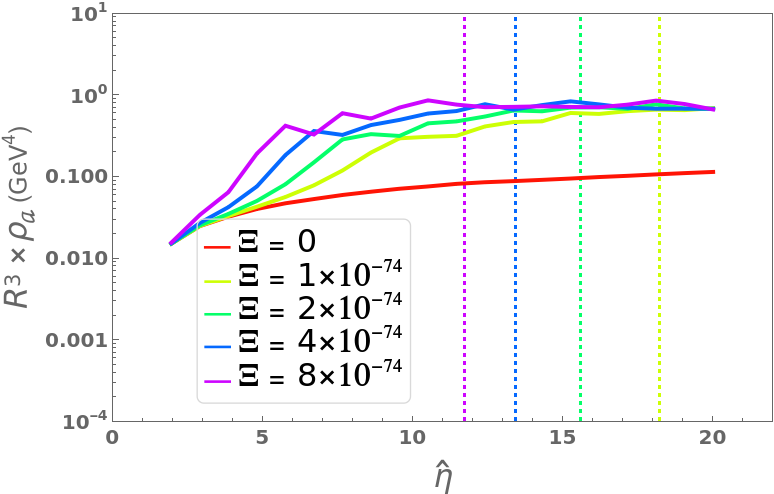}
\includegraphics[width=0.4\textwidth]{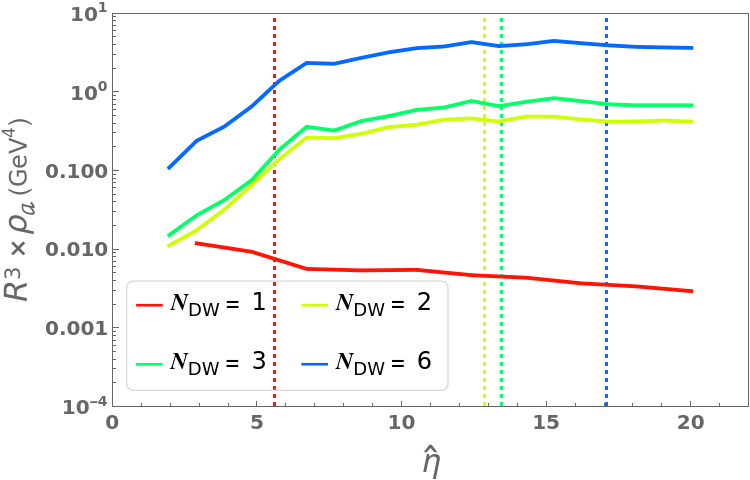}
  \vspace{-0.2cm}
  \caption{ Left: Time evolution of rescaled axion energy density $R^3 \times \rho_{a}$ with different coefficients of bias term in the case of $N_{\rm DW}=3$. The vertical lines represent the decay time of curves of the same color in the corresponding $\Xi$ case. Right: Time evolution of rescaled axion energy density $R^3 \times \rho_{a}$ with different $N_{\rm DW}$. We set $\Xi=0$ for $N_{\rm DW}=1$ and $\Xi=4\times 10^{-74}$ for $N_{\rm DW}>1$. Note, in the $N_{\rm DW}=1$ case, we set $\hat{\lambda}=13$, since after $\hat{\eta} \simeq 1$, the relation $\sigma_{\rm wall} > \mu_{\rm string}(t)/t$ is satisfied, the system dynamics will rapidly dominated by DWs, so the artificial breaking of the ratio of string
width to physical lattice spacing will not have a significant
impact on simulation results.}
  \vspace{0.05cm}
\label{fig:QCD_axion_energy}
\end{figure*}
 In FIG.~\ref{fig:QCD_axion_energy}, we measured the time evolution of rescaled axion energy density $R^3 \times \rho_a$ with different $\Xi$ and $N_{\rm DW}$.
It can be seen that the axion energy tends to be a constant at the final moment, and the energy density of the radiated free axion is (almost) proportional to the bias term and the $N_{\rm DW}^4$.


Considering the evolution of DWs energy density $\rho_{\rm wall}$ and the radiated axions energy density from which~\cite{Kawasaki:2014sqa}, one can estimate the free axion energy density at the DWs decay time
$\rho_{a}(t_{\rm dec})$ as 
\begin{equation}
\rho_a(t_{\rm dec})\simeq\frac{2(2p-1) f_a^4 N_{\rm DW}^4}{(3-2p)C_d} \Xi \sin^2(\pi/N_{\rm DW})\;,
\end{equation}
which applies to the  scenarios where $1/2<p<3/2$. Generally, to avoid affecting the BBN, one needs to require
$T_{\rm dec}> 1$ MeV. The present energy density of the axions radiated from the DWs is 
\begin{equation}
\rho_a(t_0)=\frac{R( t_{\rm dec})^3}{R(t_0)^3}\rho_a(t_{\rm dec})=\big(\frac{R( t_{\rm dec})}{R(t_{\rm eq})}\big)^3 \big(\frac{R( t_{\rm eq})}{R(t_{0})}\big)^3 \rho_a(t_{\rm dec})\;.
\end{equation}
Where, $R( t_{\rm eq})/R(t_{0})=4.15\times 10^{-5}(\Omega_{\rm CDM}h^2)^{-1}$, $R(t_{\rm dec})/R(t_{\rm eq})=(H( t_{\rm eq})^2/(2H( t_{\rm dec})^2))^{1/4}$ with $H(t_{\rm eq})=1.13\times 10^{-35}(\Omega_{\rm CDM}h^2)^2$ GeV~\cite{Weinberg:2008zzc}. With the critical present density $\rho_{0,c}=8.19\times 10^{-47} {\rm GeV}^4$, the dark matter relic abundance from DWs decay is obtained as:
\begin{equation}\label{eq:omega}
\Omega_a h^2=\frac{\rho_a( t_{\rm 0})}{\rho_{0,c}/h^2}=1.024\times 10^{-19}(\frac{8}{3})^{3/(2p)}(\frac{2p-1}{3-2p}) C_d^{3/(2p)-1}
  (  \frac{m}{{\rm GeV}})^{3/p-3/2}(\frac{f_a}{{\rm GeV}})^{4-3/p} \Xi^{1-3/(2p)} \bigg(\frac{\csc(\pi/N_{\rm DW})}{N_{\rm DW}^2}\bigg)^{3/p-2}\;.
\end{equation}

\begin{figure*}[!htp]
\includegraphics[width=0.4\textwidth]{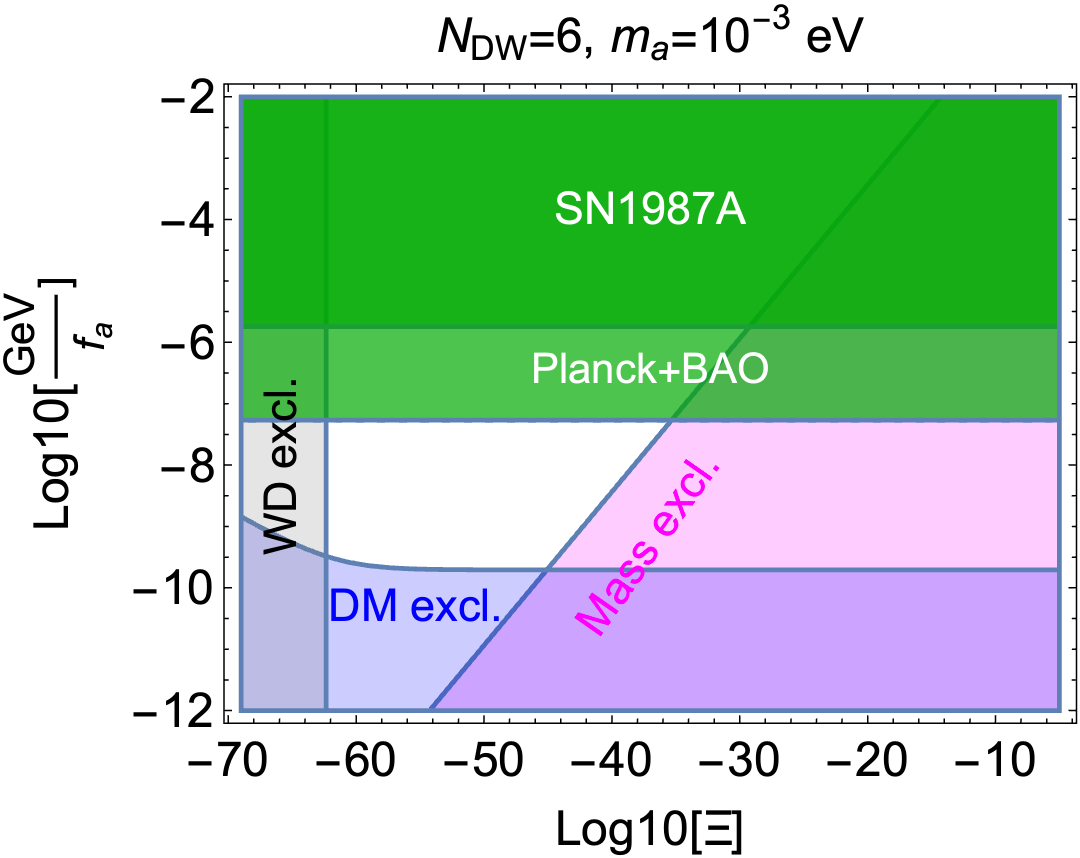} 
\includegraphics[width=0.4\textwidth]{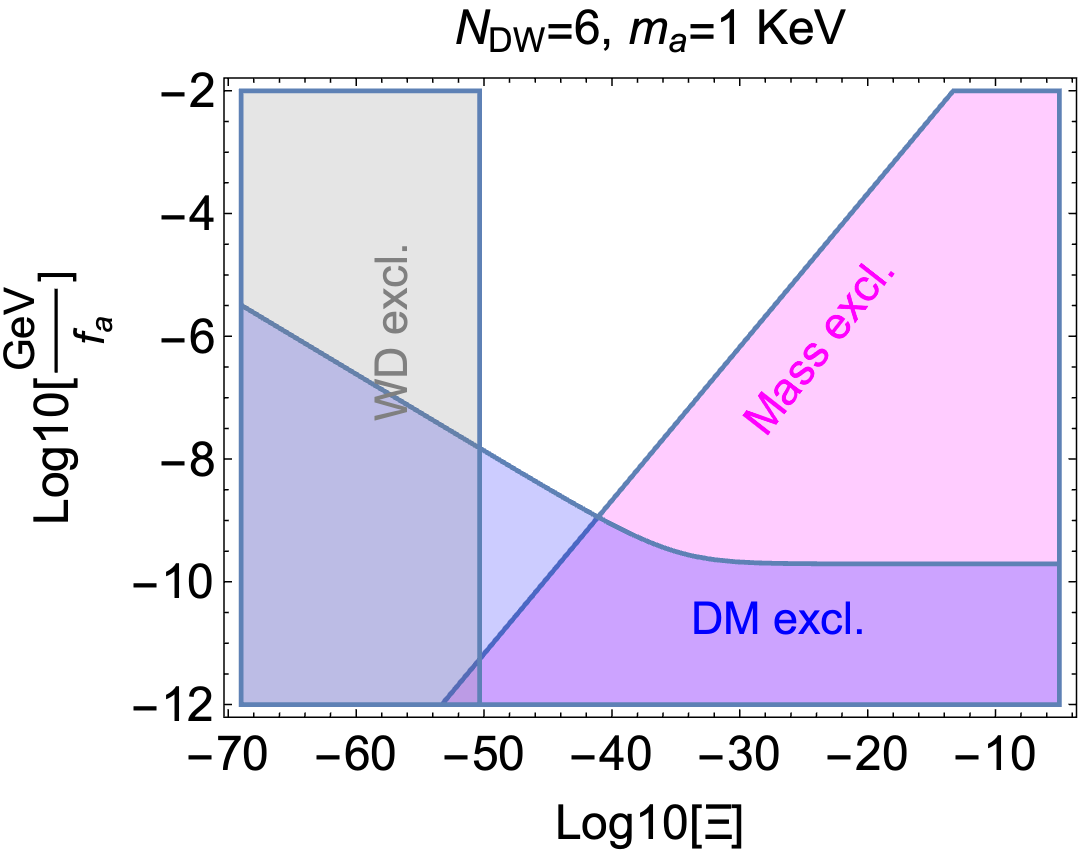}
\includegraphics[width=0.4\textwidth]{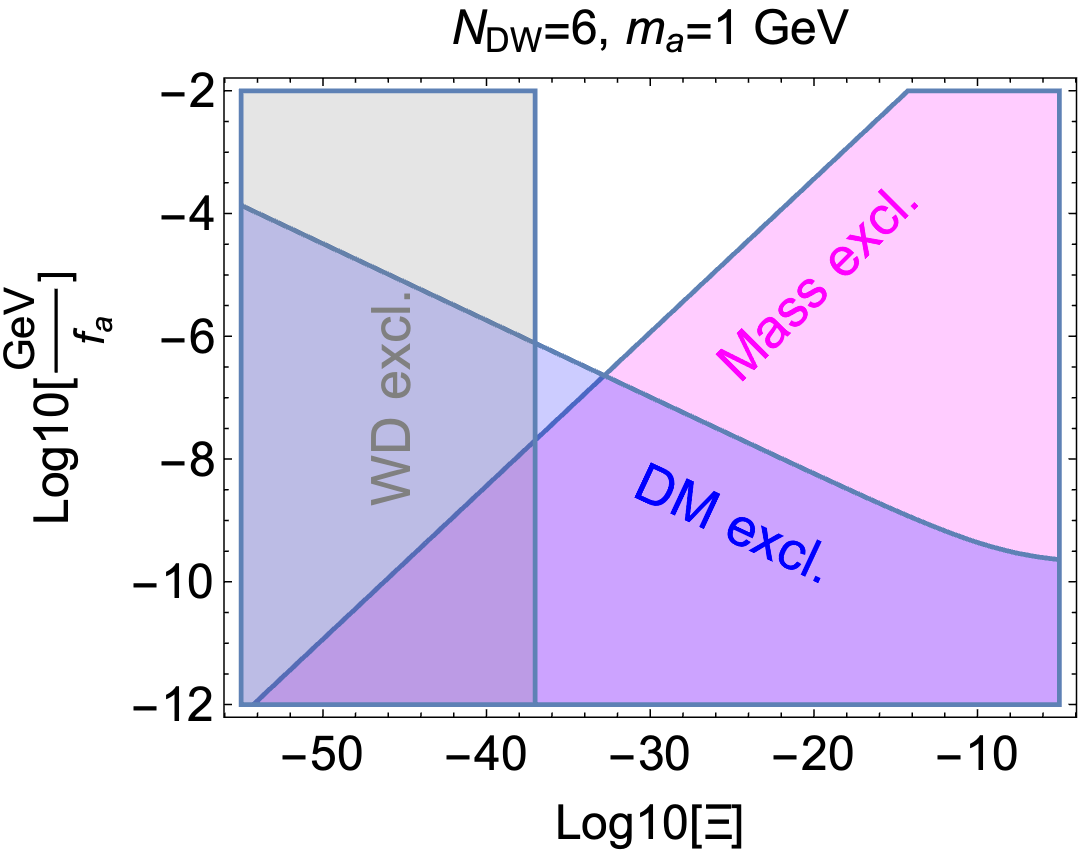} 
\includegraphics[width=0.4\textwidth]{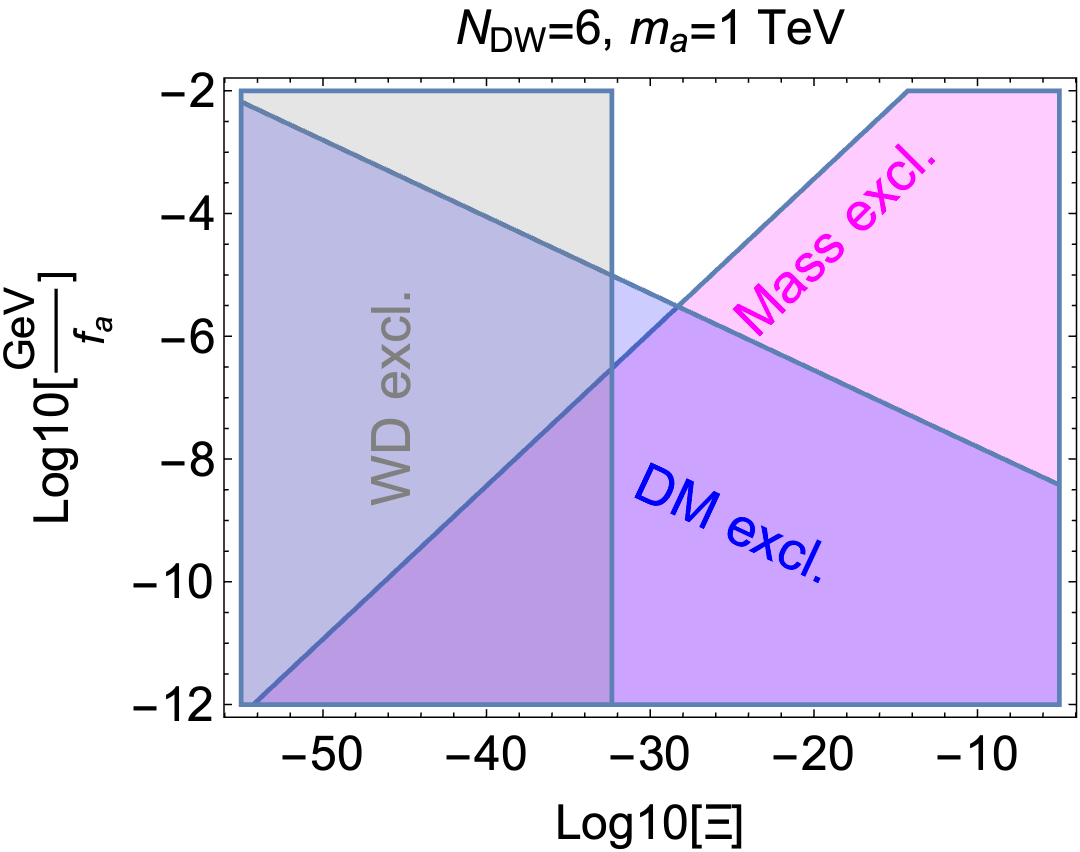}    
  \caption{Constraints on the axion (or ALP) mass versus the PQ scale $f_a$, and we place the bounds on $f_a$ in the low mass region from supernova (SN1987A) and Planck+BAO\cite{AxionLimits}.} 
  \vspace{0.1cm}  
  \label{fig:faxi}
\end{figure*}

As indicated by the Eqs.(\ref{eq:omega},\ref{eq:bam},\ref{eq:wd}), with $N_{\rm}$ increasing one would have less survival parameter spaces of ALP mass and PQ phase transition scale $f_a$. 
As an illustration, with four ALP mass benchmarks taking $p=5/4$ and $N_{\rm DW}=6$ as an example, we present the observation bounds from DM relic abundance ({\it DM excl.}), and theoretical bounds on the bias term from QCD mass ({\it Mass excl.}) and the wall domination ({\it WD excl.}) in Fig.~\ref{fig:faxi}. The figure shows that the allowed PQ scale would decrease as the ALP mass increases.

\section{GW power spectrum}

Topological defects with high energy density and non-zero quadrupole moment will cause anisotropic perturbations to the space-time background, while GWs are described by spatial metric perturbations $h_{ij}$ around FLRW background
\begin{equation}
{\rm d}s^2=-{\rm d}t^2+R^2(t)(\delta_{ij}+h_{ij}){\rm d}x^i{\rm d}x^j ,
\end{equation}
where $t$ denote the cosmic time, $h_{ij}$ are transverse and traceless with $\partial_i h_{ij}=0$ and $h_{ii}=0$. The evolution of $h_{ij}$ is obtained by solving the following equation
\begin{equation} \label{S22}
\ddot{h}_{ij}+3\frac{\dot{R}}{R}\dot{h}_{ij}-\frac{\nabla^2}{R^2}h_{ij}=\frac{16\pi G}{R^2}\Pi_{ij}^{\rm TT},
\end{equation}
with $\dot{} = {\rm d}/{\rm d}t$ and $\nabla_i={\rm d}/{\rm d}x^i$, $G$ is the Newton's gravitational constant, and $\Pi_{ij}^{\rm TT}$ is the transverse-traceless(TT) part of the anisotropic energy-momentum tensor $\Pi_{ij}$. The anisotropic tensor $\Pi_{ij}$ can be characterized by the deviation of the field energy-momentum tensor $T_{\mu \nu}$ from perfect fluid
\begin{align}
\Pi_{ij} &\equiv T_{ij}-pg_{ij}=T_{ij}-\frac{\delta_{ij}}{3} \sum_{l}T_{ll} \label{S23} \\ 
&= (\partial_i \phi_1 \partial_j \phi_1 +\partial_i \phi_2 \partial_j \phi_2) -\frac{\delta_{ij}}{3} \sum_{l}(\partial_l \phi_1 \partial_l \phi_1 +\partial_l \phi_2 \partial_l \phi_2),\label{S27}
\end{align}
where $p$ is the homogeneous background pressure and $g_{ij}=R^2(t)(\delta_{ij}+h_{ij})$. The last term in Eq.(\ref{S27}) vanishes after transverse-traceless projection.

The energy density of the stochastic GW background radiated by stochastic sources is usually defined as the 00 component of the stress-energy tensor $t_{\mu \nu}$:
\begin{align}
\rho_{\rm GW}(t) =t_{00}= \frac{1}{32\pi G}\langle \partial_{\mu}{h}_{ij}^{\rm TT}(\boldsymbol{{\rm x}},t) \partial_{\nu}{h}^{{\rm TT},ij}(\boldsymbol{{\rm x}},t) \rangle|_{\mu=\nu=0} = \frac{1}{32\pi G}\langle \dot{h}_{ij}(\boldsymbol{{\rm x}},t) \dot{h}_{ij}(\boldsymbol{{\rm x}},t) \rangle , 
\end{align}
where the bracket $\langle...\rangle$ denotes spatial average. When $kV^{\frac{1}{3}} \gg 1$ is satisfied in momentum space, the GW energy density can be expressed as
\begin{align}
\rho_{\rm GW}(t) &= \frac{1}{32\pi GV} \int_{V}{\frac{d^3 \boldsymbol{{\rm k}}}{(2\pi)^3} \dot{h}_{ij}(\boldsymbol{{\rm k}},t) \dot{h}_{ij}^*(\boldsymbol{{\rm k}},t) } \\  &=  \frac{1}{32\pi G}\frac{1}{(2\pi)^3V} \int{ |\boldsymbol{{\rm k}}|^3{\rm d(ln}k)} \int{ {\rm d} \Omega |\dot{h}_{ij}(\boldsymbol{{\rm k}},t)|^2}   \\& \equiv \int{ \frac{{\rm d} \rho_{\rm GW}}{{\rm d}{\rm ln}k} {\rm d}{\rm ln}k} , 
\end{align}
where ${\rm d}\Omega$ represents a solid angle measure in momentum space.
The GW energy density power spectrum is defined as energy density per logarithmic interval and is typically normalized by the critical energy density, $\rho_{c}\equiv3H^2/8\pi G$.  The dimensionless GW power spectrum is finally expressed as
\begin{align} 
\Omega_{\rm GW} &\equiv \frac{1}{\rho_{c}} \frac{{\rm d}\rho_{\rm GW}}{{\rm dln}k} \\ &= \frac{1}{32\pi G}\frac{1}{(2\pi)^3V} \int{ {\rm d} \Omega |\boldsymbol{{\rm k}}|^3 |\dot{h}_{ij}(\boldsymbol{{\rm k}},t)|^2}.
\end{align}
where ${\rm d}\Omega$ denotes a solid angle measure.

After getting the GW power spectrum at the end time of our simulation, we need to convert it into today's spectrum. We follow the scheme described in \cite{Price:2008hq,Dufaux:2007pt,Easther:2007vj}.  Firstly, due to the increase in spatial volume and the decrease in GW frequency, the GW energy density scales as
\begin{align} 
\frac{\rho_{\rm GW,0}}{\rho_{\rm GW,e}}=\frac{R_{\rm e}^4}{R_{0}^4}.
\end{align}
The subscripts "0" and "e" denote the quantities today and at the end of our simulation respectively, and the same is true for the following parts. We assume that entropy is always conserved, so the radiation energy density satisfies
\begin{align} 
\frac{\rho_{\rm rad,0}}{\rho_{\rm rad,e}}=\frac{R_{\rm e}^4}{R_{0}^4}\frac{g_{\rm e}^{1/3}}{g_{0}^{1/3}},
\end{align}
with $g_{\rm e}(g_{0})$ the number of relativistic degrees of freedom at the end time of simulation (today). For example, we take $g_{\rm e}=81$ at the end of the simulation in the PQ era, while $g_{0}=3.36$. In the radiation-dominated era, the radiation energy density at the end of the simulation is approximately equal to the critical energy density, i.e., $\rho_{\rm rad,e}\approx \rho_{c,\rm e}=3H_{\rm e}^2/8\pi G$. Next, we also need to know the frequency of GWs today, which is
\begin{align} 
f_{0}&=\frac{k_{\rm p,0}}{2\pi}=\frac{k_{\rm co,e}}{2\pi R_{\rm e}}\frac{R_{\rm e}}{R_{0}}=\frac{k_{\rm co,e}}{2\pi R_{\rm e}}\left(\frac{\rho_{\rm rad,0}}{\rho_{\rm rad,e}}\right)^{1/4}\left(\frac{g_0}{g_{\rm e}}\right)^{1/12} \nonumber \\ &\approx \frac{k_{\rm co,e}}{R_{\rm e}\rho_{c,\rm e}^{1/4}}\left(\frac{g_{\rm 0}}{g_{\rm e}}\right)^{1/12} \times (5.87\times10^{10} {\rm Hz})
\end{align}
where $k_{\rm p,0}$ denote physical momentum today and $k_{\rm co,e}$ denote comoving momentum at the end of our simulation. 

The GW power spectrum today can be expressed as 
\begin{align} 
\Omega_{\rm GW,0}h^2 &= \frac{h^2}{\rho_{c,\rm 0}} \frac{{\rm d}\rho_{\rm GW,0}}{{\rm dln}k_{\rm p,0}} \nonumber \\ &= \Omega_{\rm rad,0}h^2\left(\frac{g_{\rm 0}}{g_{\rm e}}\right)^{1/3} \left\{ \frac{1}{\rho_{c,\rm e}}\frac{{\rm d}\rho_{\rm GW,e}}{{\rm dln}k_{\rm co,e}} \right\},
\end{align}
where $\Omega_{\rm rad,0}h^2=h^2\rho_{\rm rad,0}/\rho_{c,\rm 0}\approx4.3\times10^{-5}$ is the abundance of radiation today. Note that we multiplied by an additional factor $h^{2}$, to take into account the uncertainty of Hubble expansion rate today.

\begin{figure*}[!htp]
\hspace{1mm}
\includegraphics[width=.4\textwidth]{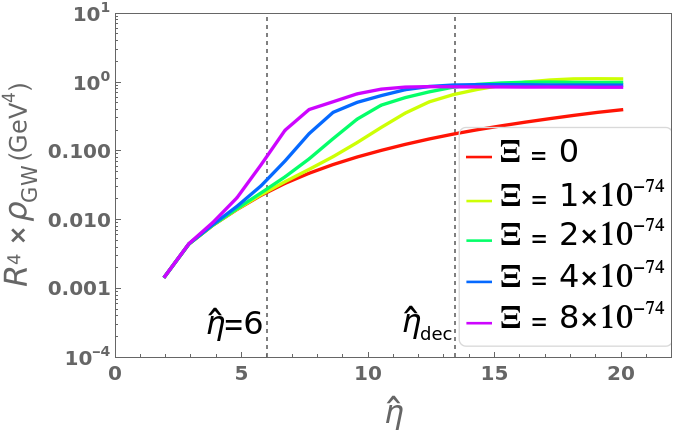}
\hspace{1mm}
\includegraphics[width=.4\textwidth]{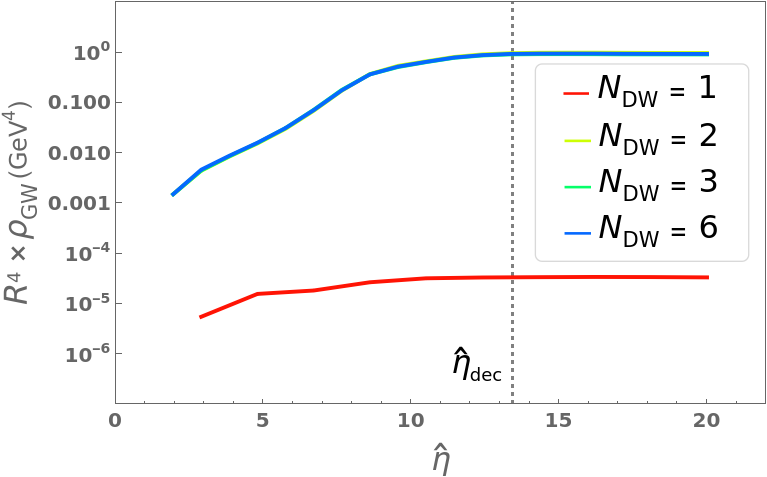}
  \caption{ Left panel: Time evolution of rescaled GW energy density $R^4 \times \rho_{\rm GW}$ with different coefficients of bias term in the case of $N_{\rm DW}=3$. Right panel: Time evolution of $R^4 \times \rho_{\rm GW}$ with different $N_{\rm DW}$. We set $\Xi=0$ for $N_{\rm DW}=1$ and $\Xi=4\times 10^{-74}$ for $N_{\rm DW}>1$.}
  \vspace{0.1cm} \label{fig:epsilongw}
\end{figure*}

\begin{figure*}[!htp]
\includegraphics[width=.4\textwidth]{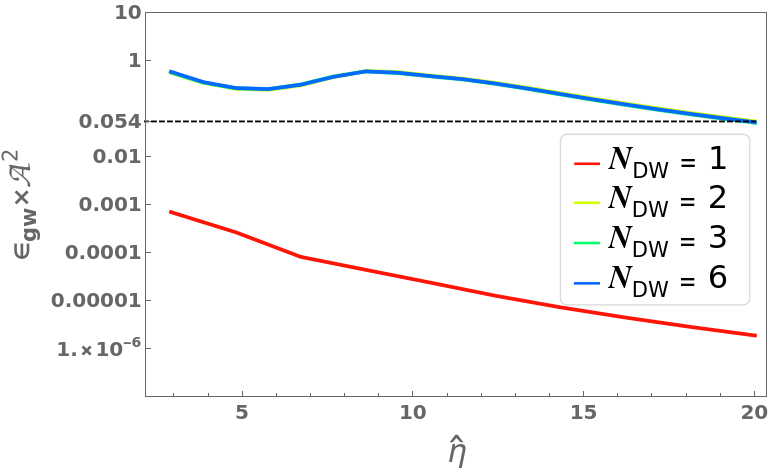}
\includegraphics[width=0.4\textwidth]{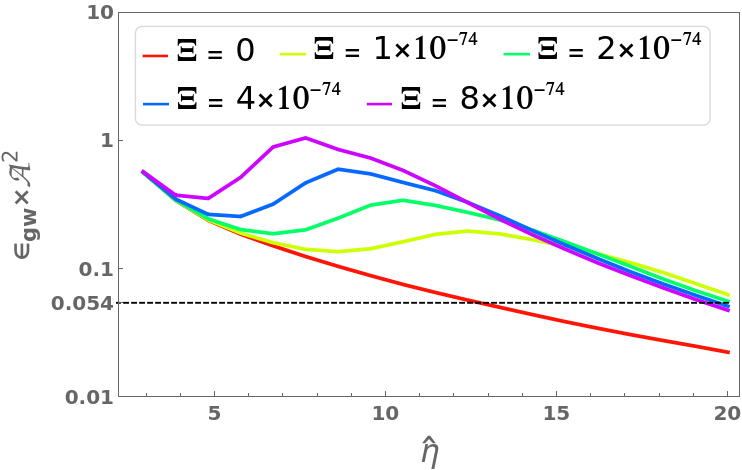}
  \caption{ Left panel: Time evolution of $\epsilon_{\rm gw } \times \mathcal{A}^2$ with different $N_{\rm DW}$. 
Right panel: Multiplication of GW efficiency parameter ($\epsilon_{\rm gw}$) and the square of area parameter of DW ($\mathcal{A}^2$) as a function of the bias term $\Xi$ in the case of $N_{\rm DW}=3$. }
  \vspace{0.1cm} \label{fig:epsilongwA2_NDW}
\end{figure*}

In Fig.~\ref{fig:epsilongw}, we plot the evolution of rescaled GW energy density, where we multiply $R^4$ to counteract the decrease in the energy density of GWs due to the expansion of the universe. Note, $\rho_{\rm GW}$ denote the dimensionless GW energy density here. We found that the energy density of GWs caused by different $\Xi$ ($\Xi>0$) will tend to the same value when the DW decays completely at $\hat{\eta}_{\rm dec}$. This means that the GW energy density at the final time of the simulation does not depend on $\Xi$ (for $\Xi>0$). In right panel of Fig.~\ref{fig:epsilongw}, we plot the time evolution of $R^4 \times \rho_{\rm GW}$ with different $N_{\rm DW}$, it can be found that the rescaled GW energy density $R^4\times \rho_{\rm GW}$ is almost independent of $N_{\rm DW}$ in the $N_{\rm DW}>1$ case. 

The results here indeed can be traced back to the behavior of the combined effects of the GW efficiency parameter and the area parameter of DW with
\begin{equation}
\epsilon_{\rm gw} \mathcal{A}^2 = \frac{\rho_{\rm gw}}{G\sigma_{\rm wall}^2},
\end{equation}
where $\rho_{\rm gw}$ denote the GW energy density, G is the Newton's gravitational constant, $\sigma_{\rm wall}=8m_0f_a^2$ is the surface mass density of DWs. We plotted the the time evolution of $\epsilon_{\rm gw}\times \mathcal{A}^2$ in different $N_{\rm DW}$ and $\Xi$ cases, see Fig.~\ref{fig:epsilongwA2_NDW}. We found that for $N_{\rm DW}>1$, the three curves in the left panel almost completely overlap, indicating that $\epsilon_{\rm gw}\times \mathcal{A}^2$ is almost independent of $N_{\rm DW}$. And that, our results show that $\epsilon_{\rm gw}\times \mathcal{A}^2$ is almost independent of the $\Xi$ for $\Xi>0$. This can explain why the GW spectrum does not depend on $N_{\rm DW}$ (for $N_{\rm DW}>1$) and $\Xi$ (for $\Xi>0$)  in FIG.~\ref{fig:QCD_GW_spectra} in the main text.

\begin{figure*}[!htp]
\includegraphics[width=0.4\textwidth]{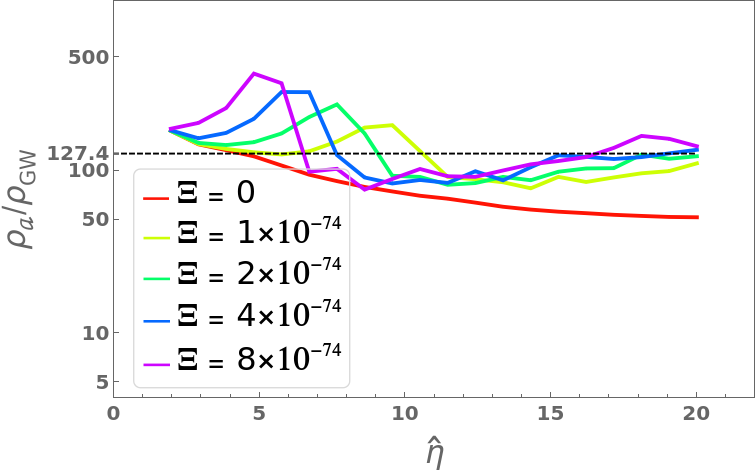}
\includegraphics[width=.4\textwidth]{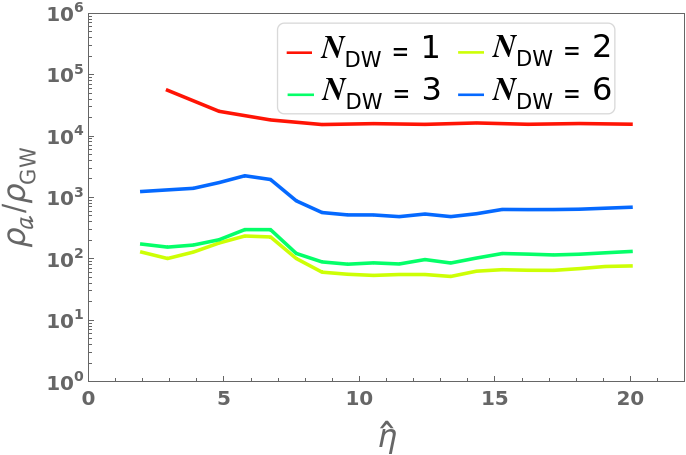}  
  \caption{ Left panel: The ratio of axion energy density $\rho_{a}$ to GW energy density $\rho_{\rm GW}$ for different bias term in the case of $N_{\rm DW}=3$; Right: Time evolution of the ratio of $\rho_a$ of $\rho_{\rm GW}$ for different $N_{\rm DW}>1$ for a fixed biased term $\Xi=4\times10^{-74}$ and $\Xi=0$ for $N_{\rm DW}=1$.}
\vspace{0.1cm}\label{fig:rho_arho_gw}
\end{figure*}

Then, we recorded the dependence of the ratio of axion energy density to GW energy density on $\Xi$ (see left panel of FIG.~\ref{fig:rho_arho_gw}). We found that for different $\Xi>0$, the final ratio of the two will tend towards a constant (127.4), indicating that the final ratio of the two is (almost) independent of $\Xi$, and the decay to axion is the main radiation channel of axion string-wall networks. Therefore, we have $\rho_a \simeq 127.4\rho_{\rm gw}$ after DWs decay.
We also found that the axion energy density is always much larger than the GW energy density throughout our simulations for different $N_{\rm DW}$. This means that for arbitrary $N_{\rm DW}$, the string-wall networks decay mainly by radiating axions.

\section{Gravitational waves in the case of $N_{\rm DW}=1$ and $N_{\rm DW}>1$ in QCD axion scenario}

In this section, we investigated the GW power spectrum obtained in the QCD axion scenario with $N_{\rm DW}=1$ and $N_{\rm DW}>1$. We found that due to stricter theoretical and experimental constraints on the parameter space of QCD axions compared to axion-like particles, GWs are too weak to be observed.  

In the $N_{\rm DW}=1$ case, the GWs are generally expected to be dominated by strings. For the sake of rigor, we investigated the GWs radiated by both pure string networks and string-wall hybrid networks, respectively. To study the GWs radiated by pure strings, we specifically prepared an additional PQ era. For this additional PQ era, we set the PQ vacuum $v_{\rm add}=6\times10^{16}$ ${\rm GeV}$ to ensure the simulation box can capture large enough Hubble volumes, and our simulation begins at an initial temperature $T_{i}=18T_{\rm c}$. The scalar fields are rescaled with the PQ vacuum $v_{\rm add}$, and the comoving lattice spacing $\delta x$ (and time-step) are rescaled by $w_*=R_iH_i$ with the $R_i$ and $H_i$ being initial scale factor and Hubble parameter. It is convenient to use the rescaled conformal time $\tilde{\eta}$ $(=\eta/\eta_i)$ with the initial conformal time being $\eta_i=1/w_*$ in our simulations. We fix $\tilde{\eta}=1$ to be the initial time at which $T=T_{i}$, so the PQ phase transition happens at $\tilde{\eta}=18$. We evolve the equations of motion in a simulation box of comoving side-length $L_{\rm PQ}=440/(R_{i}H_{i})$ and 1600 points per side. So, the dimensionless comoving lattice spacing is $\delta \Tilde{x}=0.275$, and the time-step is chosen as $\delta \tilde{\eta}=0.008$. We continued to evolve the string network until the final moment $\tilde{\eta}_{f}=270$. At the end time, the simulation box contains 4 Hubble volumes to reduce the finite volume effects, and the physical string width $\delta_{\rm st}\sim1/(\sqrt{\lambda}v_{\rm add})$
is about twice as large as the physical lattice spacing to obtain sufficient resolution (with $\lambda=0.2$).

\begin{figure*}[htb]
\includegraphics[width=.43\textwidth]{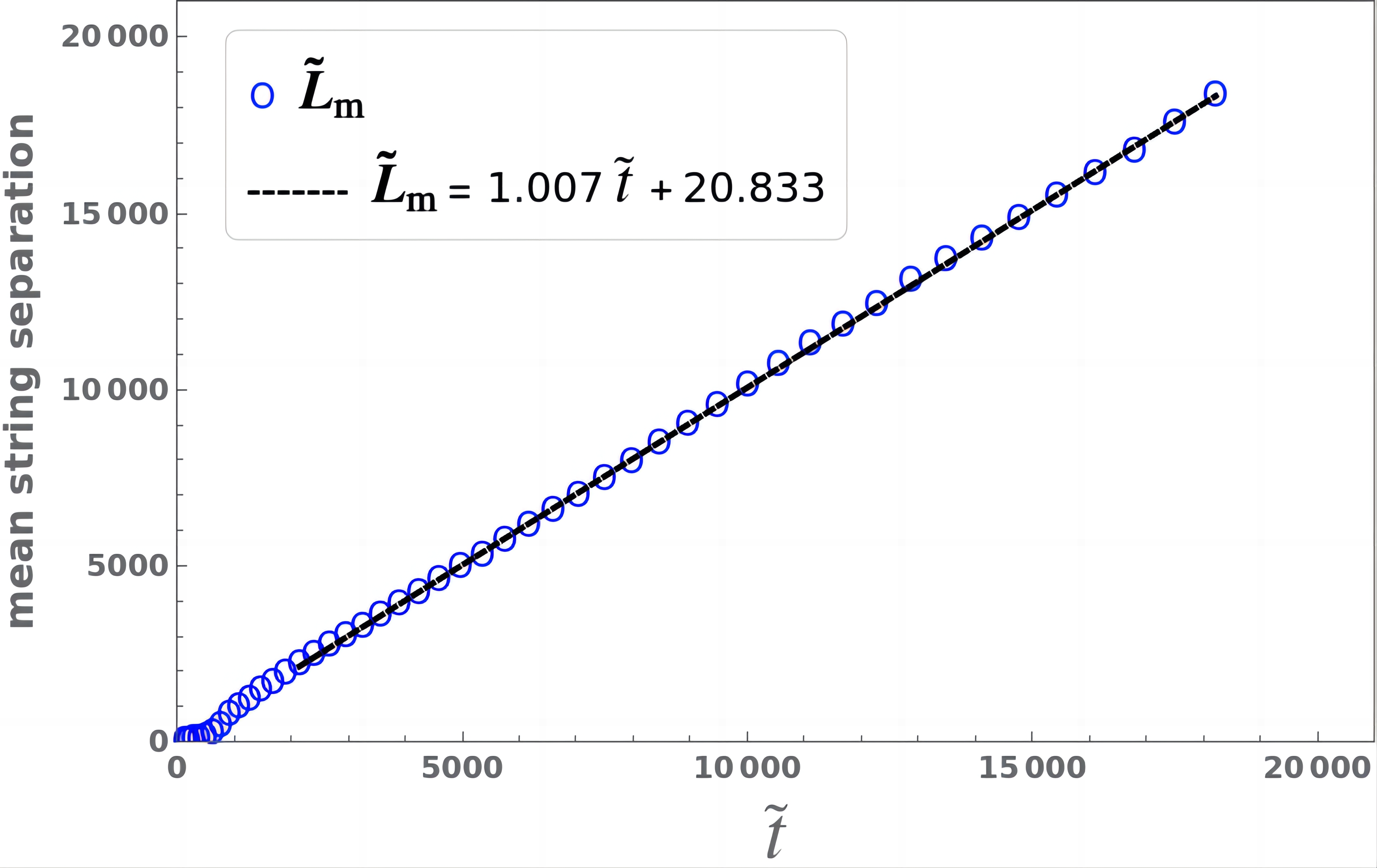} 
\hspace{10mm}
\includegraphics[width=0.42\textwidth]{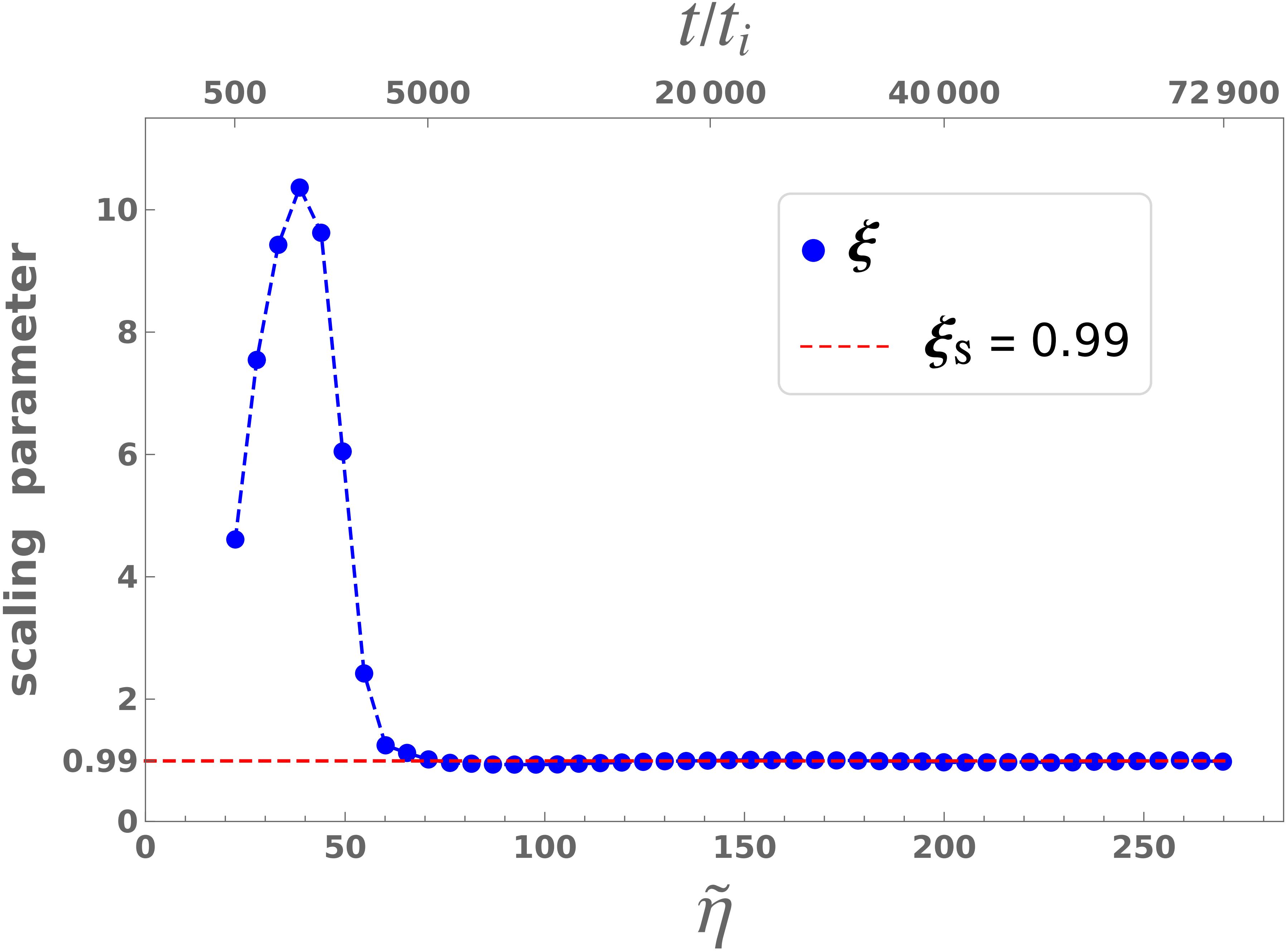}
  \caption{Left panel: Evolution of dimensionless mean string separation ${\tilde{L}_{\rm m}}$ with dimensionless cosmic time $\tilde{t}$ in the additional PQ era. The blue hollow points denote the dimensionless mean string separation calculated according to Eq.~(\ref{standard scaling model}). The black dashed line is a linear fit of the mean string separation and time after the string networks have entered the scaling regime. Right panel: Evolution of scaling parameters in rescaled conformal time $\tilde{\eta}$ and rescaled cosmic time $\tilde{t}=t/t_i$, with $t_i=1/(2H_i)$ the initial cosmic time. The blue solid points denote the scaling parameter calculated according to Eq.~(\ref{traditional way}), and the red dashed line is the scaling parameter obtained by global fitting the mean string separation with cosmic time after $\tilde{\eta}=90$ according to the standard scaling model.}
  \vspace{0.1cm}  
  \label{fig: additional PQ sc}
\end{figure*}

We use both the definition in Eq.~(\ref{traditional way}) as well as the standard scaling model Eq.~(\ref{standard scaling model}) based on mean string separation to calculate the scaling parameter and achieve consistent results. We first measured the time evolution of mean string separation during the additional PQ era, see left panel of Fig.~\ref{fig: additional PQ sc}. After a certain period of evolution, the mean string separation increases linearly with cosmic time, indicating that the string networks have entered the scaling regime. By linear fitting, the slope is $\kappa=1.007$. So the scaling parameter is $\xi_{\rm s}=1/\kappa^2=0.99$, coherent with the value of scaling parameters calculated by Eq.~(\ref{traditional way}), see right panel of Fig.~\ref{fig: additional PQ sc}. The scaling parameters measured under two measurement methods do not exhibit logarithmic increase, consistent with previous studies~\cite{Yamaguchi:1998gx,Yamaguchi:2002sh,Hiramatsu:2010yu,Hiramatsu:2012gg,Kawasaki:2014sqa,Lopez-Eiguren:2017dmc}.

\begin{figure*}[htb]
\includegraphics[width=.45\textwidth]{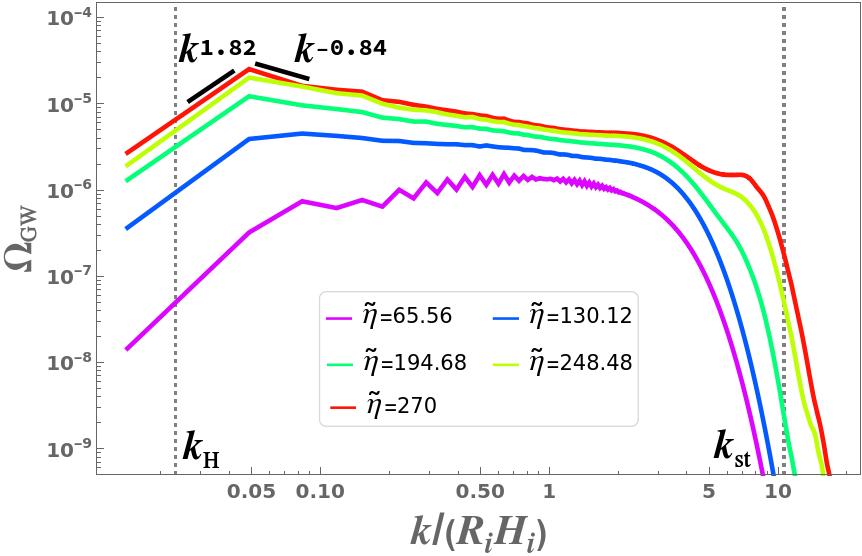} 
\hspace{8mm}
  \caption{The GW spectrum measured at five different times in the additional PQ era. }
  \vspace{0.1cm}
  \label{fig:PQGW}
\end{figure*}

We then measured the GW spectrum in the additional PQ era, see FIG.~\ref{fig:PQGW}. The GW spectrum changes its slope at two characteristic scales, the Hubble length and the string width at the final time, $k_{\rm H}=2\pi H({\tilde{\eta}_{f}})\times R({\tilde{\eta}_{f}})/(R_i H_i)$ and $k_{\rm st}=2\pi/\delta_{\rm st}(\tilde{\eta}_{f}) \times R({\tilde{\eta}_{f}})/(R_i H_i)$, with $\delta_{\rm st}(\tilde{\eta}_{f})$ the physical string width. After redshift, the peak amplitude of GW spectrum today is $(\Omega_{\rm GW}h^2)_{\rm add}=3.81\times 10^{-10}$. For the global (axion) string, the GW spectrum today will follow $\Omega_{\rm GW}h^2 \propto (v/M_{\rm pl})^4$ according to analytic calculations ~\cite{Fenu:2009qf}, with $M_{\rm pl}$ the full Planck mass. We will then compare this peak amplitude with the peak amplitude of GWs radiated by string-wall hybrid networks with $N_{\rm DW}=1$ in the QCD era, thus verifying whether GWs are dominated by pure strings.

In the QCD era with $N_{\rm DW}=1$, the string-wall network quickly collapses when $H=m(T)$ is satisfied due to the tension of domain walls. The PQ vacuum is reinterpreted as $v_{\rm re}=3.5\times 10^{16}$ GeV. Then, from the lower panel of FIG. 4 in the main text, the peak amplitude of GW spectrum today in the $N_{\rm DW}=1$ case is $(\Omega_{\rm GW}h^2)_{\rm QCD}=3.13\times 10^{-12}$. If the GW spectrum in the $N_{\rm DW}=1$ case is dominated by pure strings, then $(\Omega_{\rm GW}h^2)_{\rm add}/(\Omega_{\rm GW}h^2)_{\rm QCD}\approx (v_{\rm add}/v_{\rm re})^4$ should be satisfied. We numerically checked that $(\Omega_{\rm GW}h^2)_{\rm add}/(\Omega_{\rm GW}h^2)_{\rm QCD} \approx (v_{\rm add}/v_{\rm re})^4$, so $\Omega_{\rm GW}h^2 \propto (v/M_{\rm pl})^4$ is approximately satisfied, indicating that the GWs is dominated by strings. We require that the total axion abundance, which is contributed by the coherent oscillation of the homogeneous axion field, the production of axions from pure axion strings, and the DW decay, does not exceed the abundance of CDM observed today, $\Omega_a(t_0)h^2\leq \Omega_{\rm CDM}h^2=0.12$, this lead to an upper bound on $f_a$, $f_a\leq7.2\times 10^{10}$GeV\cite{Kawasaki:2014sqa} in the $N_{\rm DW}=1$ case. Thus the peak amplitude of GW spectrum satisfies $(\Omega_{\rm GW} h^2)_{\rm max}$$\leq(7.2\times10^{10}$GeV$/v_{\rm re})^4* 3.13\times10^{-12}$$<5.6\times10^{-35}$. So, the amplitude of GWs is too small to be detected by GW detectors in the case of $N_{\rm DW}=1$.

In the $N_{\rm DW}>1$ case, the GWs are dominated by long-lived DWs. The axion decay constant $f_a$ is reinterpreted as $f_{ a,{\rm re}} = 3.5 \times 10^{16}$ GeV. From the lower panel of FIG. 4 in the main text, the peak amplitude of GW spectrum today in the $N_{\rm DW}>1$ case is $(\Omega_{\rm GW}h^2)_{\rm QCD} \simeq 6.71\times 10^{-8}$. The GW energy density and spectrum are
proportional to the square of DW tension, $\Omega_{\rm GW} \propto \sigma_{\rm wall}^2 \propto m^2f_a^4$. For QCD axion, the axion mass reaches its maximum at 100 MeV, that is, the zero temperature mass, $m_0 = 5.707 \times 10^{-5} ( 10^{11} \, \, {\rm GeV}/ f_a)$ eV~\cite{GrillidiCortona:2015jxo}. So, we have $\Omega_{\rm GW} \propto m_0^2f_a^4 \propto f_a^2$. Correspondingly, the abundance of CDM contributed by axions does not exceed $\Omega_{\rm CDM}h^2=0.12$, resulting in a stricter upper bound on $f_a$, $f_a\leq 5.1\times 10^{9}$GeV. Therefore, the peak amplitude of GW spectrum satisfies $(\Omega_{\rm GW} h^2)_{\rm max}$$<(5.1\times10^{9}$GeV$/f_{a,{\rm re}})^2 * 6.71\times10^{-8}$$<1.42\times10^{-21}$. So, the amplitude of GWs is also too weak to be detected.

In the ALPs scenario with $N_{\rm DW}=1$, similar to the QCD axion scenario with $N_{\rm DW}=1$, the GWs are dominated by strings and we also have an upper bound on $f_a$, $f_a\leq 7.2\times 10^{10}$ GeV. So, finally, the peak amplitude of the GW spectrum satisfies $(\Omega_{\rm GW} h^2)_{\rm max}$$<5.6\times10^{-35}$, which is too weak to be probed.


\end{document}